\documentclass[fleqn,usenatbib,referee]{mnras}
\usepackage[T1]{fontenc}
\usepackage{bm}
\DeclareRobustCommand{\VAN}[3]{#2}
\let\VANthebibliography\thebibliography
\def\thebibliography{\DeclareRobustCommand{\VAN}[3]{##3}\VANthebibliography}
\usepackage[dvips]{graphicx}	
\usepackage{amsmath}	
\usepackage{amssymb}	
\usepackage{color}
\usepackage{ulem}
\newcommand{\msun}{{\rm M}_\odot}

\title[Implementation of dust particles]{
Implementation of dust particles in three-dimensional magnetohydrodynamics simulation: Dust dynamics in a collapsing cloud core 
}
\author[Koga et al.]{
Shunta Koga$^{1}$\thanks{E-mail: koga.shunta.138@s.kyushu-u.ac.jp}, Yoshihiro Kawasaki$^{1}$ and Masahiro N. Machida$^{1}$
\\
$^{1}$Department of Earth and Planetary Sciences, Faculty of Sciences, Kyushu University, Fukuoka 819-0395, Japan\\
}

\begin{document}
\label{firstpage}
\pagerange{\pageref{firstpage}--\pageref{lastpage}}
\maketitle

\begin{abstract}
The aim of this study is to examine dust dynamics on a large scale and investigate the coupling of dust with gas fluid in the star formation process.
We propose a method for calculating the dust trajectory in a gravitationally collapsing cloud, where the dust grains are treated as Lagrangian particles and are assumed to be neutral.
We perform the dust trajectory calculations in combination with non-ideal magnetohydrodynamics simulation.
Our simulation shows that dust particles with a size of $\le 10\,{\rm \mu m}$ are coupled with gas in a star-forming cloud core.
We investigate the time evolution of the dust-to-gas mass ratio and the Stokes number, which is defined as the stopping time normalized by the freefall time-scale, and show that large dust grains ($\gtrsim 100\,{\rm \mu m}$) have a large Stokes number (close to unity) and tend to concentrate in the central region (i.e., protostar and rotationally supported disk) faster than do small grains ($\lesssim 10\,{\rm \mu m}$).
Thus, large grains significantly increase the dust-to-gas mass ratio around and inside the disk. 
We also confirm that the dust trajectory calculations, which trace the physical quantities of each dust particle, reproduce previously reported results obtained using the Eulerian approach. 
\end{abstract}



\section{Introduction}
\label{sec:intro}
Stars form in molecular cloud cores composed of gas and dust grains. 
It is considered that planet embryos, which are aggregates of dust grains, appear within the rotationally supported disks that form in the star formation process. 
The star formation process has been investigated in both theoretical  and observational studies. 
Recent ALMA observations have shed new light on star and planet formation processes \citep[e.g.,][]{2014ApJ...786..114L} and imply  the onset of planet formation at the early star formation stage.
Theoretical studies on the early stage of star formation using three-dimensional magnetohydrodynamics (MHD) simulations have clarified the formation process of  protostars and circumstellar disks in molecular cloud cores \citep[e.g.,][]{2007MNRAS.377...77P, 2008A&A...477....9H, 2011MNRAS.413.2767M, 2015MNRAS.452..278T, 2015ApJ...801..117T, 2016MNRAS.457.1037W, 2016A&A...587A..32M}. 

Although gas dynamics in the star formation process has been well investigated in such studies, dust dynamics are poorly understood because of the uncertainty of dust properties. 
Some physical quantities, such as dust grain size and its distribution, determine dust properties. 
Based on interstellar extinction observations,  \citet{1977ApJ...217..425M} proposed a dust grain size distribution of $0.005 \ {\rm \mu m} \le a_{\rm d} \le 0.25 \ {\rm \mu m}$ (called the MRN size distribution), which is commonly used in star and planet formation studies. 

A very recent observation \citep{2021ApJ...915...74U} confirmed that the MRN size distribution is appropriate for the Orion A molecular cloud. 
However,  near-infrared scattered light observations  indicate that micrometer-sized dust grains may exist in molecular cloud cores \citep{2010Sci...329.1622P,2014A&A...563A.106S,2015A&A...582A..70S}.
There is also evidence for the existence of large dust grains.
The dust opacity spectral index $\beta (\equiv d \, {\rm ln} \, \kappa / d \, {\rm ln} \, \nu$, where $\kappa$ and $\nu$ are the opacity of dust grains and the observational frequency, respectively) estimated from multiple-wavelength observations was measured for molecular cloud cores \citep[e.g.,][]{2012A&A...538A.137M} and gas envelopes around Class 0/I young stellar objects (YSOs)  \citep[e.g.,][]{2009ApJ...696..841K, 2014A&A...567A..32M, 2017ApJ...840...72L, 2019A&A...632A...5G}. 
The results showed that the $\beta$ value for molecular cloud cores and Class 0/I objects is lower than that for the interstellar medium, indicating the existence of  (sub-)millimeter dust grains in star-forming clouds.
Thus, although interstellar extinction observations imply that a  dust size range of  $0.005 \ {\rm \mu m} \le a_{\rm d} \le 0.25 \ {\rm \mu m}$,  other observations indicate the existence of millimeter-sized dust grains. 
These observations may support  the possibility of dust growth  in star-forming cores.

The dust-to-gas mass ratio (hereafter $f_{\rm dg}$) is another important quantity that determines dust properties.
$f_{\rm dg}$ represents the dust concentration relative to that of gas and  is usually used as a parameter in star and planet formation studies.
For example, in some observational studies, the gas mass is estimated from the dust continuum emission \citep[e.g.,][]{1983QJRAS..24..267H} under the assumption that the gas mass is proportional to the dust mass (or the intensity of the dust emission).
In some theoretical studies that focus on dust growth in a protoplanetary disk,  $f_{\rm dg}$ is used as a crucial  parameter for determining the initial dust mass \citep[e.g.,][]{2008A&A...480..859B}.
Conventionally, $f_{\rm dg} = 0.01$ is adopted in star and planet formation studies. 
This value is supported by interstellar extinction observations \citep{1977ApJ...217..425M} and is consistent with the amount of solid matter in the solar system \citep{1981PThPS..70...35H}.

\citet{2021ApJ...908..112B} used two-dimensional MHD simulations to investigate dust dynamics in cases where large-scale filaments form just prior to the onset of star formation.
They  found  that  charged dust grains concentrate along the high-density filaments and that $f_{\rm dg}$ is almost an order of magnitude higher in such regions. 
Moreover, \citet{2016ApJ...828...46A} reported that the $f_{\rm dg}$ values obtained from the dust continuum and CO line emissions are in the range of 0.001 to 0.1 for the protoplanetary disks in their  disk survey observation in the Lupus star-forming region.
Thus, recent theoretical and observational studies, which imply dust growth in the early star formation process, do not strongly support $f_{\rm dg}=0.01$ in the protostar formation stage just prior to planet formation. 

The dust grain size distribution and dust-to-gas mass ratio significantly influence star and planet formation processes.
For example, these dust parameters determine the chemical abundances of charged species.
Dust grains absorb the charged particles produced by collisions between neutral particles and  cosmic rays and  the charge transfer between dust grains and ions.  
The grain size and dust-to-gas mass ratio are necessary for calculating the cross sections of dust grains, which determine the absorption rate. 
The chemical abundances are directly related to the resistivities of non-ideal MHD effects and determine the evolution of the magnetic field.
Therefore, dust properties are important for studying star formation.
The influences of dust properties on star and circumstellar disk formation have been investigated in  theoretical studies in terms of the magnetic dissipation process  \citep{2016A&A...592A..18M, 2016MNRAS.460.2050Z, 2018MNRAS.478.2723Z, 2021MNRAS.505.5142Z, 2017A&A...603A.105D, 2019MNRAS.484.2119K, 2020ApJ...896..158T}.
\citet{2016MNRAS.460.2050Z} showed that the dust size distribution or amount of small grains greatly affects the size and structure of the circumstellar disk.

Dust size is also important  for studying planet formation.
Dust size growth via collisions is the first step of planet formation. The growth rate is determined by the amount and size of dust.
For the classical planet formation scenario \citep[e.g.,][]{1985prpl.conf.1100H}, dust growth has been discussed theoretically in the minimum mass solar nebula \citep[][]{1981PThPS..70...35H}.
However, the planet-forming disks around Class II YSOs have complicated structures such as rings, gaps, and spiral arms, as reported by DSHARP (Disk Substructures at High Angular Resolution Project) \citep[][]{2018ApJ...869L..42H,2018ApJ...869L..43H,2018ApJ...869L..44K,2018ApJ...869L..48G,2018ApJ...869L..49I,2018ApJ...869L..50P}.
In addition, the disks around Class 0/I YSOs  have similar substructures \citep[][]{2020ApJ...902..141S}. 
These observations imply that planet formation begins in Class 0 and I stages. 
Thus, the planet formation scenario should be renewed  and considered in the framework of star formation.
Because star formation starts from a molecular cloud core, dust growth in the collapsing cloud core should be considered.
It is thus valuable to clarify the dust size (or dust growth) and dust-to-gas mass ratio  in the circumstellar disk around  very young protostars during the main accretion phase. 

Very recently, dust dynamics during the early star formation stage were investigated using multi-dimensional simulations 
 \citep[][]{2017MNRAS.465.1089B,2018A&A...614A..98V,2020A&A...641A.112L, 2021ApJ...920L..35T}.
\citet{2017MNRAS.465.1089B} calculated dust motion using three-dimensional smoothed particle hydrodynamics simulation  and found  that dust grains with a size of $\leq 10 \ {\rm \mu m}$ are coupled with gas during the protostellar collapse phase.
\citet[][]{2018A&A...614A..98V} investigated dust dynamics and growth using two-dimensional multi-fluid simulation  and showed that dust grains concentrate in viscous and self-gravitating disks.
It should be noted that the magnetic effects were ignored in these studies.

Only two studies have considered the effects of both dust (dynamics) and the magnetic field in a star formation simulation.
Using three-dimensional MHD simulations that included dust, in which fluid approximation was adopted for calculating dust dynamics,  \citet[][]{2020A&A...641A.112L} presented the condition for the decoupling of dust from gas (dust size larger than $\geq 100 \ {\rm \mu m}$). 
Their result is consistent with that of \citet{2017MNRAS.465.1089B}.
\citet{2021ApJ...920L..35T} took up the challenge of modeling dust growth in their three-dimensional MHD simulation  and considered various physical processes.
They showed that outflow helps dust growth and pointed out that dust grains grow to centimeter ($\sim$\,cm) size.
Both studies also showed  the spatial distribution of dust in a star-forming cloud. 
However, the motion of individual dust particle could not be traced because fluid approximations were adopted  \citep{2020A&A...641A.112L, 2021ApJ...920L..35T}.

The aim of  this series of studies is to trace the trajectory of individual dust particle representing a dust grain population in the protostar and disk formation processes and examine the dust dynamics  in the early star formation stage based on the trajectory history of individual dust particle. 
For this purpose, we proposed a method and implement it in our three-dimensional MHD nested grid code. 
This paper focuses on the method for treating dust dynamics in three-dimensional MHD simulation code and shows a comparison our results with those in previous studies. 
We clarify the dust spatial distribution on a large scale in the early star formation stage.
Dust growth in a circumstellar disk during the main accretion phase will be discussed in our next paper.

The remainder of this paper is organized as follows. 
We describe the method for calculating gas and dust in \S~\ref{sec:method}.  
We present the dust motion obtained from the simulation and compare it  with gas motion for various dust sizes in \S~\ref{sec:results}.
We compare our results with those in previous studies and discuss which effects should be included in future studies in \S~\ref{sec:discussion}.  We summarize our results in \S~\ref{sec:conclusion}.

\section{Method}
\label{sec:method}
\subsection{Basic equations and numerical settings of gas fluid}
\label{subsec:gas} 
The numerical settings and initial conditions are the same as those in \citet{2017ApJ...835L..11T} and \citet{2020ApJ...905..174A}. 
As described below, because we do not include the feedback effect of dust on gas (i.e. the back reaction or drag force from dust to gas), the gas dynamics shown in this paper are the same as those in our previous studies. 
Thus, we omit their details in this paper.
In this subsection, we give the numerical settings adopted in this study and our previous studies. 

The gas evolution is calculated using three-dimensional MHD simulation.
The basic equations are as follows:
\begin{equation}
\frac{\partial \rho}{\partial t}+\nabla \cdot(\rho \boldsymbol{v})=0,
\end{equation}
\begin{equation}
\label{emgas}
\rho \frac{\partial v}{\partial t}+\rho(\boldsymbol{v} \cdot \nabla) \boldsymbol{v}=-\nabla P-\frac{1}{4 \pi} \boldsymbol{B} \times(\nabla \times \boldsymbol{B})-\rho \nabla \phi,
\end{equation}
\begin{equation}
\frac{\partial \boldsymbol{B}}{\partial t}=\nabla \times(\boldsymbol{v} \times \boldsymbol{B})+\eta_{\rm O} \nabla^{2} \boldsymbol{B},
\label{eq:inductioneq}
\end{equation}
\begin{equation}
\nabla^{2} \phi=4 \pi G \rho,
\end{equation}
where $\rho, \boldsymbol{v}, P, \boldsymbol{B}, \phi, \eta_{\rm O}$ are the gas mass density, gas velocity, gas pressure, magnetic field, gravitational potential, and ohmic dissipation resistivity coefficient, respectively.
The gas pressure $P$ is given by the following barotropic equation:
\begin{equation}
\label{eq:pressure}
    P = c_{s,0}^2 \left(\rho + \rho_{\rm cri} \left(\frac{\rho}{\rho_{\rm cri}} \right)^\gamma \tanh\left(\frac{\rho}{\rho_{\rm cri}} \right)^{0.1} \right),
\end{equation}
where $c_{s,0} = 1.9 \times 10^4 \ {\rm cm \ s^{-1}}$ (speed of sound at gas temperature $T = 10 \ {\rm K}$), $\gamma = 1.4$, and $\rho_{\rm cri} = 2.0 \times 10^{-14} \ {\rm g \ cm^{-3}}$ ($n_{\rm cri} = 5.0 \times 10^9 \ {\rm cm^{-3}}$) are adopted.
The ohmic dissipation coefficient $\eta_{\rm O}$, which was formulated in \citet{2007ApJ...670.1198M} based on \citet{2002ApJ...573..199N}, is
\begin{equation}
    \eta_{\rm O} = \frac{740}{x_e} \sqrt{\frac{T}{10 {\rm K}}} \ {\rm cm^2 s^{-1}},
\end{equation}
where $x_e$ represents the gas ionization degree and is calculated using the gas number density as
\begin{equation}
    x_e = 5.7 \times 10^{-4} \left(\frac{n}{\rm cm^{-3}} \right)^{-1}.
\end{equation}
Actually, the gas ionization degree $x_e$ depends on dust properties such as dust grain size and dust chemical composition \citep[e.g.][]{2019MNRAS.484.2119K}, as discussed in \S~\ref{subsec:eta}. 
It should be noted that we ignore the dependency of the ionization on the dust grain properties. 

To perform the numerical calculation,  we use a nested grid code \citep[for details, see][]{2004MNRAS.348L...1M, 2012MNRAS.421..588M, 2013MNRAS.431.1719M}.
The cell numbers for each grid are set to ($x,y,z$) = (64, 64, 64). 
We prepare 14 grid levels ($l$=1 to 14). 
The coarsest grid ($l=1$) has a box size of $L(l=1)=1.96\times10^5$\,au and a cell width of $h(l=1)=3.07\times10^3$\,au. 
The finest grid ($l=14$) has $L(l=14)=24.0$\,au and $h(l=14)=0.374$\,au. 
We introduce a sink  at the center of the computational domain  \citep{2010ApJ...724.1006M}.
We set the sink radius $r_{\rm sink}$ and threshold number density $n_{\rm sink}$ as $r_{\rm sink} = 1 \ {\rm au}$ and $n_{\rm sink} = 10^{13} \ {\rm cm^{-3}}$, respectively.

\subsection{Dust dynamics and calculation method}
\label{subsec:dust}
We aim to understand dust dynamics and gas evolution in the star formation process.
Thus, we introduce Lagrangian dust particles in our nested grid code.
In this subsection, we describe the method used to compute the dust particle motion.
In this study, we treat dust particles as solid particles that obey the equations of motion.
Gas is treated as an Eulerian fluid and dust is treated as Lagrangian particles.
This method allows the Lagrangian physical quantities to be traced and the evolution of the size distribution, chemical reactions, and temperature of each particle to be examined.
This treatment is different from that by \citet{2020A&A...641A.112L}, who calculated the dust dynamics using a single-fluid approach that included gas and dust. 
Our treatment is also different from \citet{2007ApJ...662..627J}, \citet{ 2010ApJ...722.1437B} and \citet{2021A&A...650A.119F}.
They calculated the motion of dust particles and gas fluid in the protoplanetary disk, in order to study streaming instability which is closely related to  planet formation  and can occur due to the interplay  between the gas and dust. 
We cannot investigate such instability   because of not including the feedback (or back reaction) from dust, while such a study is beyond the scope of the present work.

\subsubsection{The equation of motion of dust particles}
\label{subsubsec:dusteom}
We describe the method used to perform the dust trajectory calculation at each time step.
We consider the gas drag force and gas self-gravity. 
The equation of motion of a dust grain can be expressed as
\begin{equation}
\label{medust}
    \frac{d \bm{v}_{\rm d}}{d t} = -\frac{\bm{v}_{\rm d} - \bm{v}}{t_{\rm s}} + \bm{g}, 
\end{equation}
where $\bm{v}_{\rm d}$ is the velocity of dust particles.
In equation~(\ref{medust}),  $t_{\rm s}$ is the time-scale called stopping time, defined as
\begin{equation}
\label{tstop}
    t_{\rm s} = \frac{a_{\rm d} \rho_{\rm s}}{v_{\rm th} \rho},
\end{equation}
where $a_{\rm d}$ and $\rho_{\rm s}$ are the dust grain size and material density of a dust grain, respectively, and $v_{\rm th}$ is the thermal velocity of molecular gas, defined as $v_{\rm th} = \sqrt\frac{8}{\pi} c_s$, in which $c_s$ is the speed of sound. 
In this study, we assume that the dust grains are composed of ice and thus adopt $\rho_{\rm s} = 1 \ {\rm g \ cm^{-3}}$, as done in  \citet{2020A&A...641A.112L}.
In addition, we assume that dust grains are spherical solid particles. 
Thus, $a_{\rm d}$ corresponds to the radius of a given dust particle.
Note that the law of gas drag varies with dust grain size $a_{\rm d}$ and mean free path.
Also note that dust grains move according to Epstein's law for the settings in this study.
With some trial calculations, we have confirmed that the dust particles adopted in this study interact with the gas according only to Epstein's law. 
The transition between the two regimes (Epstein's and  Stokes' laws) is determined by the dust grain size $a_{\rm d}$ and the mean free path $\lambda = 1 / (n \sigma_{\rm mol})$, where  $n$ and $\sigma_{\rm mol} = 2.0 \times 10^{-15} \ {\rm cm^2}$ are the number density of gas and the collisional cross section of gas molecules, respectively.
If a dust particle satisfies the condition $a_{\rm d} < \frac{9}{4} \lambda$, it obeys Epstein's law and its stopping time is given by equation~(\ref{tstop}).
In this study, we introduce the sink particle and set the threshold gas number density as $n_{\rm sink} = 10^{13} \ {\rm cm^{-3}}$, as described in \S~\ref{subsec:gas}.
The maximum gas number density never exceeds  $n_{\rm sink}$ in the whole region.
Thus, there is no region that satisfies $n > 10^{13} \ {\rm cm^{-3}}$.
Therefore, the shortest mean free path $\lambda_{\rm min}$ appeared in the calculation is 
\begin{equation}
    \lambda_{\rm min} = \frac{1}{n_{\rm sink} \sigma_{\rm mol}} \approx 50 \ {\rm cm}.
\end{equation}
The maximum dust grain size is set to be $a_{\rm d} = 1000$\,$\mu$m (= 0.1cm) in this study, as described in \S\ref{subsubsec:dust}.
Thus, the dust grain size $a_{\rm d}$ is always much smaller than $\lambda_{\rm min}$ (see Table~\ref{table:dustdistribution}). 
Therefore, the dust particles obey the condition $a_{\rm d} < \frac{9}{4} \lambda$, and thus Epstein's law is always applicable. 
Moreover, dust particles in Epstein's law should satisfy subsonic relative motions between the dust particle and  gas.
However, some particles experience the supersonic relative velocity during the calculation.
Thus, we use the correction term adopted in \citet{2012MNRAS.420.2365L} and \citet{1975ApJ...198..583K}, which is described in \S\ref{subsubsec:dustnum}. 
Thus, the gas drag term can be calculated using equations (\ref{medust}) and (\ref{tstop}), which are described in \citet{1924PhRv...23..710E}.
In addition, we assume that dust particles do not interact with each other. 

To perform the trajectory calculation, we need to determine the gas physical quantities ($\rho,\bm{v},\bm{B},\bm{g}$) at the locations of the dust particles at each time step.
Here, $\bm{g} = - \nabla \phi$ is the gas gravitational acceleration and we ignore dust self-gravity.
The local gas physical quantities are acquired from the cells surrounding the dust particle. 
In our nested grid code, the gas physical quantities (e.g., $\rho$) are defined at the center of each cell.
For example, we calculate the local gas mass density $\rho_{\rm lc}$  at the location of a dust particle using the following linear interpolation formula 
\begin{equation}
    \rho_{\rm lc} = \rho_1 + \frac{x_{\rm d} - x_1}{h_{\rm cell}} \Delta \rho_x + \frac{y_{\rm d} - y_1}{h_{\rm cell}} \Delta \rho_y + \frac{z_{\rm d} - z_1}{h_{\rm cell}} \Delta \rho_z,
\end{equation}
where $h_{\rm cell}$ is the cell width and $\rho_1$ is the density of the cell within which the dust particle is included. 
Note that the cell width is the same in each direction ($x$, $y$, and $z$ directions). 
We define $A = (x, y, z)$. $A_{\rm d}$ and $A_1$ are the locations of the dust particle and the cell that includes the dust particle, respectively.
$\Delta \rho_A$ is defined as 
\begin{equation}
    \Delta \rho_A = S(\rho_{2,A} - \rho_1, A_{\rm d} - A_1),
\end{equation}
where $\rho_{2,A}$ is the nearest cell in each direction to  the location of the dust particle along the $A$-coordinate direction and the function $S$ is defined as
\begin{equation}
    S(a,b) = a \times sgn(b),
\end{equation}
where $sgn$ is the sign function.
We calculate the local quantity of $\bm{v},\bm{B},\bm{g}$ in the same way as for $\rho$ described above. 
The interpolation method adopted in this study was used in our past studies \citep{2014ApJ...784..109T,2020ApJ...903...98H}, in which the physical quantities at arbitrary point were derived with three-dimensional linear interpolation.  
Although there are many interpolation methods to have physical quantities at arbitrary point when quantities are discretely distributed, 
we chose this method to save the computational cost and exactly reproduce the results of past numerical and analytical studies \citep{2014ApJ...784..109T}.
Hereafter, the gas physical quantities indicate those at the location of a dust particle obtained using linear interpolation and we describe the local gas physical quantities with subscript "lc".

To calculate the dust trajectory, we need to integrate equation (\ref{medust}). 
Note that ${\bm v}_{\rm lc}$ is used instead of ${\bm v}$ in equation (\ref{medust}). 
When dust is strongly coupled with gas, $\Delta t_{\rm d} / t_{\rm s}$ is close to infinity, where $\Delta t_{\rm d}$ is the numerical time step used in the trajectory calculation. 
Thus, in a strongly coupled region, the time step in the orbit (or trajectory) calculation becomes very short, which makes the calculation  very difficult.
To avoid this difficulty, we analytically determine the relative velocity between the dust and gas at the next time step, as described in the next subsection.

\subsubsection{Relative velocity and numerical implementation}
\label{subsubsec:dustnum}
According to the prescription described in \citet{2014MNRAS.440.2147L} and \citet{2021ApJ...913..148T}, the equation of the time evolution of relative velocity $\Delta \bm{v} = \bm{v}_{\rm d} - \bm{v}_{\rm lc}$ is adopted.
We can describe the equation of  the relative velocity as
\begin{equation}
\label{emrelv}
    \frac{d \Delta \bm{v}}{d t} = -\frac{\Delta \bm{v}}{t_{\rm s}} + \bm{a}_{\rm ext},
\end{equation}
where $\bm{a}_{\rm ext}$ is the external force term of the relative motion, expressed as 
\begin{equation}
\label{eq:aext}
    \bm{a}_{\rm ext} = \frac{1}{\rho_{\rm lc}} \nabla P_{\rm lc} + \frac{1}{4 \pi \rho_{\rm lc}} \boldsymbol{B}_{\rm lc} \times(\nabla \times \boldsymbol{B}_{\rm lc}).
\end{equation}
Note that the external force $\bm{a}_{\rm ext}$ only operates  the gas and does not operate the dust.
The derivation of equation~(\ref{emrelv}) is summarized in \S\ref{sec:appendix}.  
It should be noted that we need to be careful in using equation~(\ref{emrelv}).
As described in \S\ref{subsubsec:dust}, we adopt different sizes of dust grains in a wide range from 0.01\,$\mu$m to 1000\,$\mu$m. 
As explained in \S\ref{sec:appendix}, equation~(\ref{emrelv}) is applicable for the dust grain with a size of $a_{\rm d}\le 100$\,$\mu$m, while it  is not very appropriate for the dust grains with a size of  $a_{\rm d} \ge 1000$\,$\mu$m especially in the low density gas region of $n\lesssim10^6$\,cm$^{-3}$. 
Although we have to keep this in mind, we do not step into the use of equation~(\ref{emrelv})  in more detail in this study. 
The validity of equation~(\ref{emrelv}) will be tested in our future study. 
As mentioned in \S\ref{subsubsec:dusteom}, we use  the corrected  $t_{\rm s} $ to avoid the supersonic relative motion, which is described as 
\begin{equation}
\label{corrtstop}
    t_{\rm s} = \frac{a_{\rm d} \rho_{\rm s}}{v_{\rm th} \rho} \frac{1}{\sqrt{1+\frac{9 \pi}{128}\left(\frac{\Delta v}{c_{s}}\right)^{2}}},
\end{equation}
where $\Delta v = | \bm{v}_{\rm d} - \bm{v}_{\rm lc} |$ is the relative velocity between dust and gas.

In this study, dust particles are assumed to be electrically neutral.
However, if dust grain is charged, the dust particles feel an additional Lorentz force $\bm{f}_{\rm cd}$, where $\bm{f}_{\rm cd}$ is defined as
\begin{equation}
    \bm{f}_{\rm cd}=\frac{Z_{\rm d} e}{m_{\rm d}}\left(\bm{E} + \frac{\bm{v}_{\rm d}  \times \bm{B}}{c} \right),
\end{equation}
where $Z_{\rm d}e, m_{\rm d}, \bm{E}, c$ are the charge of the grain, the grain mass, the electric field and the speed of light, respectively.
Since the neutrally charged dust is assumed in this study, $Z_{\rm d}$ is equal to zero.
Thus, the Lorentz force term $\bm{f}_{\rm cd}$ which only charged dust grains feel is not included in equation~(\ref{medust}) and does not appear in equation~(\ref{eq:aext}).
Actually, equations (\ref{medust})--(\ref{eq:aext}) mean that neutral dust grains feel the Lorentz force ($- \frac{1}{4 \pi \rho} \boldsymbol{B} \times(\nabla \times \boldsymbol{B})$) through the gas fluid. 
The validity of the above assumption is discussed in \S~\ref{subsec:chargedust}.

Next, we describe the scheme of the numerical calculation.
Hereafter, superscripts $n$ and  $n+1$ stand for the physical quantities at time $t$ and $t + \Delta t_{\rm d}$, respectively, where $\Delta t_{\rm d}$ is the time increment and defined at the end of this subsection.
The location of a dust particle at the next step ($t = t + \Delta t_{\rm d}$) is calculated with second-order accuracy using 
\begin{equation}
    \bm{x}_{\rm d}^{n+1} = \bm{x}_{\rm d}^{n} + \bm{v}_{\rm d}^{n} \Delta t_{\rm d} + \Big(\frac{d \bm{v}_{\rm d}}{d t} \Big)^{n} \frac{(\Delta t_{\rm d})^2}{2}.
\end{equation}
The relative velocity between dust and gas in the next step $\Delta \bm{v}^{n+1}$ is calculated by integrating equation~(\ref{emrelv}), which yields
\begin{equation}
\label{nextdv}
    \Delta \bm{v}^{n+1} = \Delta \bm{v}^n e^{-\frac{\Delta t_{\rm d}}{t_{\rm s}}} + \bm{a}_{\rm ext}^{n+\frac{1}{2}} t_{\rm s} (1 - e^{-\frac{\Delta t_{\rm d}}{t_{\rm s}}}),
\end{equation}
where 
\begin{equation}
\bm{a}_{\rm ext}^{n+\frac{1}{2}} = \frac{1}{2} \Big( \bm{a}_{\rm ext}^{n} + \bm{a}_{\rm ext}^{n+1} \Big)
\end{equation}
is obtained with equation~(\ref{eq:aext}). 
In addition, we rewrite equation (\ref{nextdv}) using Taylor expansion, and thus $\Delta \bm{v}^{n+1}$ is given by
\begin{equation}
\label{nextdvapp}
\Delta \bm{v}^{n+1}=\left\{
  \begin{aligned}
  &\Delta \bm{v}^{n} \Big(1 - \frac{\Delta t_{\rm d}}{t_{\rm s}} \Big) + \bm{a}_{\rm ext}^{n+\frac{1}{2}} \Delta t_{\rm d} & {\rm for} \ \ \frac{\Delta t_{\rm d}}{t_{\rm s}} &< 10^{-12},\\
  &\Delta \bm{v}^n e^{-\frac{\Delta t_{\rm d}}{t_{\rm s}}} + \bm{a}_{\rm ext}^{n+\frac{1}{2}} t_{\rm s} \Big(1 - e^{-\frac{\Delta t_{\rm d}}{t_{\rm s}}} \Big) & {\rm for} \ \  10^{-12} &\leq \frac{\Delta t_{\rm d}}{t_{\rm s}} \leq 500, \\
  &\bm{a}_{0,{\rm ext}} t_{\rm s} & {\rm for} \ \ \frac{\Delta t_{\rm d}}{t_{\rm s}} &> 500.
  \end{aligned}
\right.
\end{equation}
For $\frac{\Delta t_{\rm d}}{t_{\rm s}} < 10^{-12}$, $t_{\rm s}$ is very long, indicating that dust and gas are weakly coupled.
To avoid the round-off error of double precision in our code, we substitute $e^{-\frac{\Delta t_{\rm d}}{t_{\rm s}}} \approx 1 - \frac{\Delta t_{\rm d}}{t_{\rm s}}$ into equation~(\ref{nextdv}).
Similarly, for $\frac{\Delta t_{\rm d}}{t_{\rm s}} > 500$, $t_{\rm s}$ is very short, indicating that dust and gas are strongly coupled.
To avoid underflow error, we use the approximation $e^{-\frac{\Delta t_{\rm d}}{t_{\rm s}}} \approx 0 $ in equation (\ref{nextdv}).
Finally, with the relative velocity at $t + \Delta t_{\rm d}$, we can update the dust velocity as
\begin{equation}
\label{eq:nextvd}
    \bm{v}_{\rm d}^{n+1} = \bm{v}^{n+1}_{\rm lc} + \Delta \bm{v}^{n+1}.
\end{equation}
The trajectory calculation described above is conducted at each time step of gas evolution.

We describe the method used for setting the timestep for dust dynamics $\Delta t_{\rm d}$. 
When the dust velocity is larger  than the gas velocity or the Alfv\'en veclocity, the time step for the dust motion $\Delta t_{\rm d}$ should be shorter than that for the gas fluid $\Delta t_{l,{\rm gas}}$, where $\Delta t_{l,{\rm gas}}$ is the time step of the gas fluid at grid level $l$ within which the dust particle exists and is determined during MHD calculation. 
In such a case, we calculate the dust motion with a subcycled time step.  
If this is not the case, we use $\Delta t_{l,{\rm gas}}$ as the time step for the dust motion. 
In summary, $\Delta t_{\rm d}$ is given by
\begin{equation}
    \Delta t_{\rm d} = {\rm Min}\Big(\Delta t_{l,{\rm gas}}, {\rm Min}[\Delta x_{\rm cell}/|v_{x,{\rm d}}|,\Delta y_{\rm cell}/|v_{y,{\rm d}}|,\Delta z_{\rm cell}/|v_{z,{\rm d}}|] \Big).
\end{equation}
$\Delta A_{\rm cell}$ and $v_{A,{\rm d}}$ ($A=x,y,z$) are the cell width and dust velocity, respectively, in the $A$ direction. 
In this study, we set $\Delta x_{\rm cell} = \Delta y_{\rm cell} = \Delta z_{\rm cell}$.
The dust trajectory calculation is synchronized with the MHD calculation.
Thus, for $\Delta t_{\rm d} < \Delta t_{l,{\rm gas}}$, we use equations (\ref{medust}) - (\ref{eq:nextvd}) to update the location and velocity of dust particles until the summation of $\Delta t_{\rm d}$ reaches $\Delta t_{l,{\rm gas}}$.

Finally, we mention the treatment of dust particles after falling onto the sink.
When a particle reaches the sink or the region within $r_{\rm sink} \le 1\,{\rm au}$, we stop the trajectory calculation of the particle.

\subsection{Initial conditions}
\subsubsection{MHD (gas fluid) calculation}
\label{subsubsec:gas}
As described in \S\ref{subsec:gas},  the initial condition is identical to that adopted in  \citet{2017ApJ...835L..11T} and \citet{2020ApJ...905..174A}. 
Thus, we simply describe the initial condition of our MHD calculation. 

As the initial condition, we adopt a critical Bonnor-Ebert density profile with an isothermal temperature of $10$\,K and a central density of $6\times10^5$\,cm$^{-3}$. 
The density is increased by a factor of $2$ to promote contraction. 
The mass $M_{\rm cl}$ and radius  $R_{\rm cl}$ of the initial cloud are $M_{\rm cl}=1.25\,\msun$ and $R_{\rm cl}=6.13\times10^3$\,au, respectively. 
Uniform magnetic field $B_0=5.1\times10^{-5}$\,G and rigid rotation $\Omega_0=2\times10^{-13}$\,s$^{-1}$ are adopted for the initial  cloud. 
The ratios of the thermal $\alpha_0$, rotational $\beta_0$, and magnetic $\gamma_0$ energies with respect to the gravitational energy of the initial cloud are $\alpha_0=0.42$, $\beta_0=0.02$, and $\gamma_0=0.1$, respectively. 
The mass-to-flux ratio normalized by the critical value $(2\pi G^{1/2})^{-1}$ is $\mu_0=3$.

\subsubsection{Dust trajectory calculation}
\label{subsubsec:dust}
We distribute dust particles in the initial cloud. 
The spatial distribution of dust particles in spherical coordinates is given in Table~\ref{table:dustdistribution}.
The radius of the initial cloud (or Bonnor-Ebert sphere) is 6130 au, as described in \S\ref{subsubsec:gas}.
The dust particles are distributed every 10\,au in the range of 10--6130\,au.
Thus, they are placed at 613 locations in the radial direction.
In the azimuthal ($\phi$) direction, the dust particles are placed  every 90$^\circ$ in the range of $\phi=$ 0--270$^\circ$ (4 locations).
In the zenith  ($\theta$) direction, they are placed every 15$^\circ$ in the range of $\theta=$ 0--90 $^\circ$ (7 locations). 
The dust particles are distributed so that the whole region of the initial cloud is spatially covered.  
The small number of particles in the $\phi$ direction is sufficient for this analysis because the symmetry along the $z$-axis is mostly maintained during the calculation (see \S\ref{sec:results}).

The dust grains are prepared to have six different sizes in the range of $a_{\rm d}=$0.01--1000\,${\rm \mu m}$, as shown in Table~\ref{table:dustsize}.
We adopt a wide range of grain sizes because there is  no conclusive evidence for the size of dust grains  in molecular cloud cores, as mentioned in \S\ref{sec:intro}. 
In total,  102,984 dust particles are included in the initial cloud for the MHD calculation. Dust grains with six different sizes (Table~\ref{table:dustsize}) are located at the locations listed in Table~\ref{table:dustdistribution}.

\begin{figure*}
\includegraphics[width=0.9\linewidth]{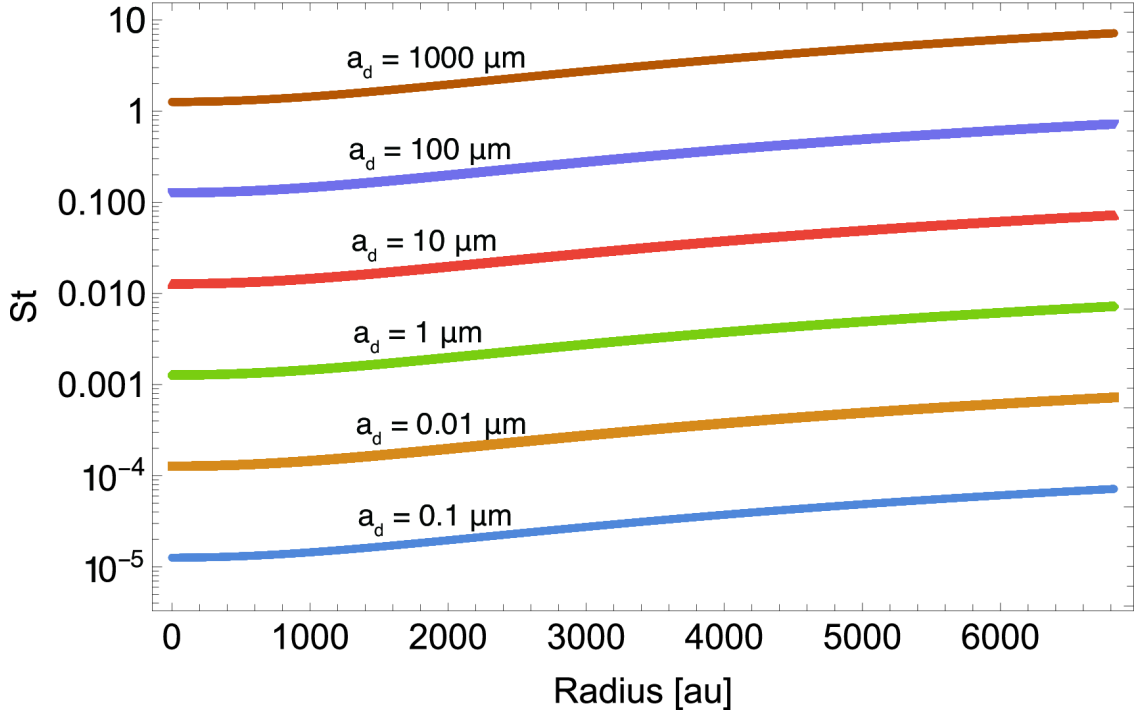}
\caption{Initial Stokes number St ($=t_{\rm s} / t_{\rm ff}$) of dust particles distributed in the cloud versus radius.}
\label{fig:initialst}
\end{figure*}

To evaluate how strong a dust particle is initially coupled with the gas, the Stokes number (St) is adopted here. 
St is defined as the stopping time normalized by a dynamical time-scale. 
In previous works that investigated dust evolution in a protoplanetary disk \citep[e.g.,][]{1977MNRAS.180...57W}, St is  normalized by the Keplerian time-scale $\Omega_{\rm Kep}^{-1}$, where $\Omega_{\rm Kep} = \sqrt{GM_*/r^3}$.
In this study, we mainly focus on the dust motion on a scale larger than the (Keplerian) disk scale.
Thus, instead of the Keplerian time-scale,  the freefall time-scale $t_{\rm ff} = \sqrt{3 \pi/32 G \rho}$ is used as the dynamical time-scale. 
Therefore, in this study, we define  the Stokes number as St $= t_{\rm s} / t_{\rm ff}$. 
Fig.~\ref{fig:initialst} plots the initial Stokes number for all the  dust particles distributed in the Bonnor-Ebert sphere against the radius.
Since $t_{\rm s}$ is proportional to the dust grain size, St becomes 10 times larger for 10 times larger $a_{\rm d}$.
Except for the dust particles of $a_{\rm d} = 1000\,{\rm \mu m}$, St < 1 is fulfilled within the initial cloud (or the Bonnor-Ebert sphere).

Initially, the velocity of dust grains is set to be equal to the gas velocity at the position of the dust grain.
In other words, all dust grains are perfectly coupled with gas at the beginning of the calculation.
Fig.~\ref{fig:initialst} indicates that dust grains of $a_{\rm d} = 1000\,{\rm \mu m}$ satisfies ${\rm St} > 1$ and these particles could be initially decoupled from the gas.
To more realistically set the initial conditions for large-sized grains,  we would need to begin the simulations from the stage of the formation of molecular cloud cores.

\begin{table}
 \caption{Initial spatial distributions of dust particles}
 \label{table:dustdistribution}
 \centering
  \begin{tabular}{cl}
   \hline
   Coordinate & Initial particle locations \\
   \hline \hline
   $r$ & 10--6130\,au \ (every 10\,au, 613 locations) \\
   $\phi$ & 0$^\circ$, 90$^\circ$, 180$^\circ$, 270$^\circ$ \ (every 90$^\circ$, 4 locations) \\
   $\theta$ & 0$^\circ$, 15$^\circ$, 30$^\circ$, 45$^\circ$, 60$^\circ$, 75$^\circ$, 90$^\circ$ \ (every 15$^\circ$, 7 locations) \\
   \hline
  \end{tabular}
\end{table}

\begin{table}
 \caption{Dust grain sizes used in calculation}
 \label{table:dustsize}
 \centering
  \begin{tabular}{cl}
   \hline
   Dust grain size $a_{\rm d}$ [$\rm \mu$m] \\
   \hline \hline
   0.01, 0.1, 1, 10, 100, 1000  \\
   \hline
  \end{tabular}
\end{table}

\subsection{Method for calculating dust-to-gas mass ratio}
\label{subsec:particleweight}
One of the aims of this study is to calculate the spatial distribution and time evolution of the dust-to-gas mass ratio $f_{\rm dg}$.  
The dust mass density $\rho_{\rm d}$ is required for estimating $f_{\rm dg}$ because $f_{\rm dg} \equiv \rho_{\rm d} / \rho$ for a given spatial scale.
In the calculation, however, we treat the dust grains as particles.
Thus, $\rho_{\rm d}$ cannot be simply defined. 
To estimate the dust mass density, we introduce gas particles as well as dust particles and calculate how $f_{\rm dg}$ changes from the initial state by weighting the mass on dust and gas particles.
In this subsection, we describe the method used for weighting the dust and gas particles.

To calculate the trajectory of gas tracer particles, the local physical quantities of the gas fluid are used along with those of the dust particles, as described in $\S$\ref{subsec:dust}.
This is the same procedure as that used in \citet{2012ApJ...758...86F}, but without the chemical reaction with tracers.
The initial locations ($r, \phi, \theta$) of the gas particles are set according to Table~\ref{table:dustdistribution}. 
Thus, seven kinds of particle (six different-sized dust grains and one gas particle) are placed  at a  given location.  
\begin{figure}
    \centering
    \includegraphics[width=\linewidth]{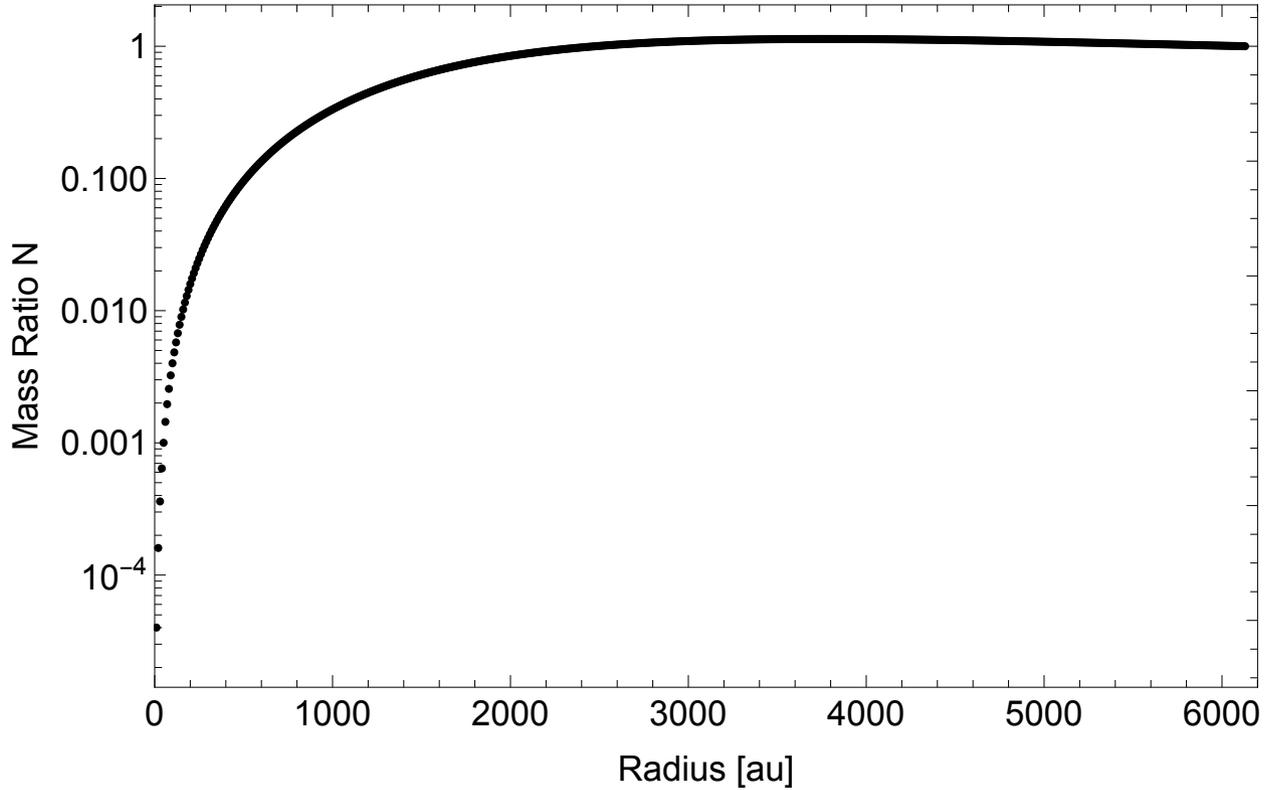}
    \caption{Shell mass ratio $N$ of initial cloud (or Bonnor-Ebert sphere). 
    The  ratio $N$ is defined as the shell mass normalized by the mass of the outermost shell and is used when weighting the dust and gas particles to derive the dust-to-gas mass ratio $f_{\rm dg}$.
    }
    \label{fig:rbemassratio}
\end{figure}
\begin{figure*}
    \centering
    \includegraphics[width=\linewidth]{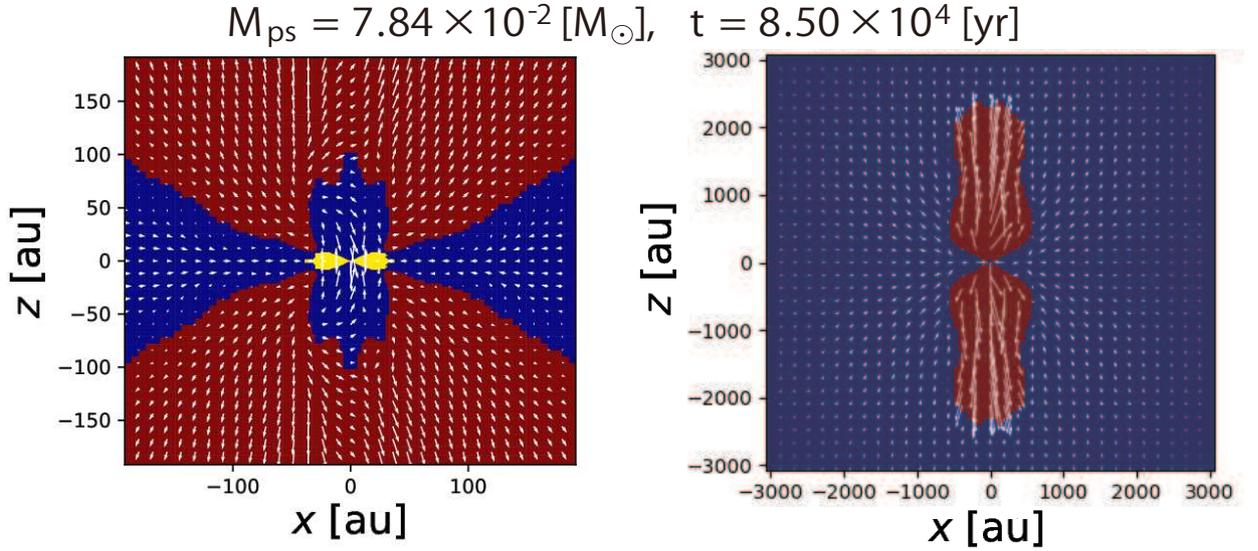}
    \caption{Envelope (dark blue),  disk (yellow), outflow (red), and protostar regions classified based on our criteria on the $y=0$ plane. 
    The protostar region is not visible due to its small size compared to the scale of the panels. Arrows indicate the velocity at a given point.
    The whole disk region is shown in the left panel and the whole outflow region is shown in  the right panel. 
    The protostellar mass (or sink mass) $M_{\rm ps}$  and the elapsed time $t$ after the cloud begins to collapse are given at the top of each panel. 
    }
    \label{fig:l6and10}
\end{figure*}

Next, we describe the weighting of the mass with dust and gas particles. 
We assume that the weighted mass corresponds to the gas shell mass.
Note that the particles are uniformly placed in the initial cloud core, as described in \S\ref{subsubsec:dust}.
The gas mass of the shell $dm_r$, which is distributed on the radius $r$,  is given by
\begin{equation}
    dm_r = 4 \pi \rho_r r^2 dr,
\end{equation}
where $\rho_r$ is the gas mass density at radius $r$ and $\rho_r$ is determined from the initial gas cloud, as described in \S\ref{subsubsec:gas}.
With $dr = 10\,{\rm au}$,  the radial direction is discretized according to the radial coordinates listed in  Table \ref{table:dustdistribution}.
Then, each shell mass is normalized by the mass of the outermost shell, which is located at $r = 6130 \ {\rm au}$.
The shell mass ratio $N$ is given by  
\begin{equation}
   N = \frac{dm_r}{dm_{\rm 6130\,au}}.
\end{equation}
To estimate $f_{\rm dg}$, the shell mass ratio $N$ is adopted for all particles (six different-sized dust grains and one gas particle) for weighting the particles, meaning that all particles has the internal parameter $N$. 
Fig.~\ref{fig:rbemassratio} plots the shell mass ratio $N$ against the radius.
Note that the weighting is the same as long as the initial location $r$ is the same, even when  the initial dust particle has different $\phi$ and $\theta$.

In this study, we use  the dust-to-gas mass ratio normalized by the initially and spatially uniform value of $f_{\rm dg}$, denoted as $\delta f_{\rm dg}$, for each dust grain size $a_{\rm d}$.
The change in the dust-to-gas mass ratio $\delta f_{{\rm dg}}$ for each dust grain size represents the change in $f_{\rm dg}$ from the initial value and is given by 
\begin{equation}
\label{eq:deltafdg}
    \delta f_{{\rm dg}} = \frac{\sum\limits_{i} N_{i,{\rm d}}}{\sum\limits_{j} N_{j,{\rm g}}}, 
\end{equation}
where $N_{i,  {\rm d}}$ and $N_{j,{\rm g}}$ are the mass shell ratio $N$ of the $i-$th dust particle of a certain dust grain size and the $j-$th gas particle, respectively. The summation is performed only on the particles that satisfy the requirement described below. 
Note that since we prepare six different sized dust grains as listed in Table~\ref{table:dustsize}, the summation in equation~(\ref{eq:deltafdg}) is done  every size of dust grain ($a_{\rm d}=0.01, 0.1, 1, 10, 100, 1000$\,$\mu$m).

With this method, it is difficult to estimate $f_{\rm dg}$ at each point. 
Thus, instead of estimating $f_{\rm dg}$,  we calculate  $\delta f_{\rm dg}$ for either four distinct regions (envelope, protostar, disk, and outflow) or various spatial scales (for details, see \S\ref{subsec:criteria} and $\S$\ref{subsec:fdg}).
In equation~(\ref{eq:deltafdg}), the summation is performed only on the particles that satisfy the imposed conditions. 
The details of the conditions are described in \S\ref{subsec:fdg}.
It should be noted  that an initially and spatially uniform $f_{\rm dg}$ is not necessary for this study.
However, almost all research has adopted $f_{\rm dg} = 0.01$ (e.g., \citealt{1977ApJ...217..425M} and \citealt{1981PThPS..70...35H}).
In this study, we do not focus on the validly of $f_{\rm dg} = 0.01$; instead, we discuss the time evolution of the change in $f_{\rm dg}$ from the initial value using $\delta f_{\rm dg}$.

\subsection{Criteria of characterizing each region }
\label{subsec:criteria}
We classify the computational domain within the star-forming core (or Bonnor-Ebert sphere)  into four regions (envelope, protostar, disk, and outflow) based on the following criteria:
\begin{itemize}
    \item Protostar:  the region inside the sink particle accretion radius, in which the sink radius is $r_{\rm sink} = 1$ au.
    \item Disk: the region where the gas rotational velocity is much faster than the radial velocity ($v_\phi > 2 |v_r|$) and is supported by rotation to some extent ($v_\phi > 0.6\,v_{\rm K}$, where $v_{\rm K}$ is the Keplerian velocity).
    \item Outflow: the region where the radial velocity of the gas is faster than the speed of sound $c_{s,0}$ ($v_r > c_{s,0}$) defined in \S\ref{subsec:gas}.\footnote{We confirmed that the outflow region does not change significantly when a strict criterion (for example, $v_r > 2 c_{s,0}$) is adopted.}
    \item Envelope: the rest of the computational domain within the star-forming cloud (or Bonnor-Ebert sphere).
\end{itemize}
We determined these criteria through trial and error referring to \citet{2012A&A...543A.128J}.
We discuss the gas evolution in \S \ref{subsec:gasevolution} and the time evolution of the dust-to-gas mass ratio in \S \ref{subsec:fdg} based on these criteria.

\section{Results}
\label{sec:results}
\subsection{Evolution of gas fluid}
\label{subsec:gasevolution}
We describe the time evolution of  the gas fluid in this subsection. 
Fig.~\ref{fig:l6and10} plots the disk, outflow, and envelope regions, as determined using the criteria  in \S \ref{subsec:criteria}, at different spatial scales, where the whole disk region is shown in the left panel and the whole outflow region is shown in the right panel.  
The figure indicates that the disk, outflow, and envelope regions are clearly divided.
The disk radius is about 40\,au and the outflow reaches to about 3000\,au at the end of the simulation.
The opening angle of the outflow is about 60$^\circ$.

\begin{figure}
    \centering
    \includegraphics[width=\linewidth]{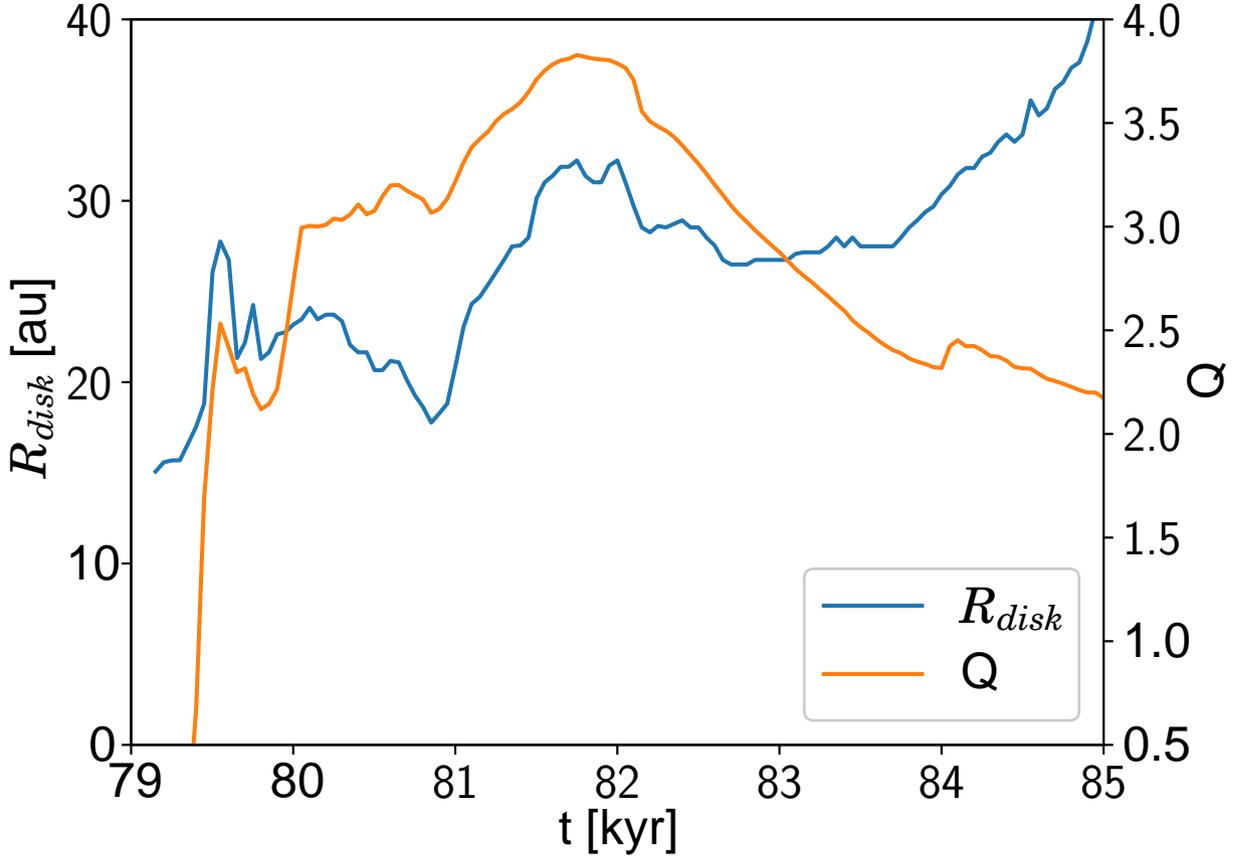}
    \caption{Time evolution of disk radius $R_{\rm disk}$ (blue, left axis) and Toomre $Q$ parameter (orange, right axis) versus elapsed time. }
    \label{fig:tdiskrq}
\end{figure}
Fig.~\ref{fig:tdiskrq} shows the time evolution of the disk radius $R_{\rm disk}$  and Toomre $Q$ parameter  \citep[][]{1964ApJ...139.1217T}.
$R_{\rm disk}$ is taken as the distance from the center to the farthest cell that satisfies the disk criteria on the equatorial plane.
The Toomre $Q$ parameter is defined as
\begin{equation}
\label{eq:qvalue}
    Q=\frac{\int_{\rho>\rho_{\mathrm{crit}}} \frac{c_s \kappa}{\pi G \Sigma} \Sigma d S}{\int_{\rho>\rho_{\mathrm{crit}}} \Sigma d S},
\end{equation}
where $\rho_{\rm crit}$ is the critical density, which is the minimum gas mass density in the disk region, $c_s$ is the local speed of sound, $\kappa$ is the epicyclic frequency, $G$ is the gravitational constant, and $\Sigma$ is the gas surface density of the disk (for details, see \citealt{2017ApJ...835L..11T}).
We adopt $\kappa = \Omega_{\rm K}$, where $\Omega_{\rm K}$ is the local Keplerian frequency. 
Note that \citet{2011PASJ...63..555M} showed that the rotaion velocity is roughly approximated by  the Keplerian velocity even during the main accretion phase \citep[see also][]{2017ApJ...835L..11T}.  
The $Q$ parameter given in equation (\ref{eq:qvalue}) is the mass-weighted average value over the disk.
As time goes by, the gas continues to be supplied from the envelope and the disk gradually grows.
During the calculation, the disk becomes gravitationally unstable when the $Q$ parameter becomes close to 2. 

\begin{figure*}
    \centering
    \includegraphics[width=\linewidth]{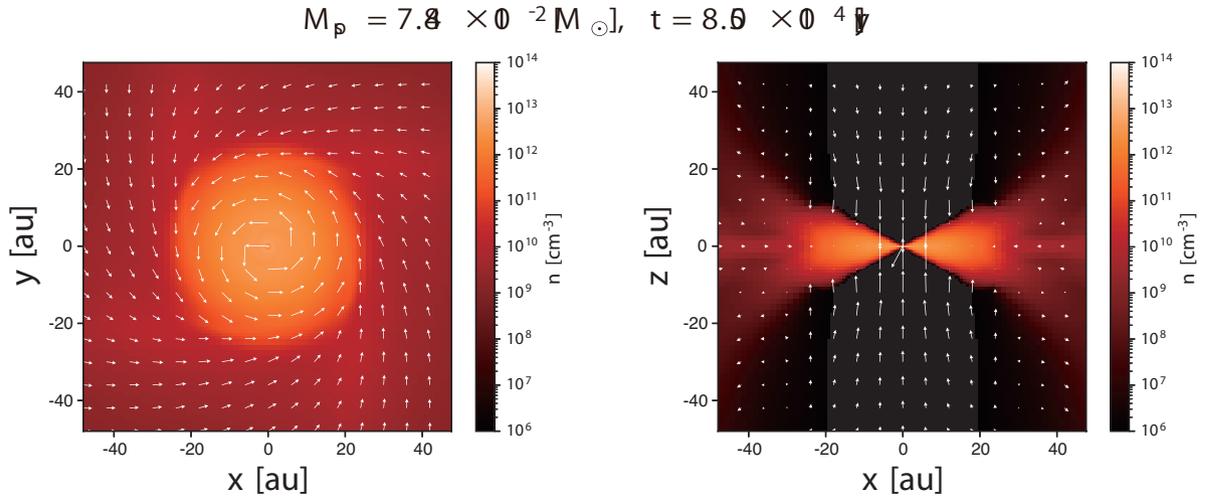}
    \caption{Gas number density (color) and velocity (arrows) on $z=0$  (left) and $y=0$ (right)  planes at end of simulation.
    The mass of the protostar (or sink particle) and the elapsed time are given at the top of each panel.
}
    \label{fig:l12gas}
\end{figure*}

Fig.~\ref{fig:l12gas} plots the  density and velocity distributions of the gas at the end of the simulation. 
The figure shows that a rotationally supported disk forms at the center  and that the gas of the disk is supplied  from the envelope (Fig.~\ref{fig:l12gas} left).
In addition, the outflow is driven from the surface of the disk (Fig.~\ref{fig:l12gas} right).
We stopped the simulation at $t =$ 85000 years after the start of the cloud collapse, at which time the mass of the sink cell is $M_{\rm sink} = 0.0784\,\msun$ and the disk has a radius of about 40\,au.
Hereafter, the discussed dust dynamics are those at this stage.

\begin{figure}
\begin{center}
\includegraphics[width=\linewidth]{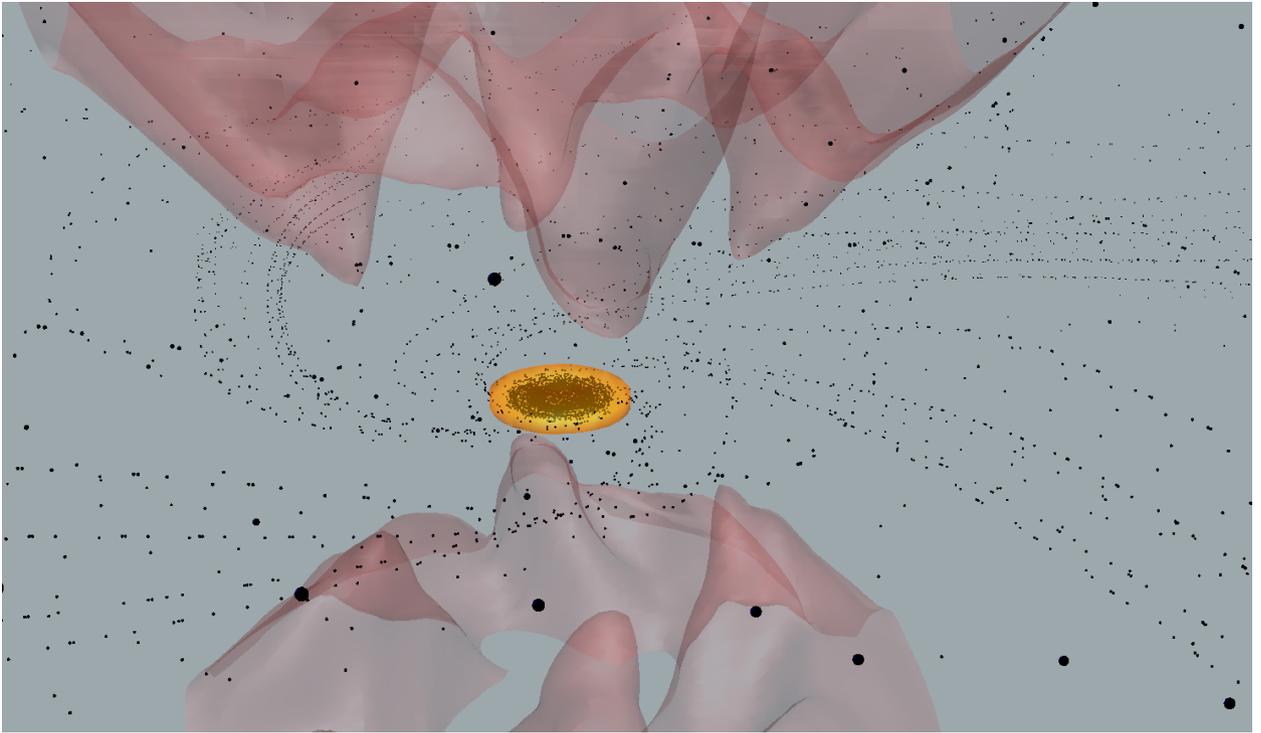}
\caption{
Three-dimensional structures of gas fluid and dust particles.
The spatial scale is about 300 au.
The yellow surface corresponds to the rotationally supported disk  (iso-density surface of $n = 5 \times 10^{11}\,{\rm cm^{-3}}$) and the red surface indicates the outflow (iso-velocity surface of $v_r = 2 \ {\rm km\,s^{-1}}$).
Black points are dust particles of various sizes.
Please see movie l93D.mp4 to trace the time sequence.
It should be noted that dust grains are described in blue and plotted with only $a_{\rm d} = 0.01 \ {\rm \mu m}$ in the movie.
}
\label{fig:middle3d}
\end{center}
\end{figure}
\begin{figure}
\begin{center}
\includegraphics[width=\linewidth]{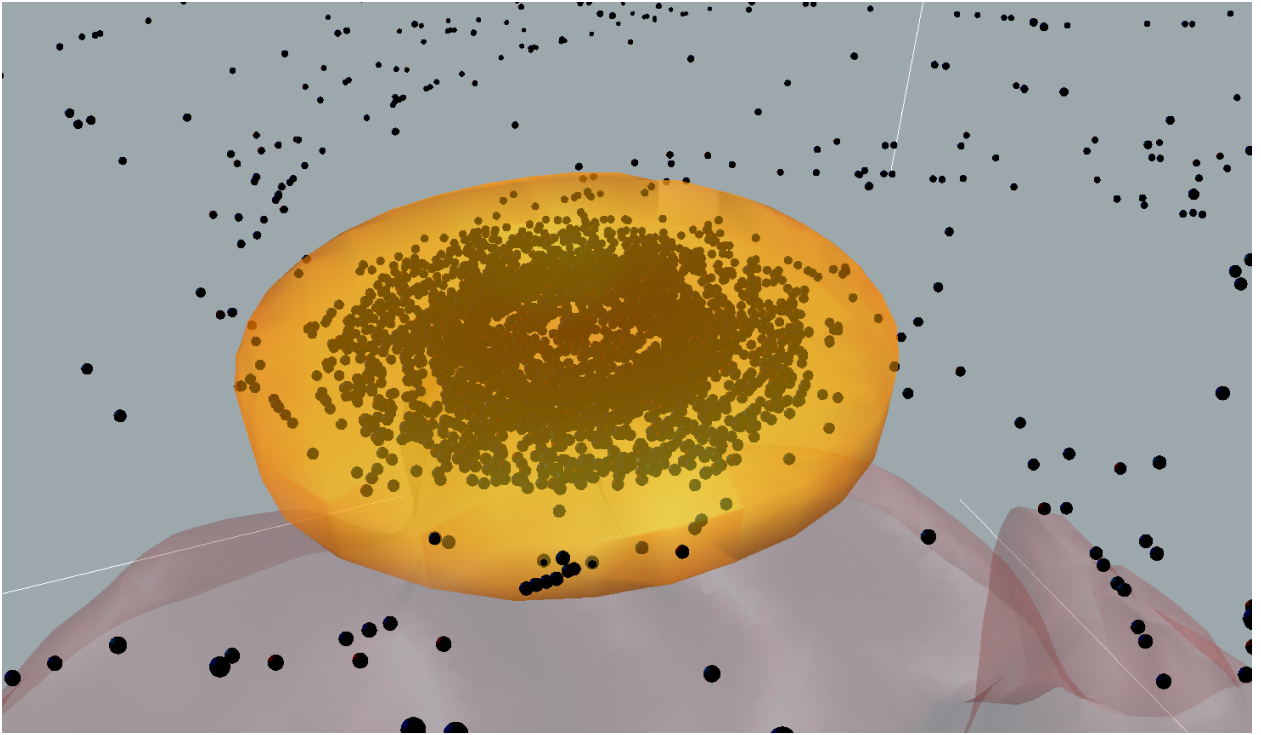}
\caption{Same as Fig.~\ref{fig:middle3d} but with a different spatial scale (about 100 au).
Please see the movie l123D.mp4 to trace the time sequence. 
}
\label{fig:near3d}
\end{center}
\end{figure}
\subsection{Dust displacement}
\label{subsec:dustpos}
Figs.~\ref{fig:middle3d} and \ref{fig:near3d} show the spatial distributions of the gas fluid and dust particles in three dimensions. The yellow contour is the high-density or disk region ($n = 5 \times 10^{11}\,{\rm cm^{-3}}$) and the red contour is the outflow region ($v_r = 2 \ {\rm km\,s^{-1}}$). 
Note that only the inner region of the rotationally supported disk, which has a minimum density of $\sim10^{10}\,{\rm cm^{-3}}$, is represented by the yellow contour. 
In the figures, each black point corresponds to a dust particle, and these particles are distributed all over the initial cloud (or Bonnor-Ebert sphere). 
Fig.~\ref{fig:middle3d} has a scale of about 300\,au and Fig.~\ref{fig:near3d} is a close-up view of the area around the center.
These figures indicate that many dust particles create a disk-like structure around the sink and that some dust particles are swept up by the gas outflow.
The dust motion is similar to the gas fluid motion.

\begin{figure*}
\includegraphics[width=0.9\linewidth]{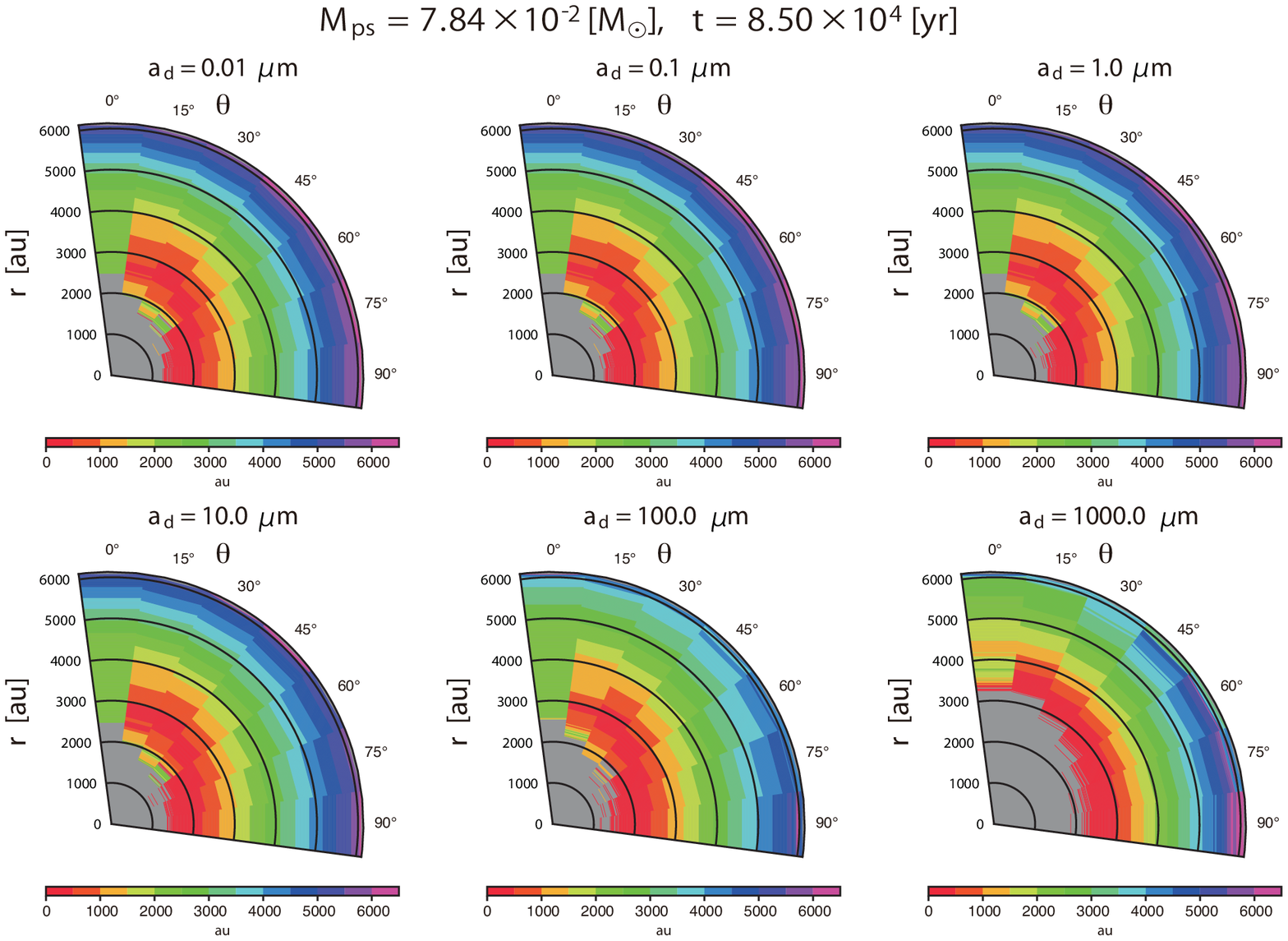}
\caption{
Distances of dust particles from center at end of simulation (color) plotted on initial $r$ and $\theta$ location planes placed according to Table \ref{table:dustdistribution}.
The particles distributed in the gray region fall onto the sink by the end of  the simulation.
The dust grain size $a_{\rm d}$, given at the top of each panel, differs in each panel. 
}
\label{fig:dustpos}
\end{figure*}
\begin{figure*}
\includegraphics[width=0.9\linewidth]{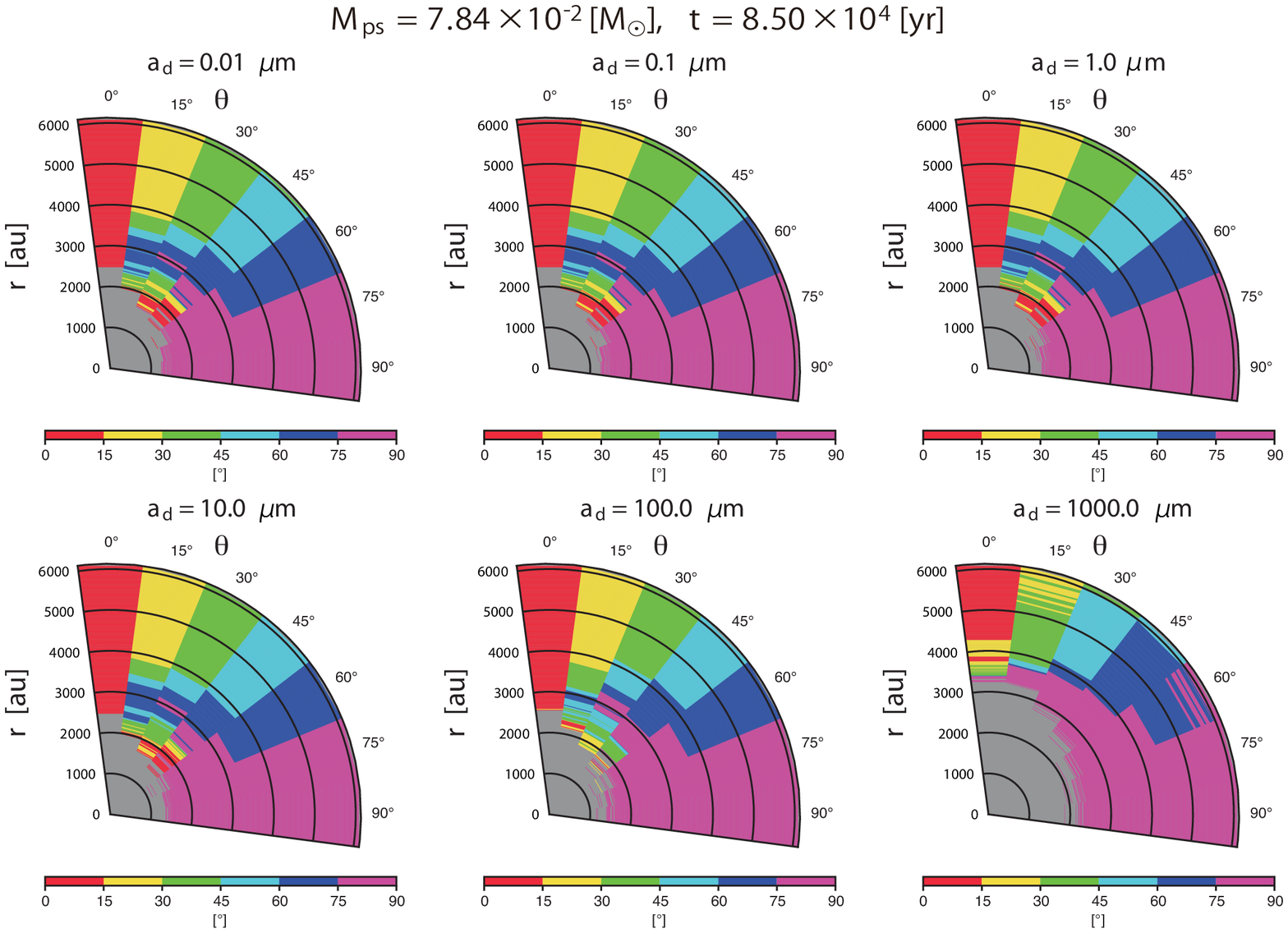}
\caption{Same as Fig.~\ref{fig:dustpos} but color indicates the zenith angle $\theta_{\rm end}$ at the end of the simulation.}
\label{fig:dustpostheta}
\end{figure*}
\begin{figure*}
\includegraphics[width=0.9\linewidth]{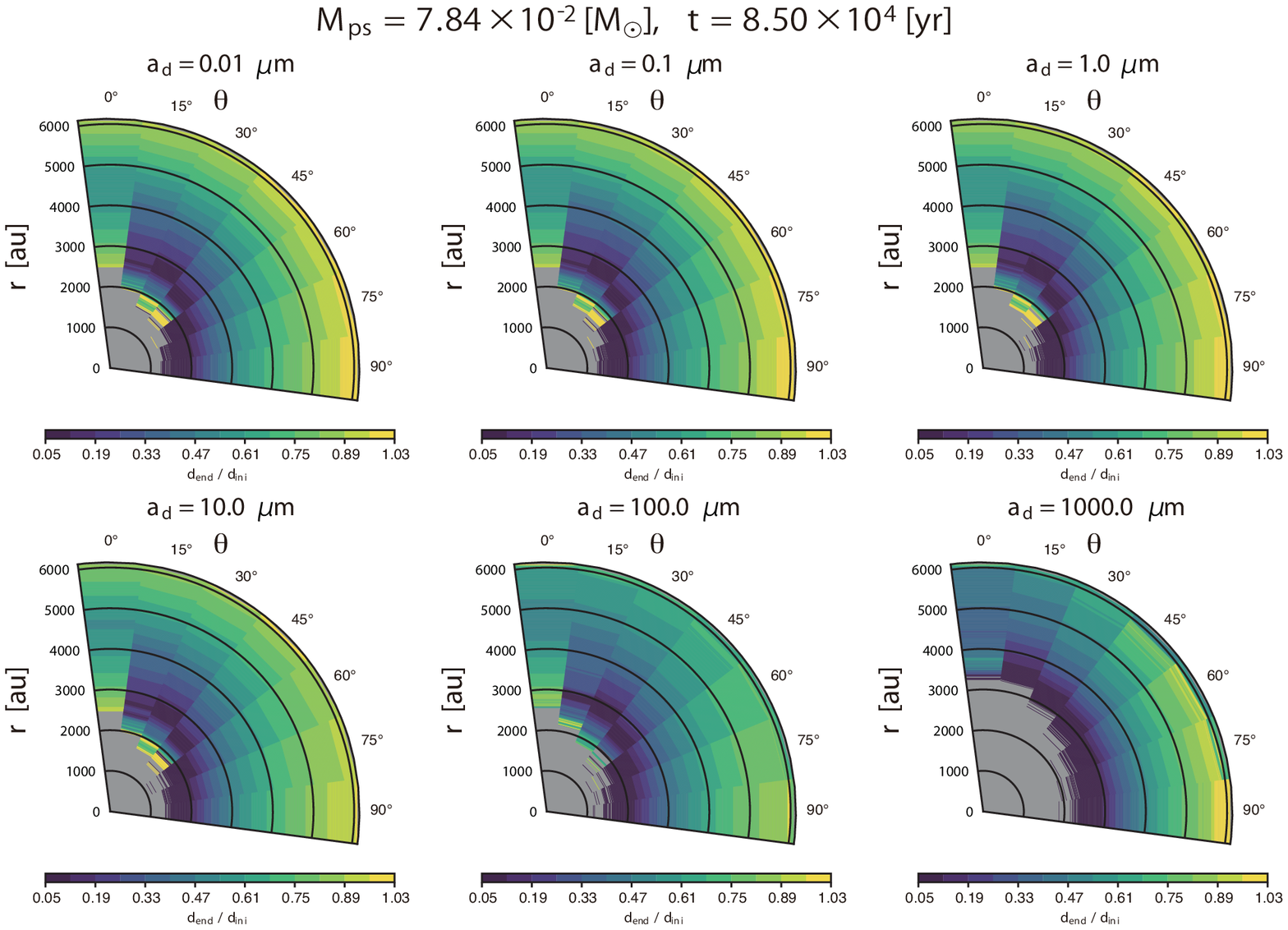}
\caption{Same as Fig.~\ref{fig:dustpos} but color indicates ratio of distances of dust particles during epochs of end of simulation to those in initial state.}
\label{fig:dustposrelinit}
\end{figure*}
\begin{figure*}
\includegraphics[width=0.9\linewidth]{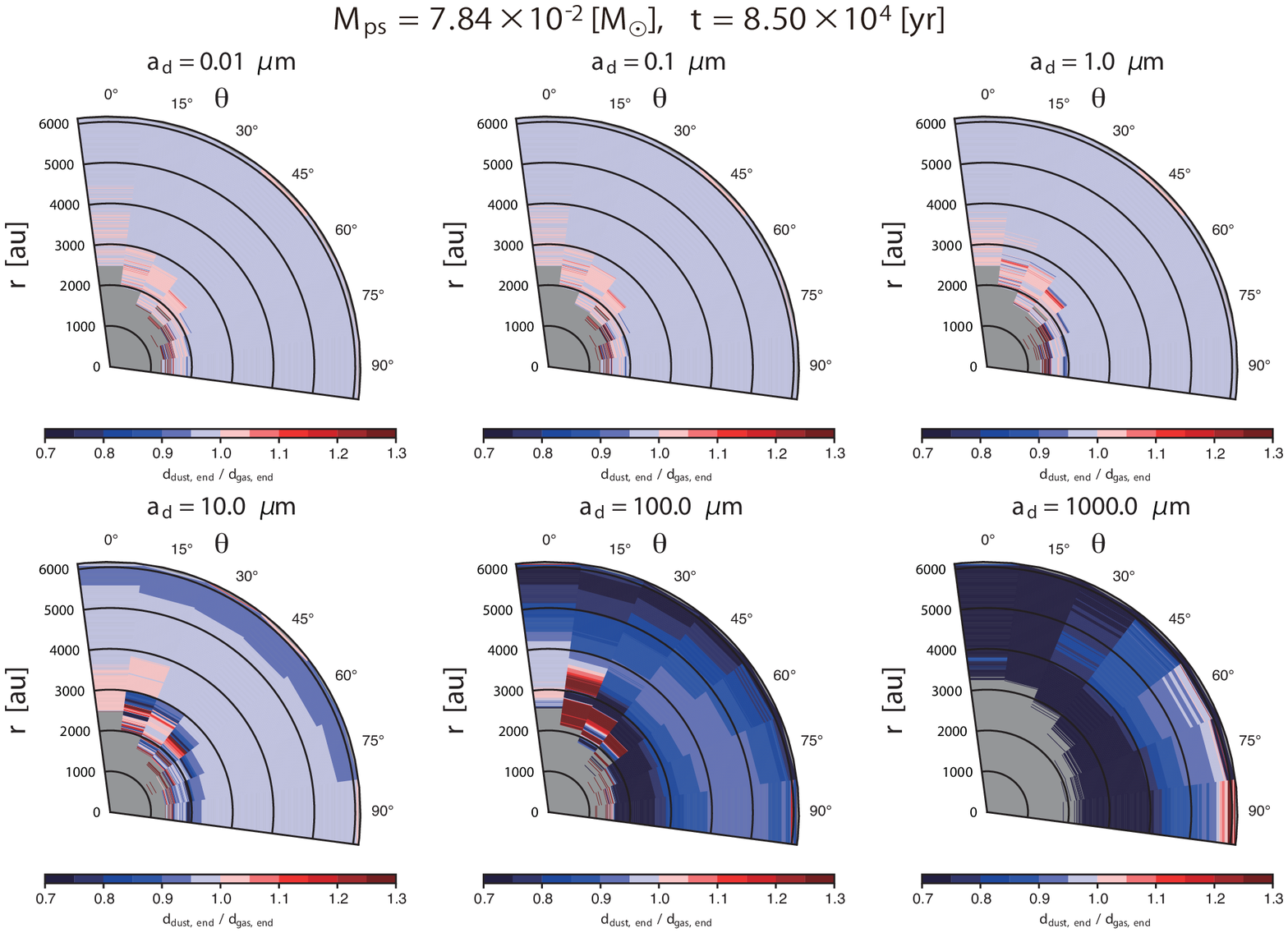}
\caption{Same as Fig.~\ref{fig:dustpos} but color indicates ratio of distances of dust particles to those of gas particles initially located at same location ($d_{\rm dust, end}/d_{\rm gas, end}$).}
\label{fig:dustposrelgas}
\end{figure*}

Fig.~\ref{fig:dustpos} shows the distances from the center of the dust particles at the end of the simulation.
Each panel shows the result for a given dust grain size $a_{\rm d}$.
Each location ($r$ and $\theta$) corresponds to the initial location of a dust particle. The color indicates the distance from the center at the end of the simulation.
The particles in the gray area have fallen into the sink.
For simplicity, we plot only the particles with $\phi = 0^\circ$; similar results were obtained for $\phi = 90^\circ, 180^\circ$, and $270^\circ$. 
Our simulations show that  the dust particles initially located at large $\theta$, which have a large specific angular momentum, tend to later fall into the center.
At the end of the simulation, the dust grains initially distributed in the  $\theta \simeq 0^\circ$ direction with a size of  $0.01\,{\rm \mu m} \leq a_{\rm d} \leq 100\,{\rm \mu m}$ are rolled up to $r > 1000\,{\rm au}$ by the gas outflow, except for the particles that have already fallen onto the sink.
The large particles ($a_{\rm d} = 1000\,{\rm \mu m}$, Fig.~\ref{fig:dustpos} bottom right) show  different behavior from that of the relatively small grains.
Overall, dust grains with a size of $a_{\rm d} = 1000\,{\rm \mu m}$ tend to rapidly fall into the center due to their longer stopping time (\S\ref{sec:stokesnumber}).

Fig.~\ref{fig:dustpostheta} shows the zenith angle of the dust particles $\theta_{\rm end}$ at the end of the calculation plotted on initial $r$ and $\theta$ location planes. 
We find that some dust grains with $15^\circ \leq \theta \leq 75^\circ$ tend to have a larger zenith angle $\theta_{\rm end}$ than the initial angle.
This tendency is stressed for the particles located relatively close to the center.
In addition to gravity, the gas motion is governed also by Lorentz force and the magnetic field lines have  an hourglass shape. 
Thus, the dust grains coupled with gas move along magnetic field lines and reach the region with a large zenith angle (the purple region in Fig.~\ref{fig:dustpostheta}). 
Almost all the dust grains initially distributed around  the $z$-axis  (or grains with $\theta = 0^\circ$) maintain their initial zenith angle, while some of the grains with $a_{\rm d}=1000\,{\rm \mu m}$ are disrupted in their trajectory by the outflow.
In addition, the figure indicates that the dust particles initially distributed near the equatorial plane are not swept up by the outflow.

Fig.~\ref{fig:dustposrelinit} shows the distance ratio of the dust particles during the period between the end of the simulation and the initial state ($d_{\rm end}/d_{\rm ini}$, where $d_{\rm end}$ and $d_{\rm ini}$ are the distances of a particle from the center  at the end of the simulation and the initial state, respectively).
In the figure, color indicates how close a particle is to the center of the gravitationally collapsing cloud.
The figure indicates that  the particles initially placed around the center ($r\lesssim 2000$\,au) tend to rapidly fall onto the sink or protostar, whereas the dust particles swept up by the outflow move to the outer region once they approach the central region.
Thus,  the distance ratio $d_{\rm end}/d_{\rm ini}$ in such dust particles is larger in other particles  in Fig.~\ref{fig:dustposrelinit}.
The distance ratios are less than unity (i.e., $d_{\rm end}/d_{\rm ini}<1$) for all particles, indicating that  no particle moves outward from the initial location within the simulation time.
However,  it is expected that with further time integration the ratio $d_{\rm end}/d_{\rm ini}$  will eventually exceed unity and that the dust particles will be ejected from the collapsing cloud core with the gas outflow.

Fig.~\ref{fig:dustposrelgas} shows the ratio of the distances of the dust particles to those of the gas particles at the end of the calculation ($d_{\rm dust, end}/d_{\rm gas, end}$).
We estimated the ratio for each dust and gas particle located at a given initial location.
Thus, the  figure indicates the separation of each dust particle from each gas particle at the end of the simulation (they were initially placed at the same location).
A dust grain is well coupled with the gas  when the ratio is almost unity, but decoupled otherwise.   
This figure shows that dust particles with a size of $0.01\,{\rm \mu m} \leq a_{\rm d} \leq 10\,{\rm \mu m}$ are broadly coupled with the gas.
For $a_{\rm d} = 1000\,{\rm \mu m}$, the dust grains are decoupled from the gas. 
Such dust grains are closer to the center than is the gas, indicating that the dust grains are more concentrated than the gas at the center. 
Dust grains with a size of $a_{\rm d} = 100\,{\rm \mu m}$ are also noticeably decoupled from the gas.
Interestingly, for $a_{\rm d} = 100\,{\rm \mu m}$, some dust particles, which are distributed in the range of $2000\,{\rm au} \leq r \leq 3000\,{\rm au}$ with $15^\circ \leq \theta \leq 45^\circ$, are more distant from the center than is the gas because they reached the driving area of the gas outflow earlier than did the gas particles located at the same initial location.

\begin{figure*}
\includegraphics[width=0.9\linewidth]{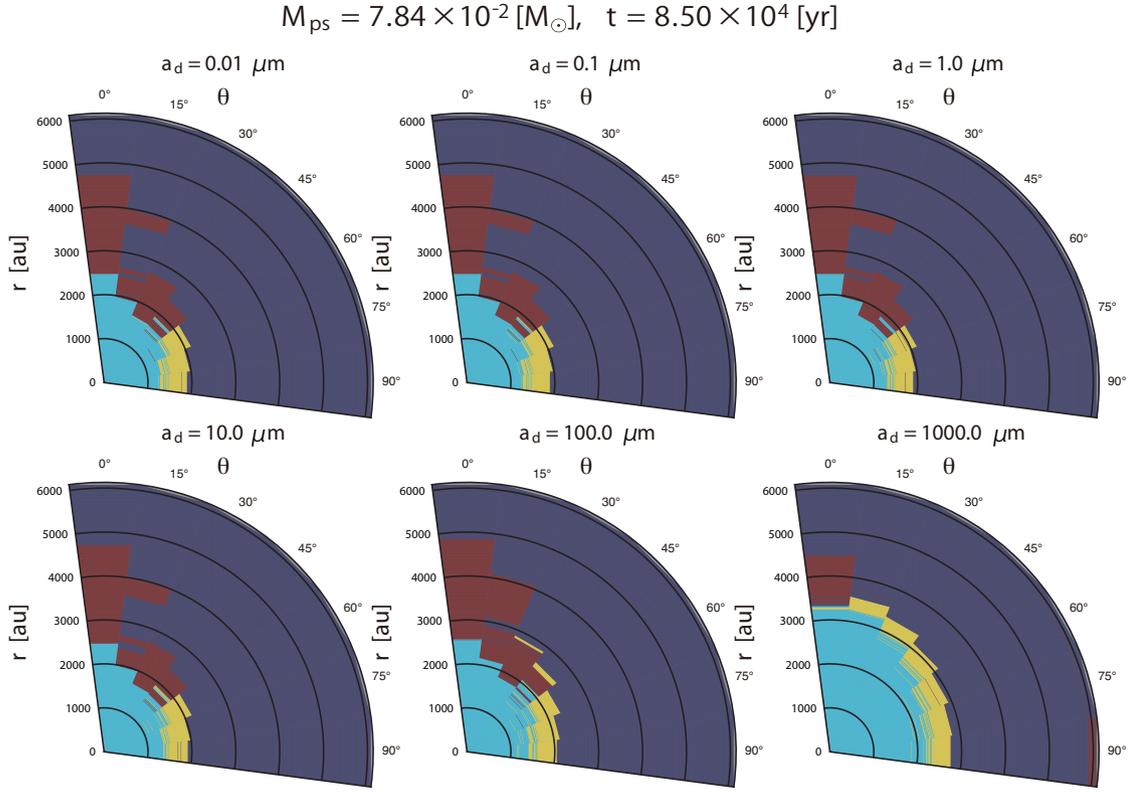}
\caption{Same as Fig.~\ref{fig:dustpos} but color indicates envelope (dark blue), protostar (light blue), disk (yellow), and  outflow (red) regions.
}
\label{fig:dustregion}
\end{figure*}
\subsection{Dust initial location and gas regions}
\label{subsec:dustposgasregion}
In this study, dust is introduced as particles, whose motion can be traced.
In \S\ref{subsec:gasevolution}, we classified the computation domain into four regions (envelope, protostar, disk, and outflow) based on physical criteria.
In this subsection, we specify the regions to which the dust particles belong in terms of the initial location plane ($r$, $\theta$) and dust size ($a_{\rm d}$).

Fig.~\ref{fig:dustregion} shows the regions where the dust (color) is located on the initial location ($r$, $\phi$) plane, on which the regions of the envelope (dark blue),  protostar (light blue),  disk (yellow),  and outflow (red) are plotted. 
First, we focus on the particles that belong to the outflow (red) region.
The dust grains placed within the outflow are those initially placed at $0^\circ \leq \theta \leq 45^\circ$ when the dust size is in the range of  $0.01 \ {\rm \mu m} \leq a_{\rm d} \leq 100 \ {\rm \mu m}$. 
In the figure,  dust with a size of $a_{\rm d}=100\,{\rm \mu m}$ occupies a larger area of the outflow region than that with a size of $a_{\rm d}<100\,{\rm \mu m}$ ($a_{\rm d}=0.01$. 0.1, 1.0, 10\,${\rm \mu m}$), indicating that dust grains with a size of $a_{\rm d}=100\,{\rm \mu m}$ are preferentially ejected by the outflow.  
In contrast, dust grains with a size of $a_{\rm d}=1000\,{\rm \mu m}$ are swept up by the outflow only when they are initially placed along the $z-$axis with $\theta = 0^\circ$.

Next, we focus on the particles within the disk. 
When the dust grain size satisfies $a_{\rm d} \leq 100 \ {\rm \mu m}$, only the dust grains initially located in the range of $60^\circ \leq \theta \leq 90^\circ$ can enter the disk and rotate with the gas.
On the other hand, dust grains with  the size of $a_{\rm d} = 1000 \ {\rm \mu m}$ reach the disk with  a wide range of $\theta$ 
( $15^\circ \leq \theta \leq 90^\circ$).

\begin{figure*}
\begin{tabular}{ccc}
\begin{minipage}{0.33\hsize}
\includegraphics[width=\columnwidth]{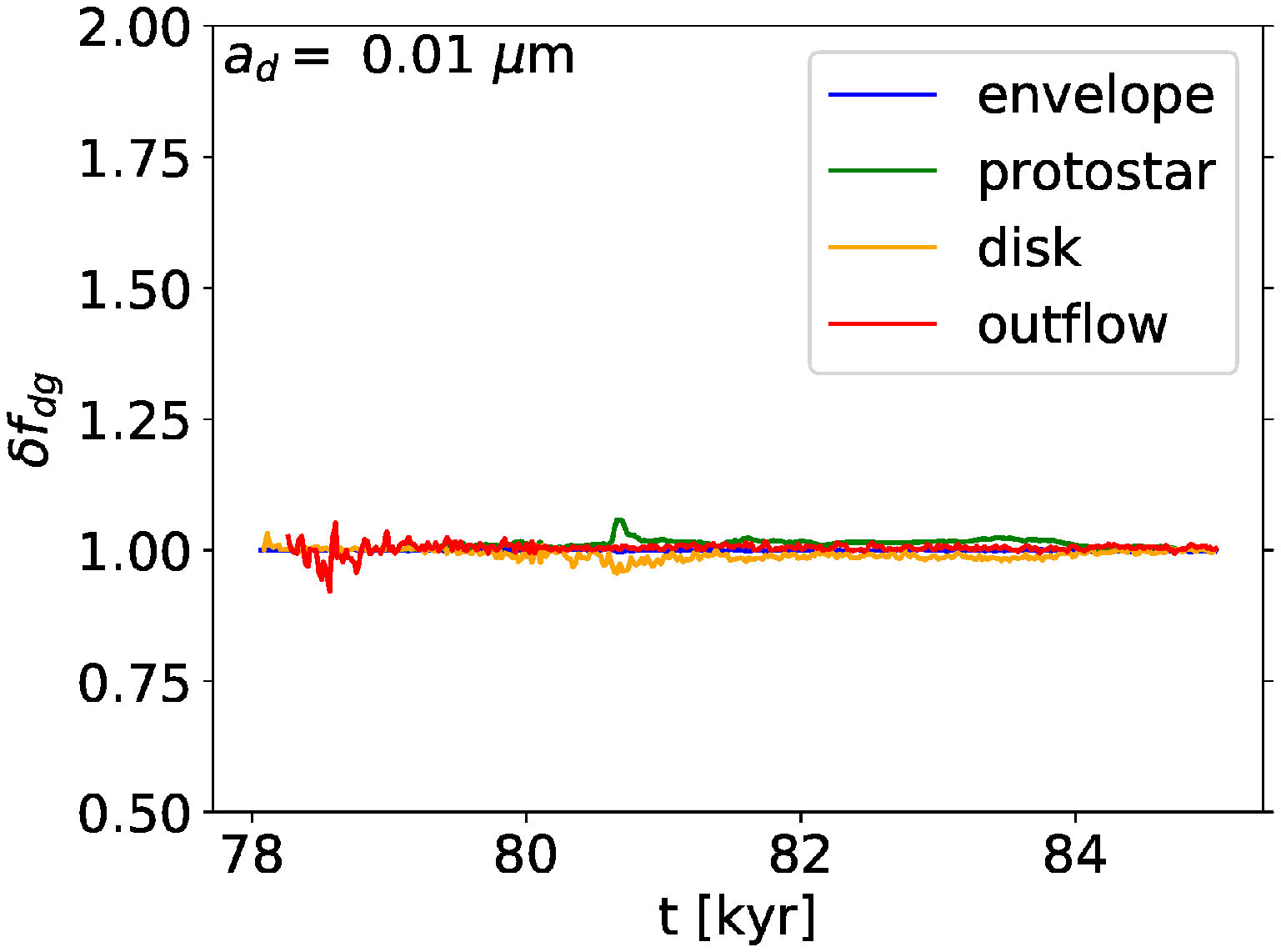}
\end{minipage}
\begin{minipage}{0.33\hsize}
\includegraphics[width=\columnwidth]{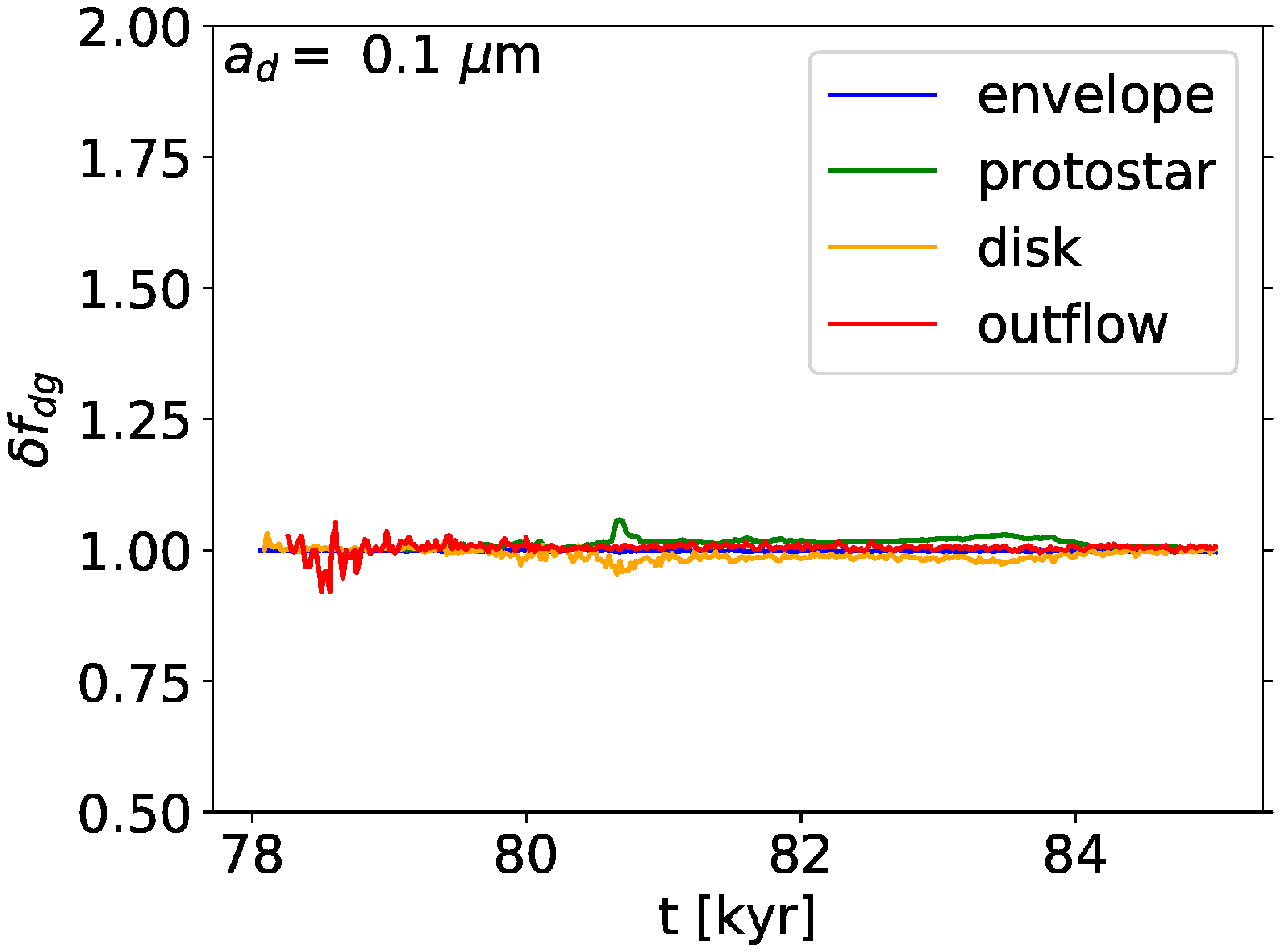}
\end{minipage}
\begin{minipage}{0.33\hsize}
\includegraphics[width=\columnwidth]{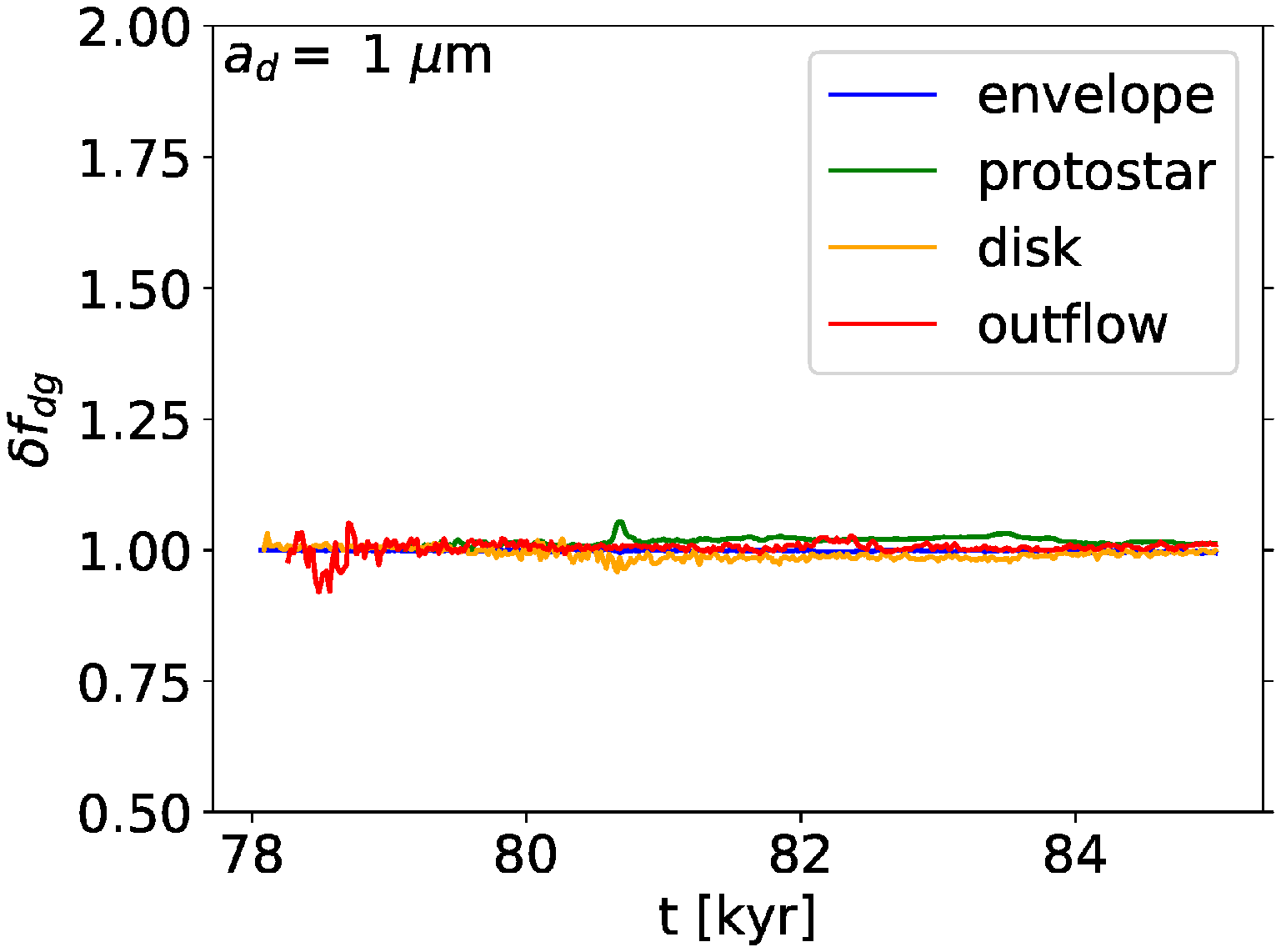}
\end{minipage} 
\\ 
\begin{minipage}{0.33\hsize}
\includegraphics[width=\columnwidth]{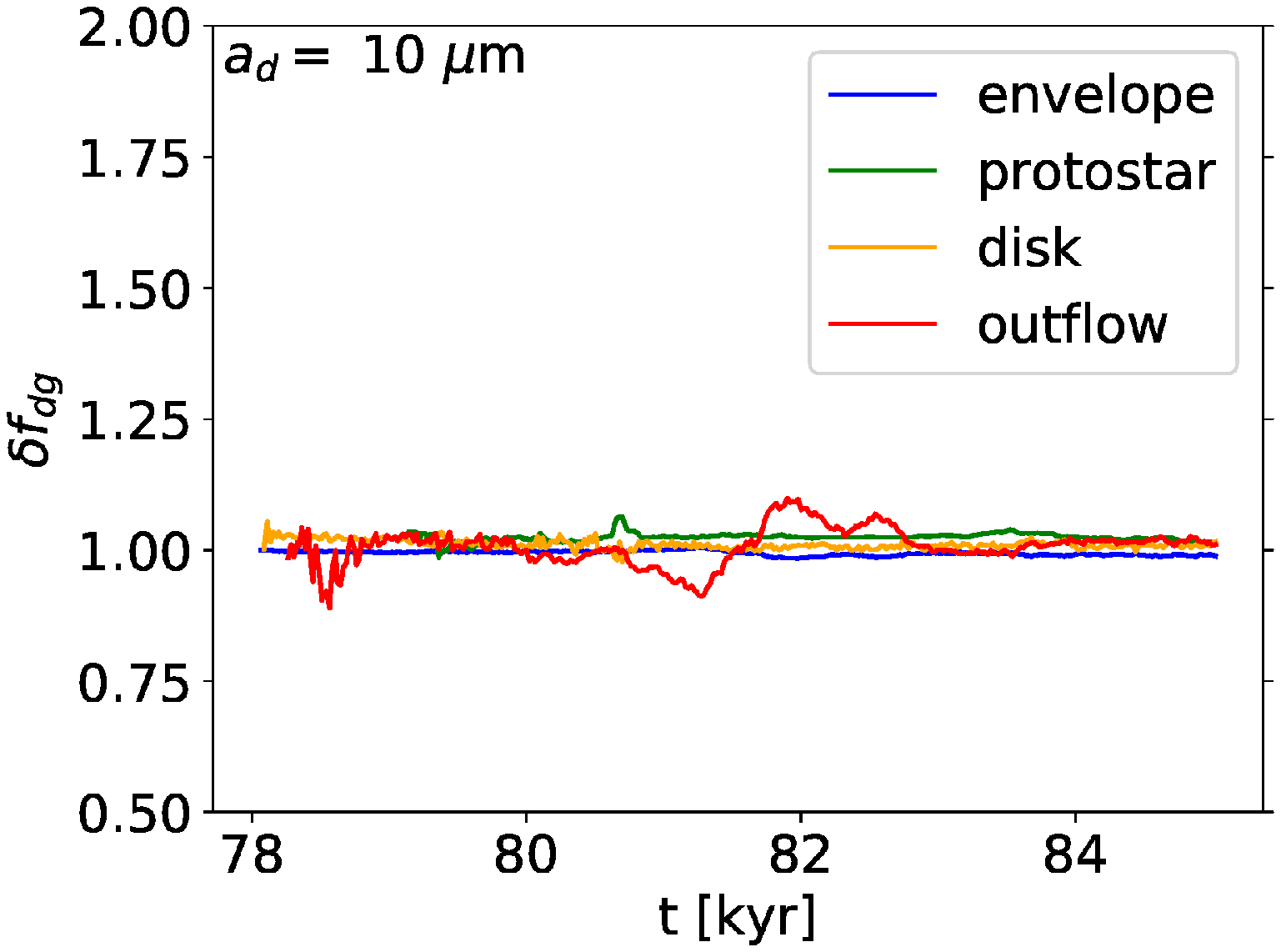}
\end{minipage}
\begin{minipage}{0.33\hsize}
\includegraphics[width=\columnwidth]{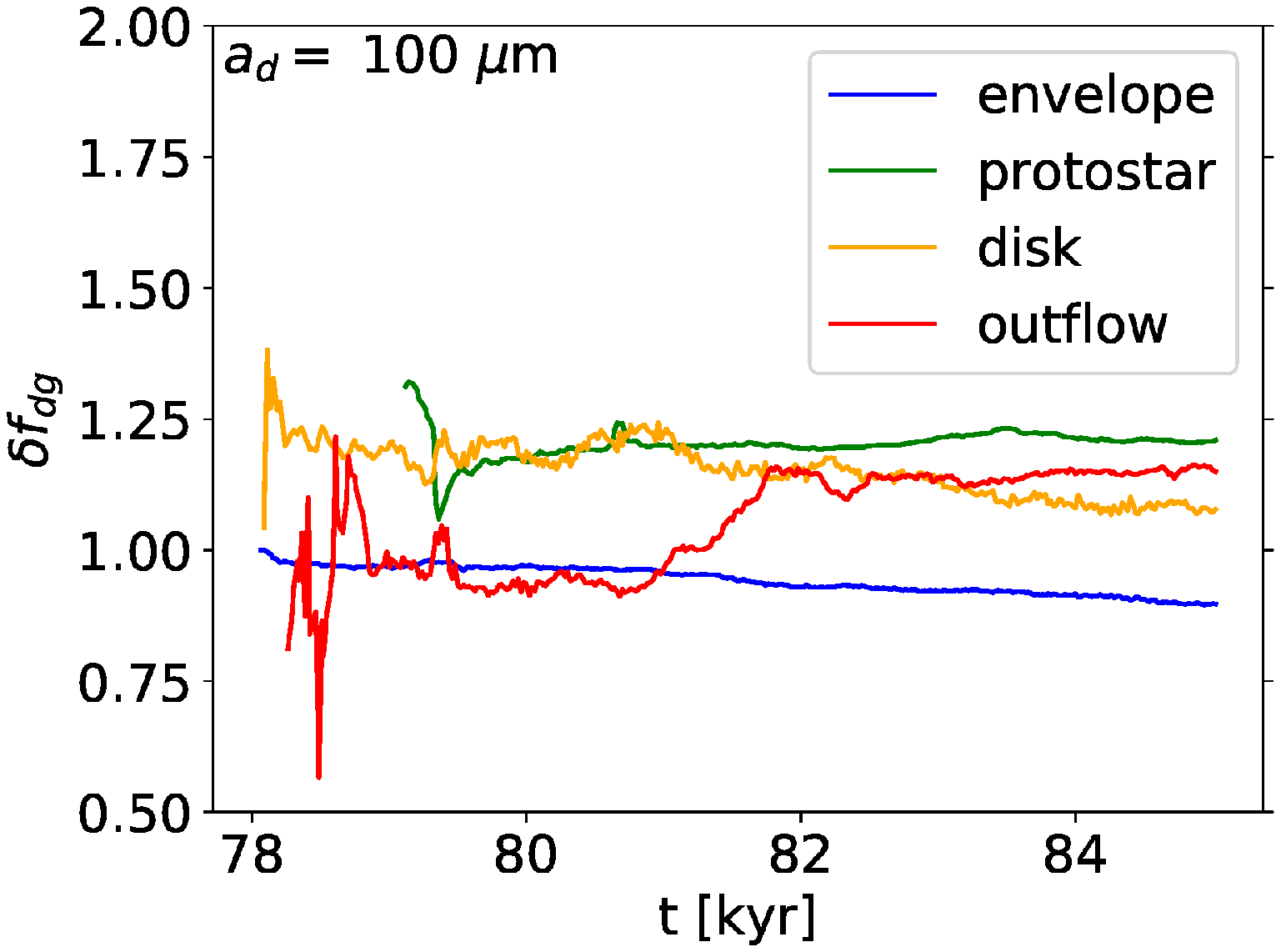}
\end{minipage}
\begin{minipage}{0.33\hsize}
\includegraphics[width=\columnwidth]{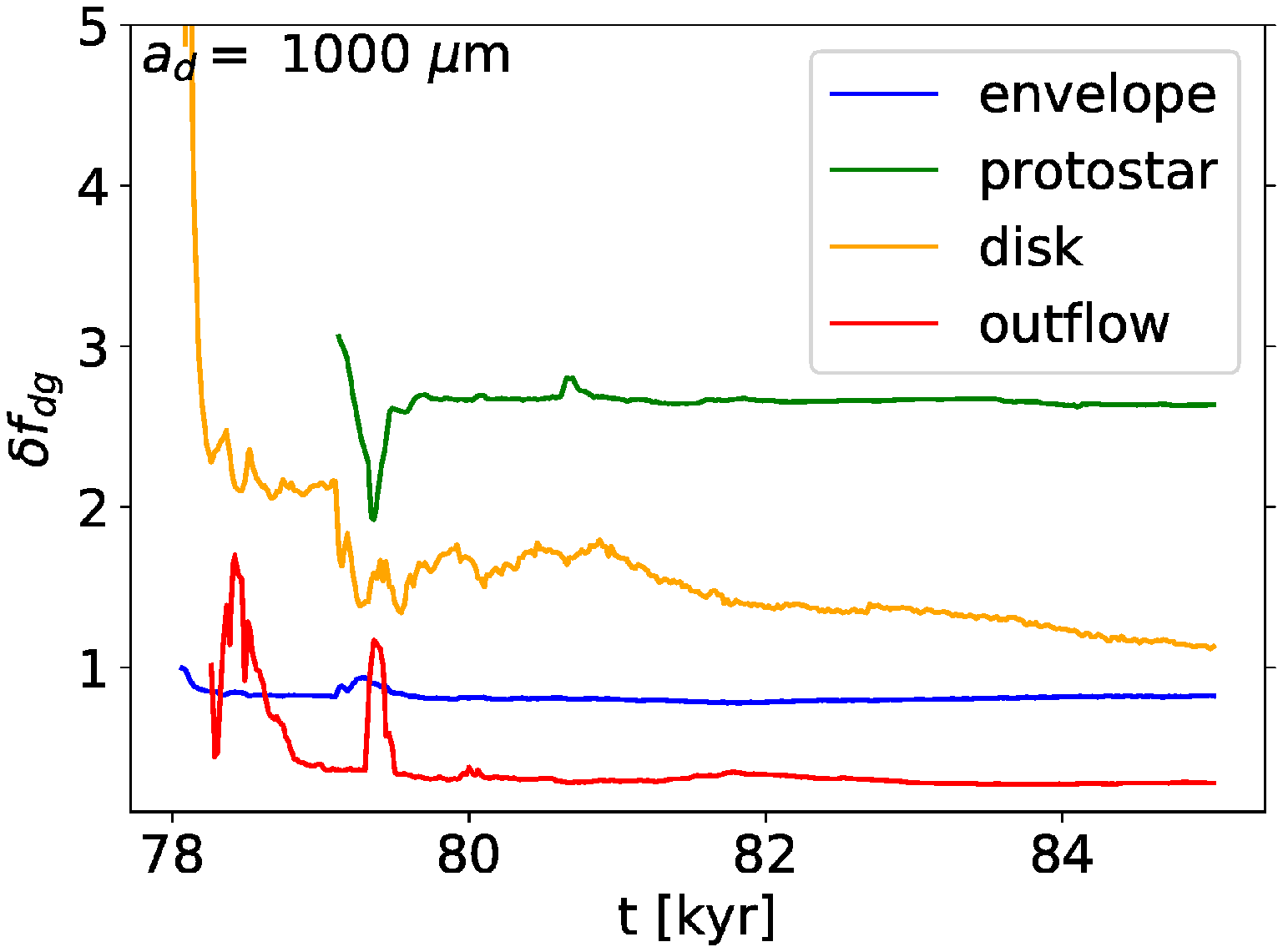}
\end{minipage}
\end{tabular}
\caption{Time evolution of change in dust-to-gas mass ratio $\delta f_{\rm dg}$ for various dust sizes ($a_{\rm d}=0.01$--1000\,$\mu{\rm m}$). 
In each panel, dust grains belonging to envelope (blue), protostar (green), disk (yellow), and outflow (red) regions are shown.
}
\label{fig:fdgregion}
\end{figure*}

\subsection{Dust-to-gas mass ratio}
\label{subsec:fdg}
\begin{figure*}
\begin{tabular}{ccc}
\begin{minipage}{0.33\hsize}
\includegraphics[width=\columnwidth]{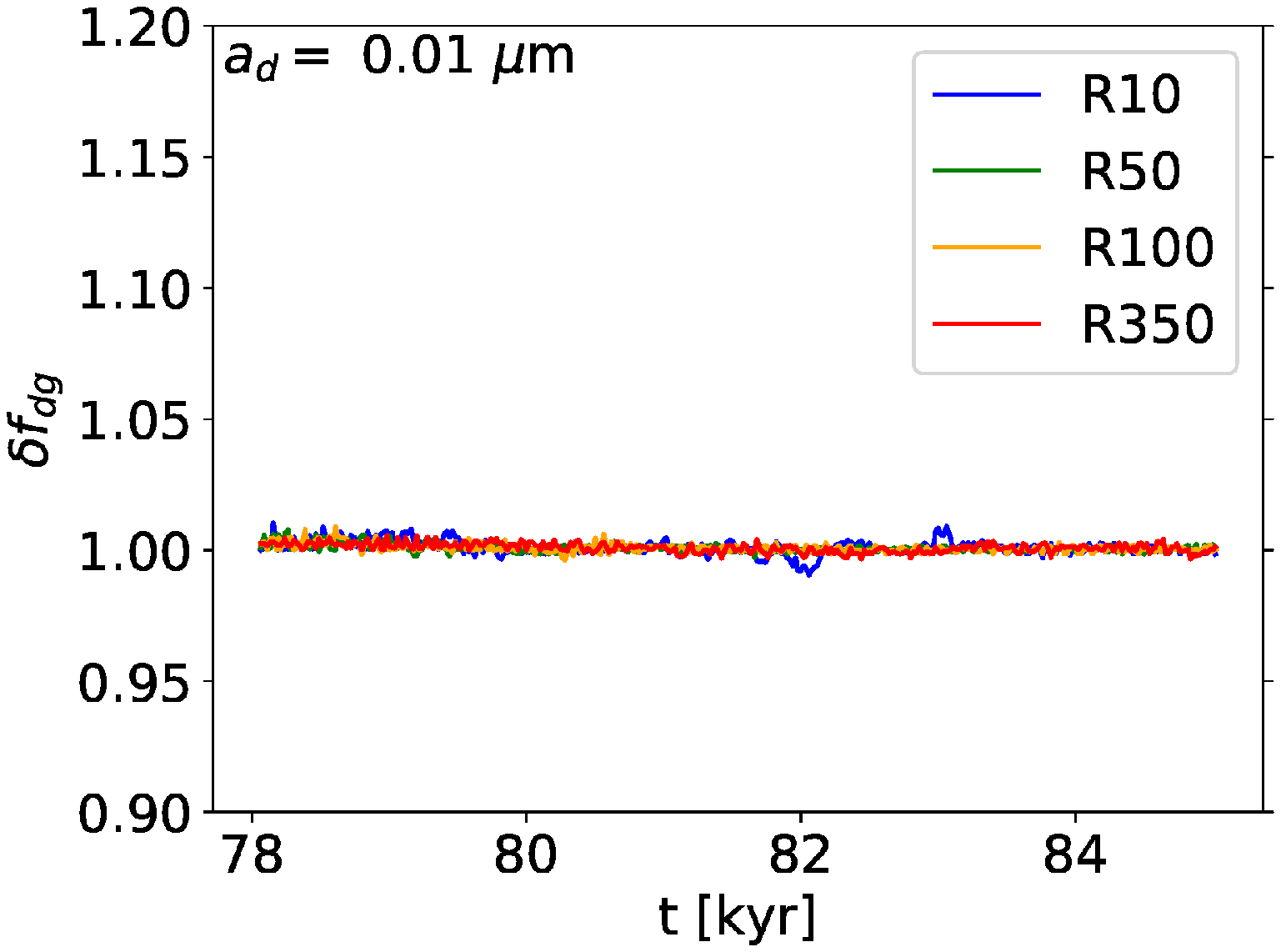}
\end{minipage}
\begin{minipage}{0.33\hsize}
\includegraphics[width=\columnwidth]{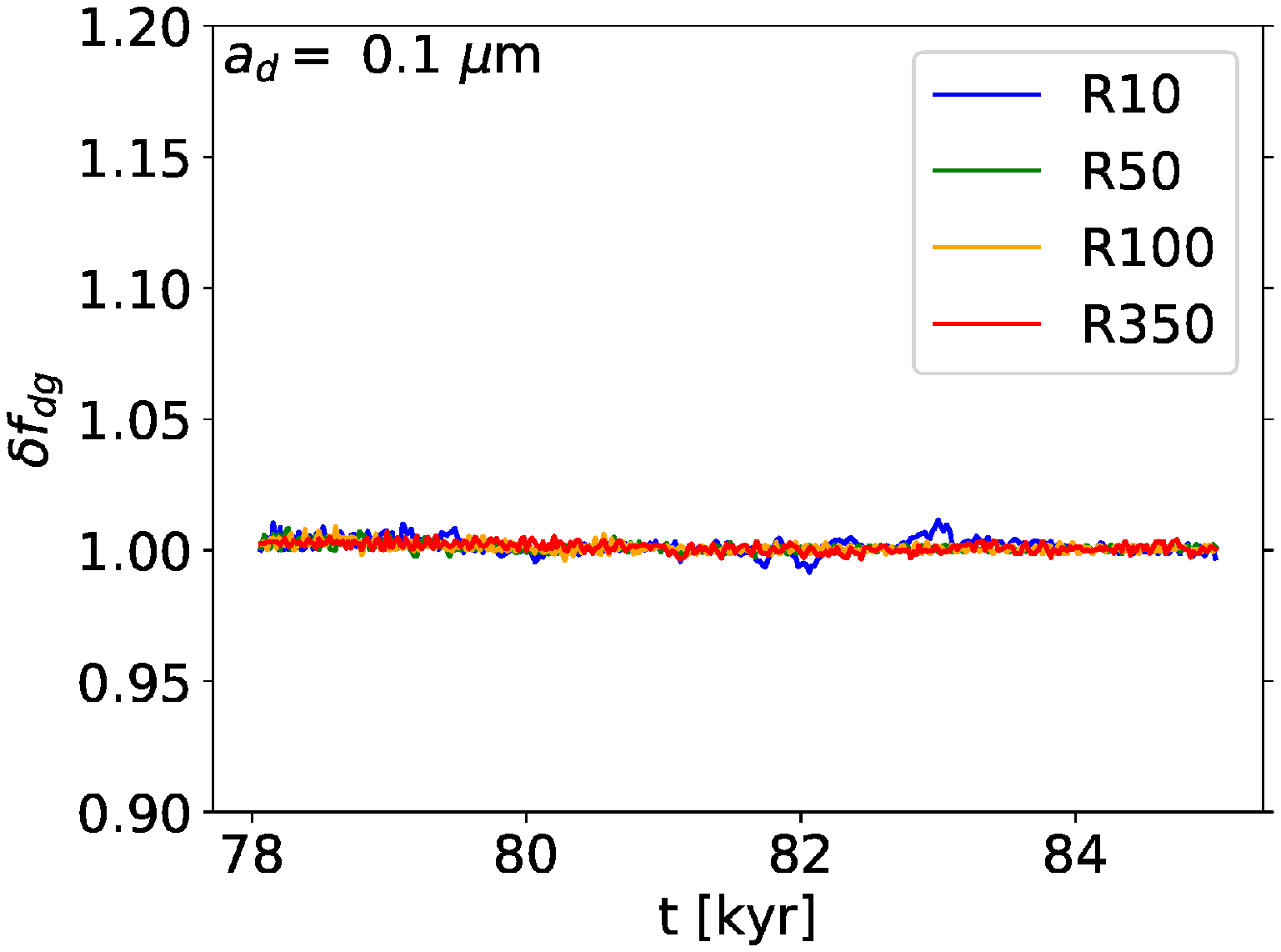}
\end{minipage}
\begin{minipage}{0.33\hsize}
\includegraphics[width=\columnwidth]{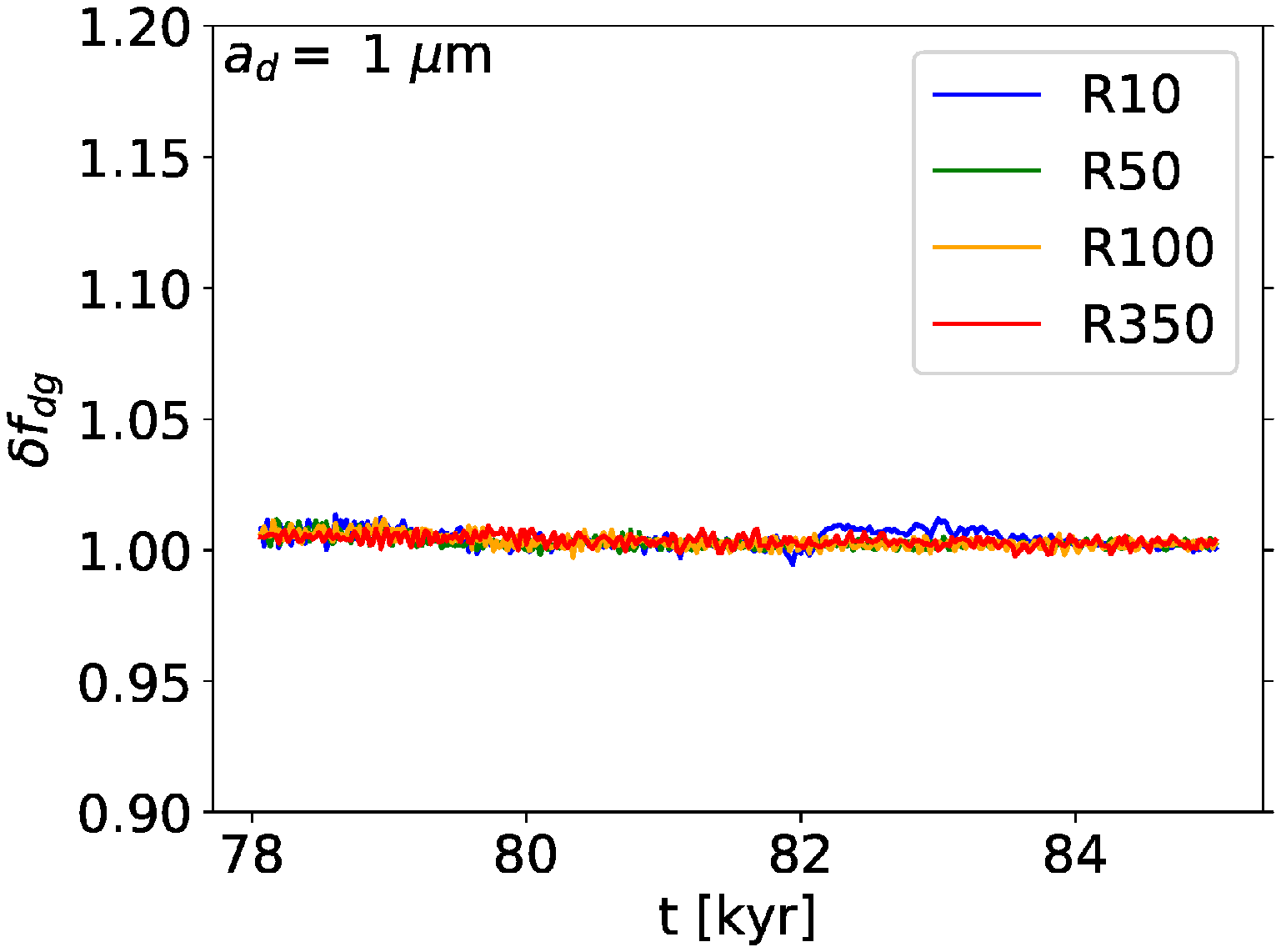}
\end{minipage}
\\
\begin{minipage}{0.33\hsize}
\includegraphics[width=\columnwidth]{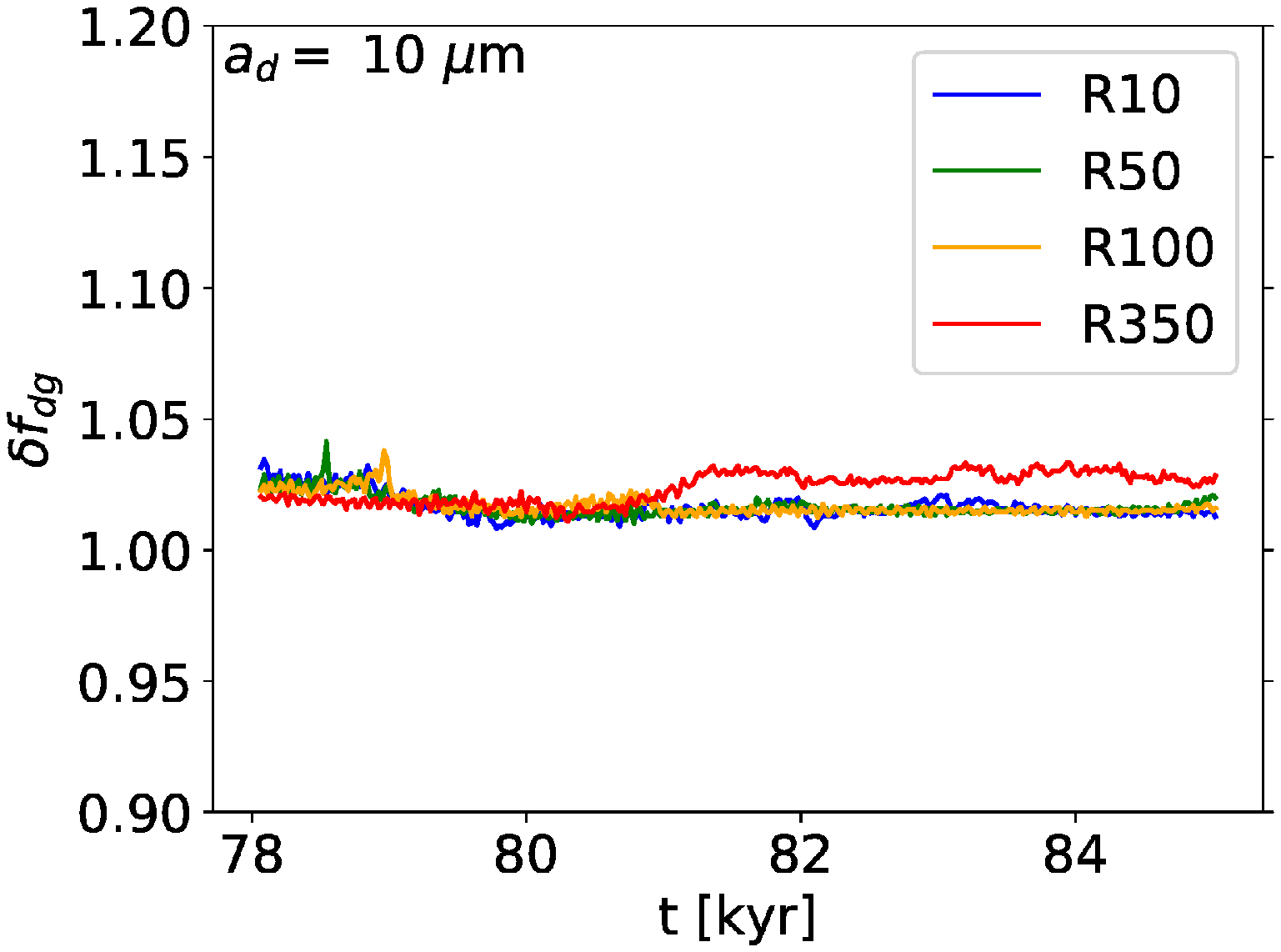}
\end{minipage}
\begin{minipage}{0.33\hsize}
\includegraphics[width=\columnwidth]{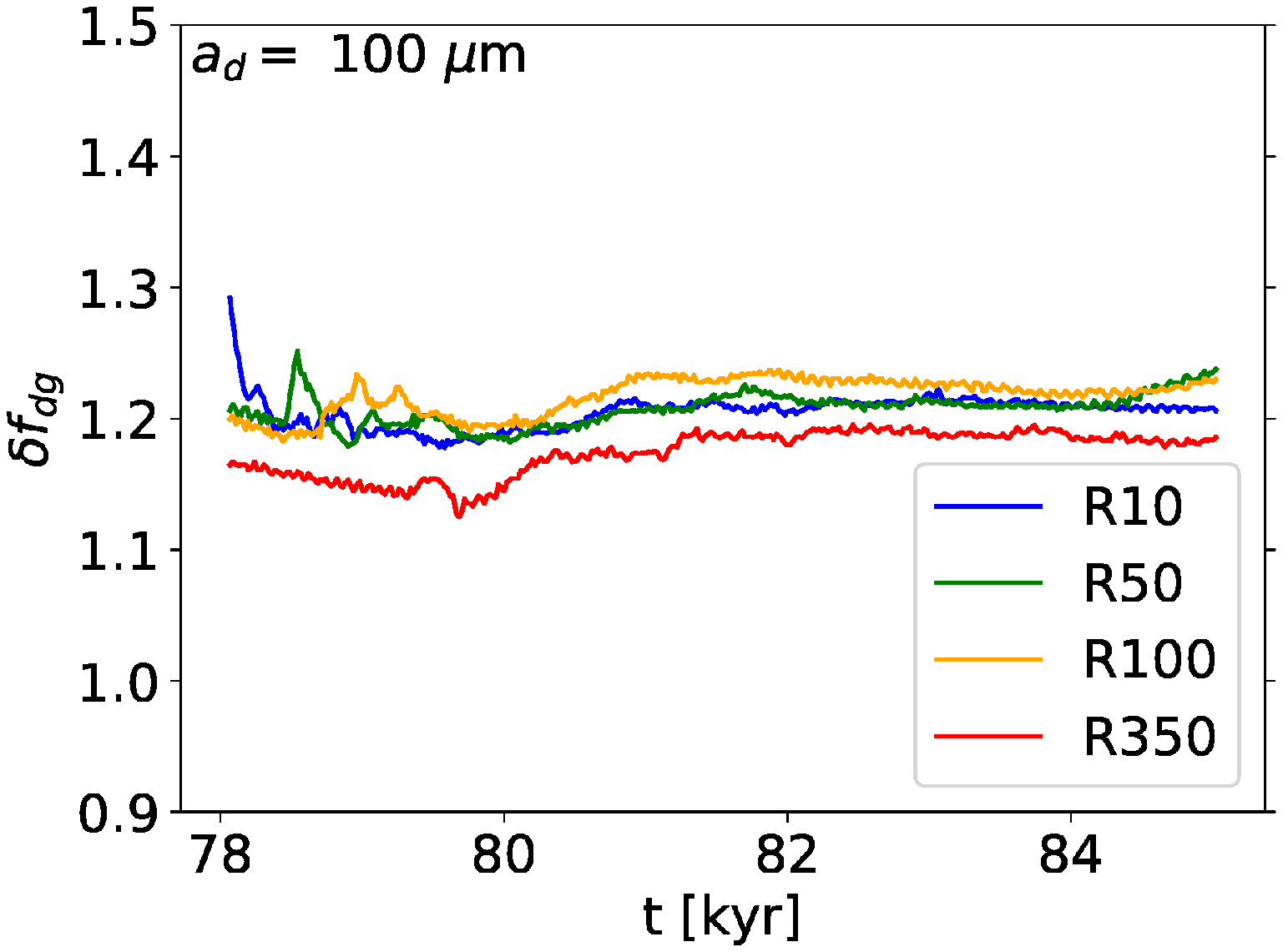}
\end{minipage}
\begin{minipage}{0.33\hsize}
\includegraphics[width=\columnwidth]{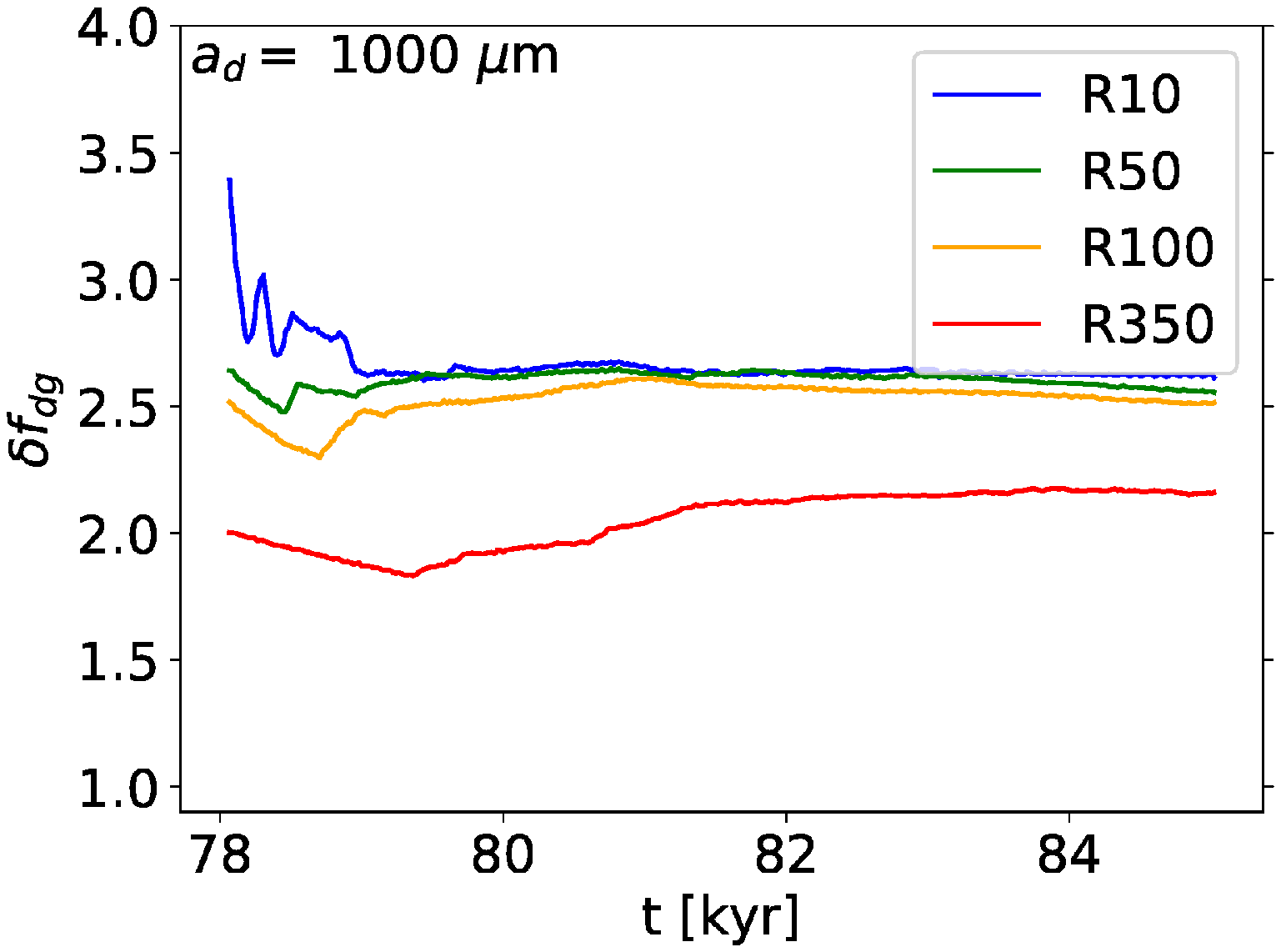}
\end{minipage}
\end{tabular}
\caption{Same as Fig.~\ref{fig:fdgregion} but curves represent scales, namely spheres within 10 (blue), 50 (green), 100 (yellow), and 350 (red)\,au. 
}
\label{fig:fdgscale}
\end{figure*}

\begin{figure*}
\begin{tabular}{ccc}
\begin{minipage}{0.33\hsize}
\includegraphics[width=\columnwidth]{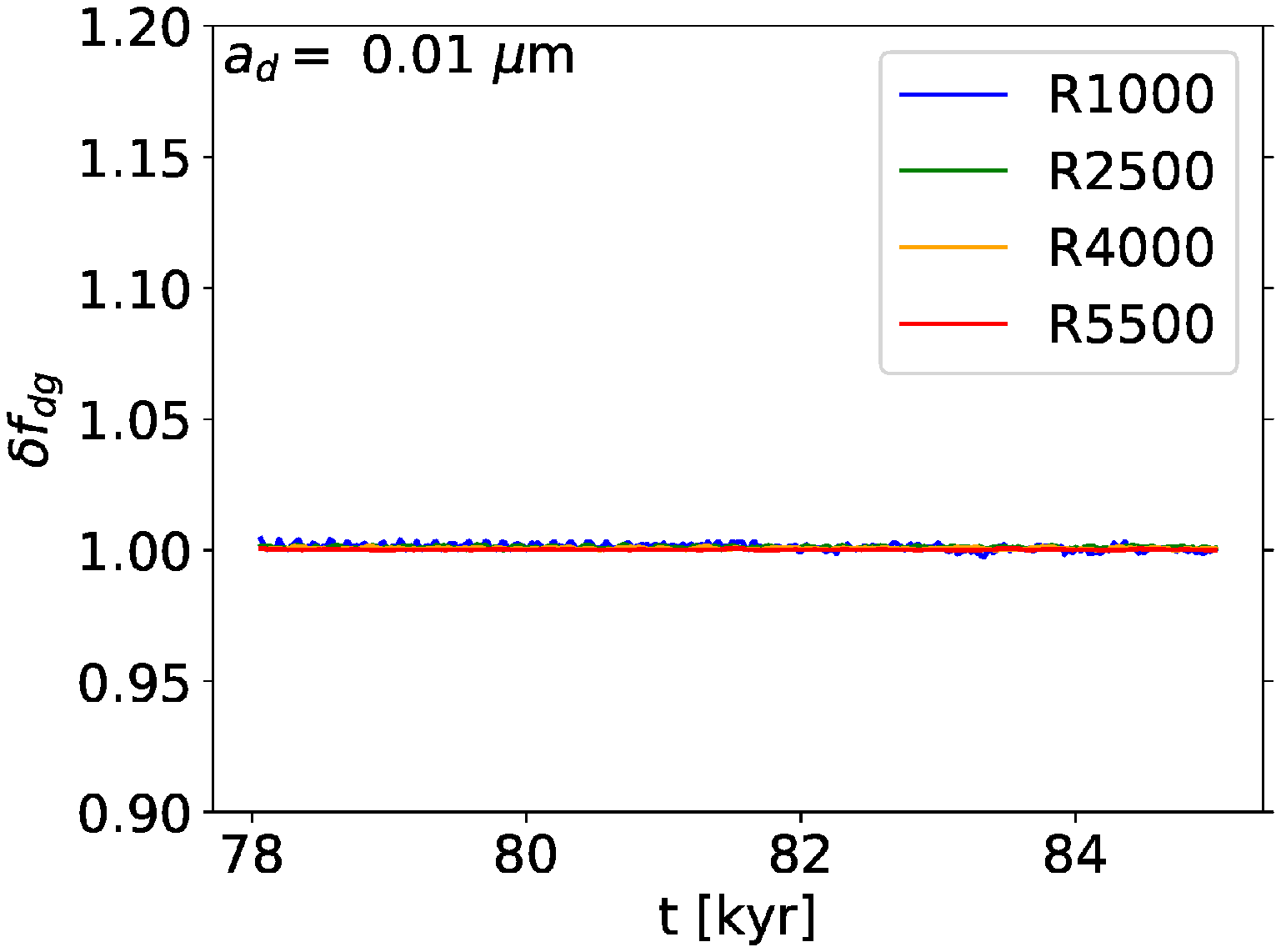}
\end{minipage}
\begin{minipage}{0.33\hsize}
\includegraphics[width=\columnwidth]{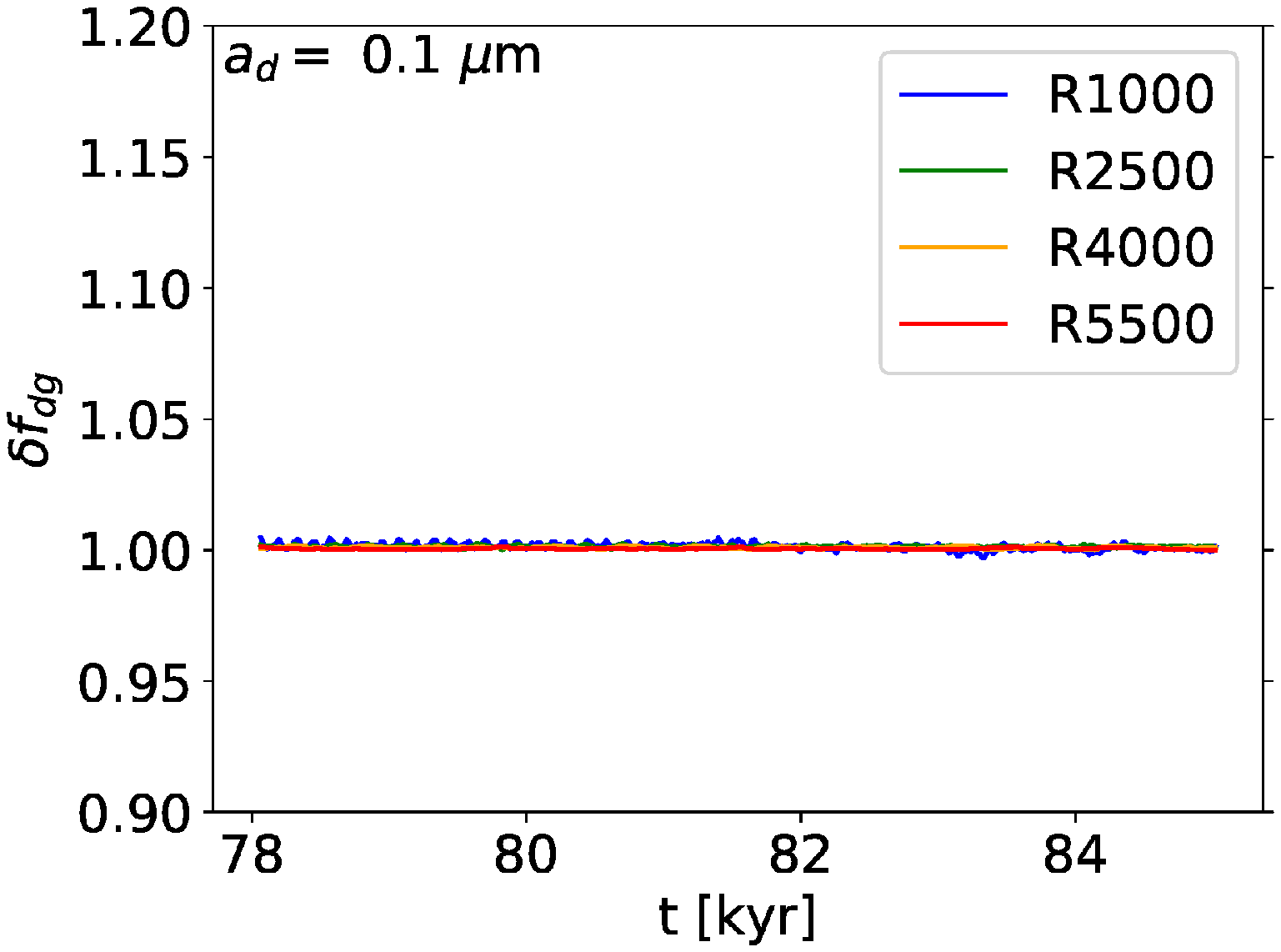}
\end{minipage}
\begin{minipage}{0.33\hsize}
\includegraphics[width=\columnwidth]{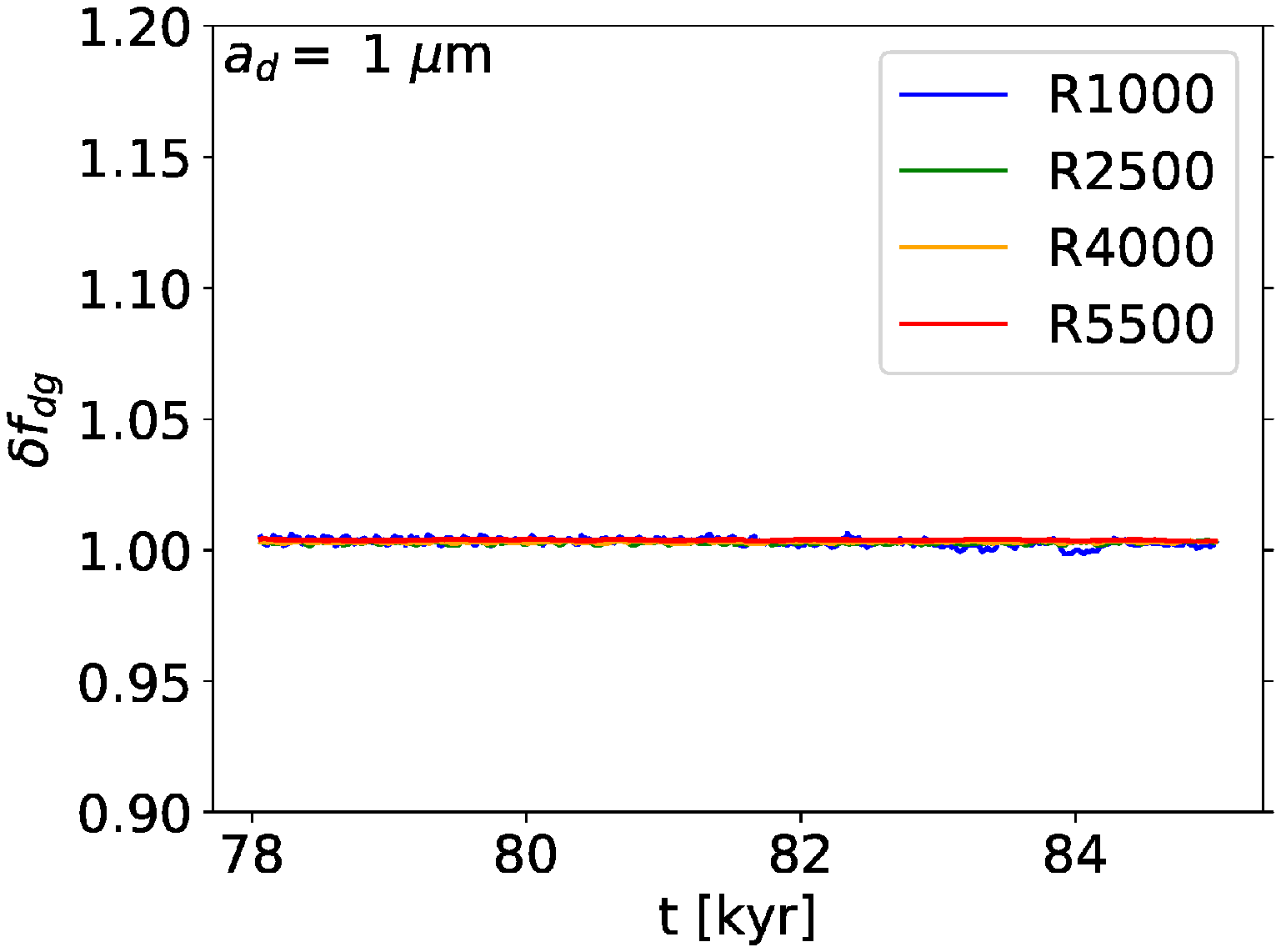}
\end{minipage}
\\
\begin{minipage}{0.33\hsize}
\includegraphics[width=\columnwidth]{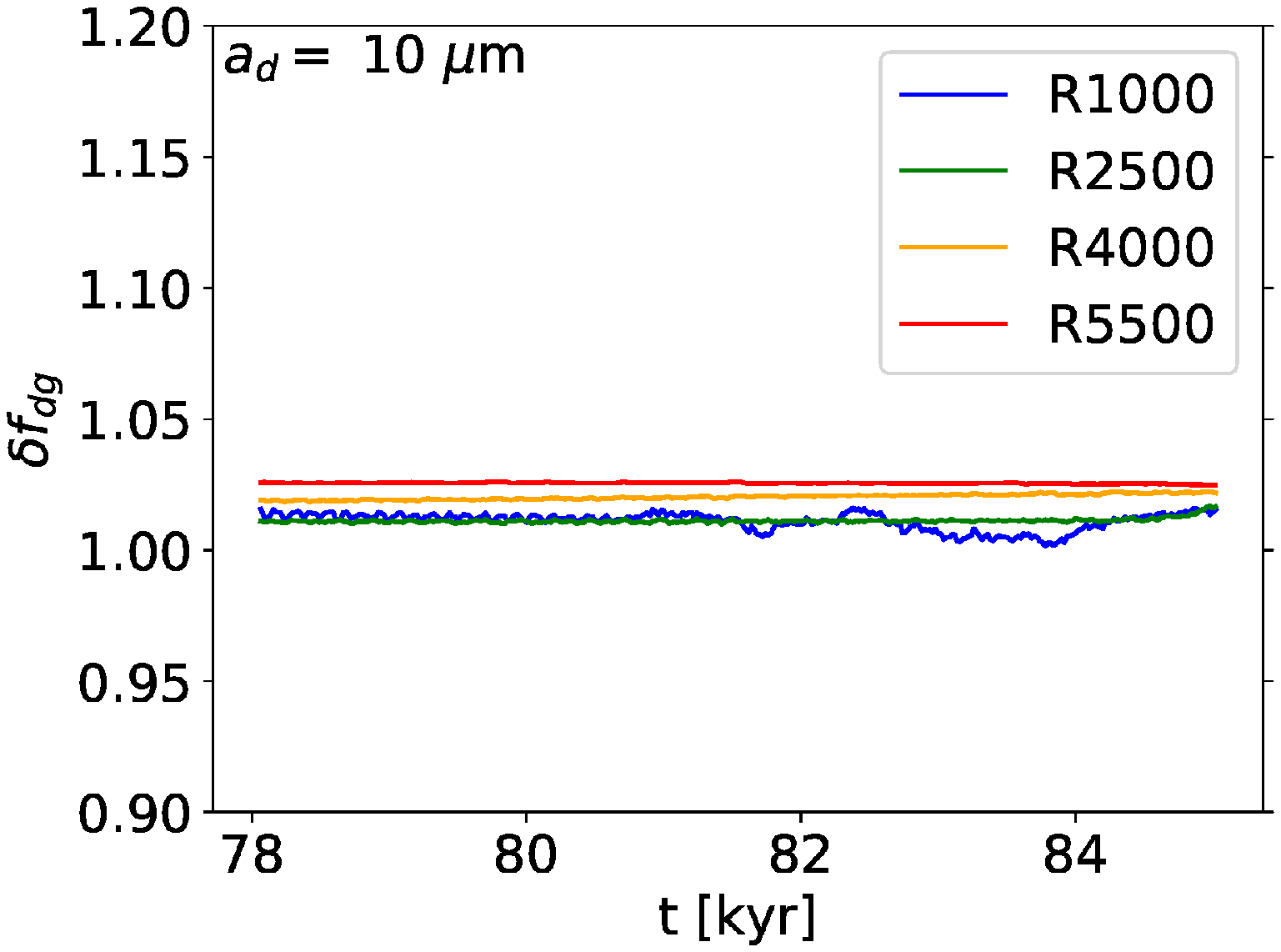}
\end{minipage}
\begin{minipage}{0.33\hsize}
\includegraphics[width=\columnwidth]{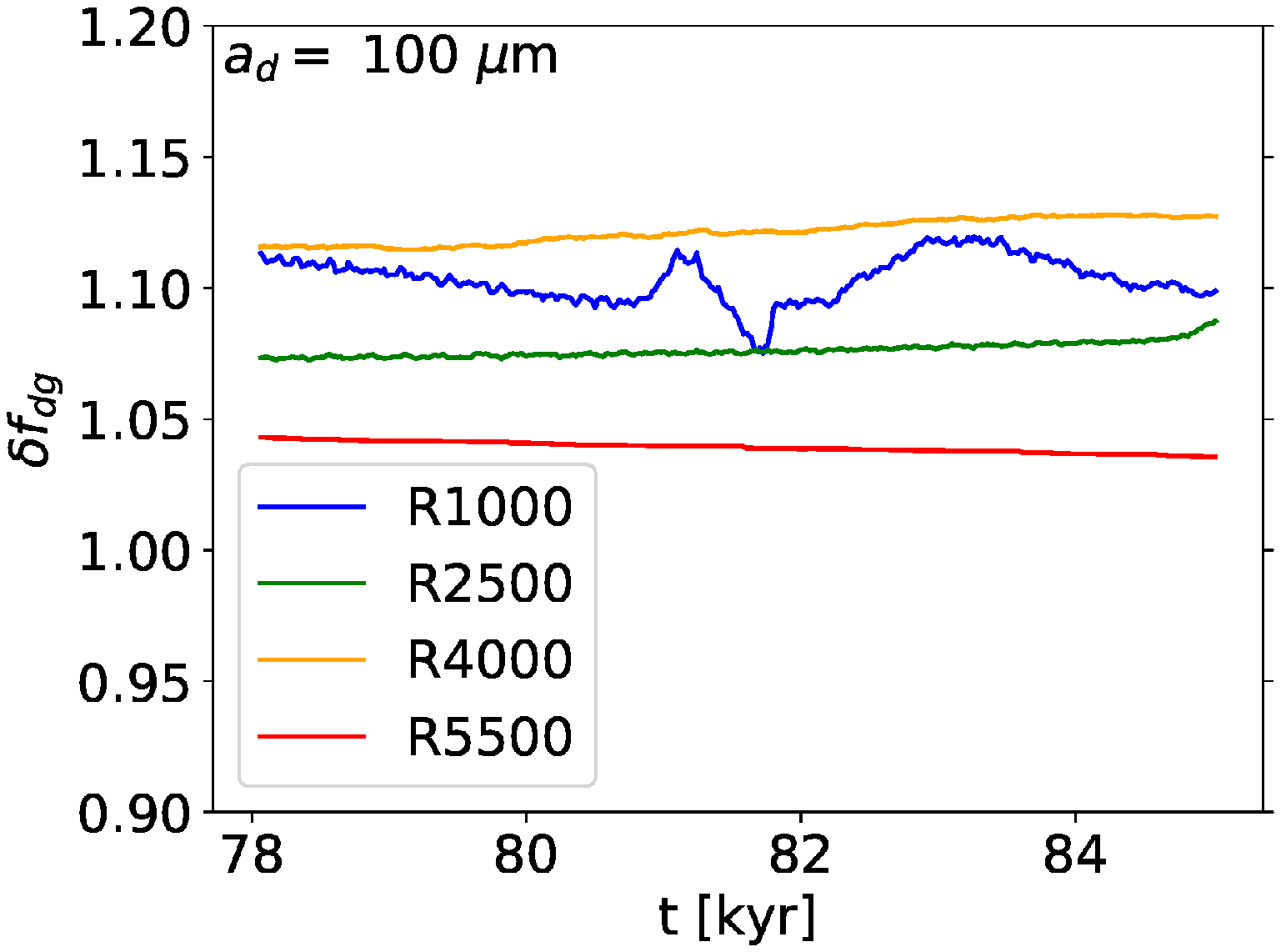}
\end{minipage}
\begin{minipage}{0.33\hsize}
\includegraphics[width=\columnwidth]{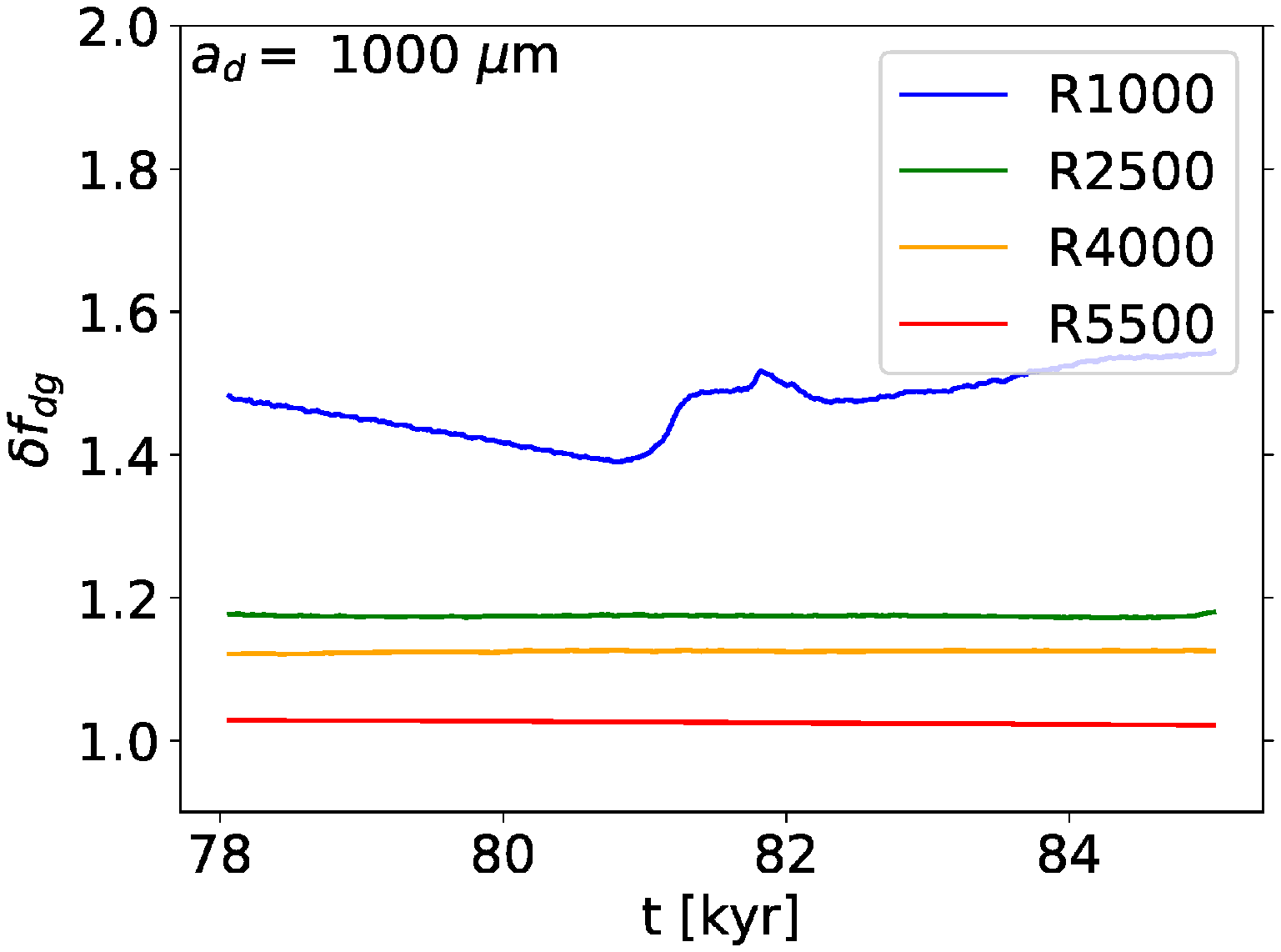}
\end{minipage}
\end{tabular}
\caption{Same as Fig.~\ref{fig:fdgregion} but curves represent scales, namely 1000 (blue), 2500 (green), 4000 (yellow), and 5500 (red)\,au. 
}
\label{fig:2fdgscale}
\end{figure*}

Figs.~\ref{fig:fdgregion}, \ref{fig:fdgscale}, and \ref{fig:2fdgscale} show the time evolution of the change in the dust-to-gas mass ratio $\delta f_{\rm dg}$ defined in \S\ref{subsec:particleweight}. 
Fig.~\ref{fig:fdgregion} plots $\delta f_{\rm dg}$ for the envelope, protostar, disk, and outflow regions (for their definitions, see \S\ref{subsec:criteria}). 
For dust with a size of $0.01\,{\rm \mu m} \leq a_{\rm d} \leq 10\,{\rm \mu m}$, $\delta f_{\rm dg}$ is in the range of 0.9--1.1.
Thus, $f_{\rm dg}$ changes  within 10 \% from the initial value.
Dust grains with a size of $a_{\rm d} \geq 100 \ {\rm \mu m}$ have a  noticeable difference from those with a size of $a_{\rm d} \leq 10 \ {\rm \mu m}$  in terms of the time evolution of $f_{\rm dg}$.
Dust particles with a size of $a_{\rm d} = 100\,{\rm \mu m}$ are concentrated in not only the relatively high-density regions (protostar and disk) but also the low-density region (outflow) because they are stirred up by the outflow.
$\delta f_{\rm dg}$ for dust particles with a size of $a_{\rm d} = 100\,{\rm \mu m}$  decreases to $f_{\rm dg}=0.8$ in the envelope region  at the end of the simulation.

The bottom-right panel of Fig.~\ref{fig:fdgregion} indicates that dust grains with a size of $a_{\rm d} = 1000 \ {\rm \mu m}$ tend to rapidly fall into the center and behave very differently from those with a size of $a_{\rm d} = 100 \ {\rm \mu m}$ in the outflow region.
Dust particles with a size of $a_{\rm d} = 1000\,{\rm \mu m}$ are exhausted in the low-density gas regions (envelope and outflow) because they are decoupled from the lower-density gas.
For dust particles with a size  of $a_{\rm d} = 1000 \ {\rm \mu m}$, $\delta f_{\rm dg}$ in the outflow region decreases to 0.3 and that in the protostar region increases to 3.0.

Figs.~\ref{fig:fdgscale} and \ref{fig:2fdgscale} plot $\delta f_{\rm dg}$ with different spatial scales, where we estimated the change in the dust-to-gas mass ratio within a sphere with radius $R$. 
These figures indicate that  the time evolution of the spatial distribution of $\delta f_{\rm dg}$ strongly depends on the dust size. 
As shown in Fig.~\ref{fig:fdgregion}, dust particles with a size of $0.01\,{\rm \mu m} \leq a_{\rm d} \leq 10\,{\rm \mu m}$ are coupled with gas due to their short stopping time.
Particles with a size larger than $100\,{\rm \mu m}$ fall into the center significantly faster than does the gas; this is exaggerated at smaller scales.
Thus, large dust particles should tend to be more concentrated in the center in the early stages of the simulation.
Dust particles with a size of $100\,{\rm \mu m}$ behave differently from those with a size of $1000\,{\rm \mu m}$ because the former are effectively swept up by the gas outflow.

\subsection{Stokes number}
\label{sec:stokesnumber}
\begin{figure*}
\includegraphics[width=\linewidth]{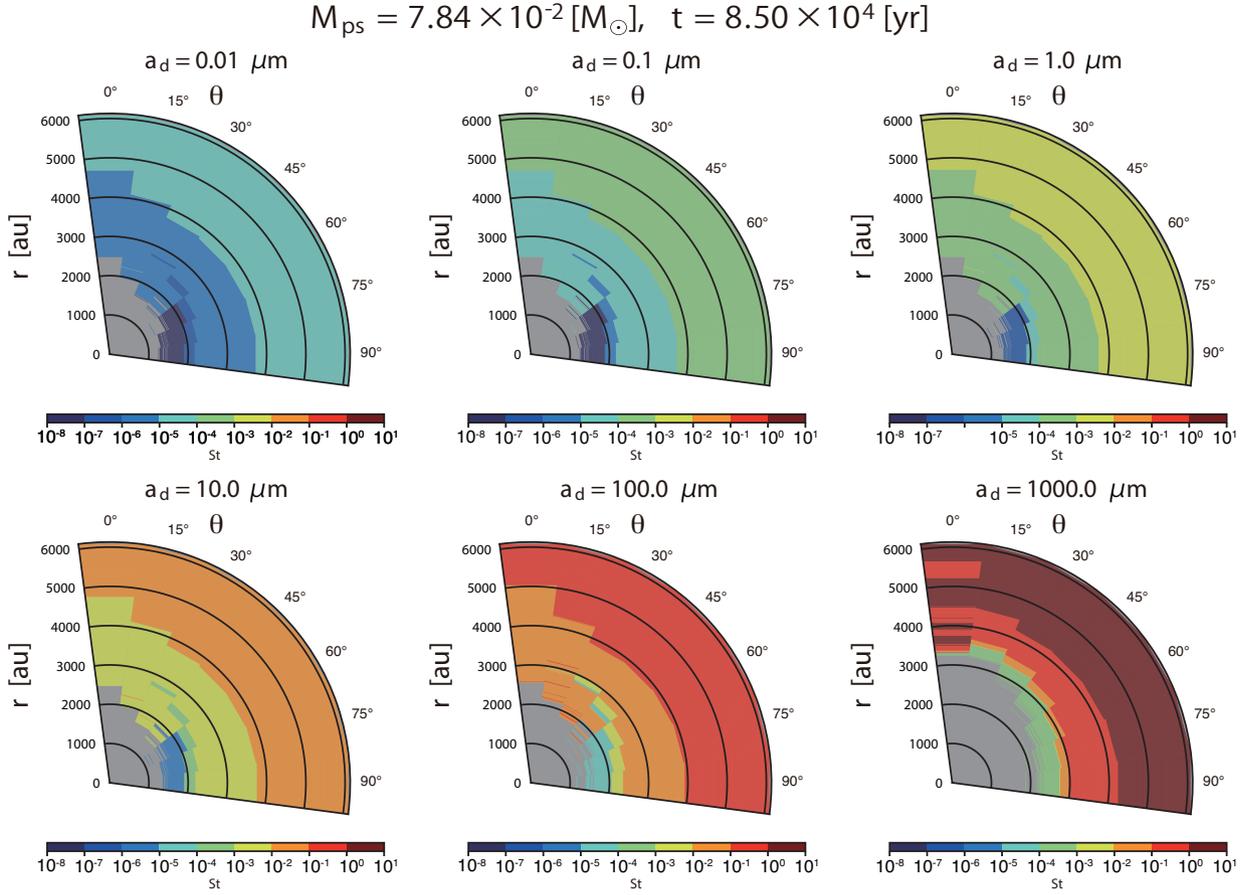}
\caption{
Same as Fig.~\ref{fig:dustpos} but color indicates  Stokes number St ($\equiv t_{\rm s} / t_{\rm ff}$) of  dust particle at each location.
}
\label{fig:dustposst}
\end{figure*}

In this section, we discuss how strong the dust particles are coupled with gas.
To evaluate the coupling strength, the Stokes number defined in \S \ref{subsubsec:dust} is used. 
Fig.~\ref{fig:dustposst} shows the Stokes number St  of dust particles at each location.
Large grains tend to have a large St value because the stopping time $t_{\rm s}$ is proportional to grain size (see equation~(\ref{tstop})).
In the figure, the St value of dust particles with a size of $0.01 \,{\rm \mu m} \leq a_{\rm d} \leq 100 \,{\rm \mu m}$ is always below unity (i.e., St $<1$).
Thus, these grains are coupled with the gas.
Conversely, the St value of some dust grains with a size of $a_{\rm d} = 1000 \,{\rm \mu m}$ can exceed unity (i.e., St $>1$). 
St increases in the low density gas region because the stopping time $t_{\rm s}$ becomes long.
In other words, the momentum transfer from gas to dust grains becomes inefficient in this region.
This is confirmed by equation~(\ref{tstop}), where the stopping time $t_{\rm s}$ is inversely proportional to the gas density.
It should be noted that the equation of motion for dust particles used in this study may not be fully appropriate when the Stoke number exceeds unity, as described in \S\ref{sec:appendix}.  
The Stokes number can exceed unity when the dust particles have a size of $a_{\rm d} = 1000 \,{\rm \mu m}$, as described above. 
Although we do not further step into the formulation of the equation of motion for large-sized dust grains in this study, we need to care about the behavior of such grains. 
We will focus on the treatment of large-sized grains in our future study. 

\begin{figure*}
\begin{tabular}{ccc}
\begin{minipage}{0.33\hsize}
\includegraphics[width=\columnwidth]{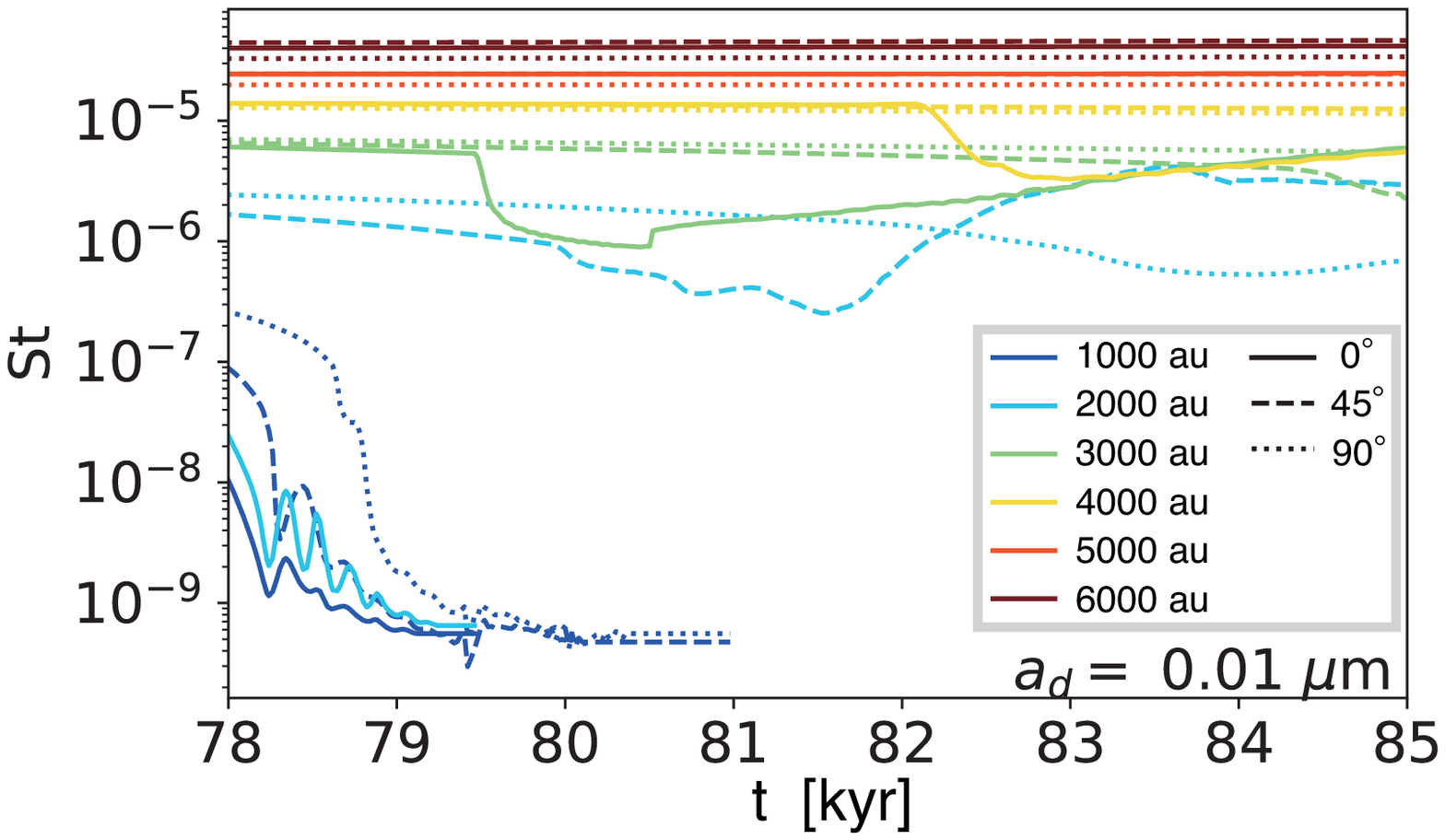}
\end{minipage}
\begin{minipage}{0.33\hsize}
\includegraphics[width=\columnwidth]{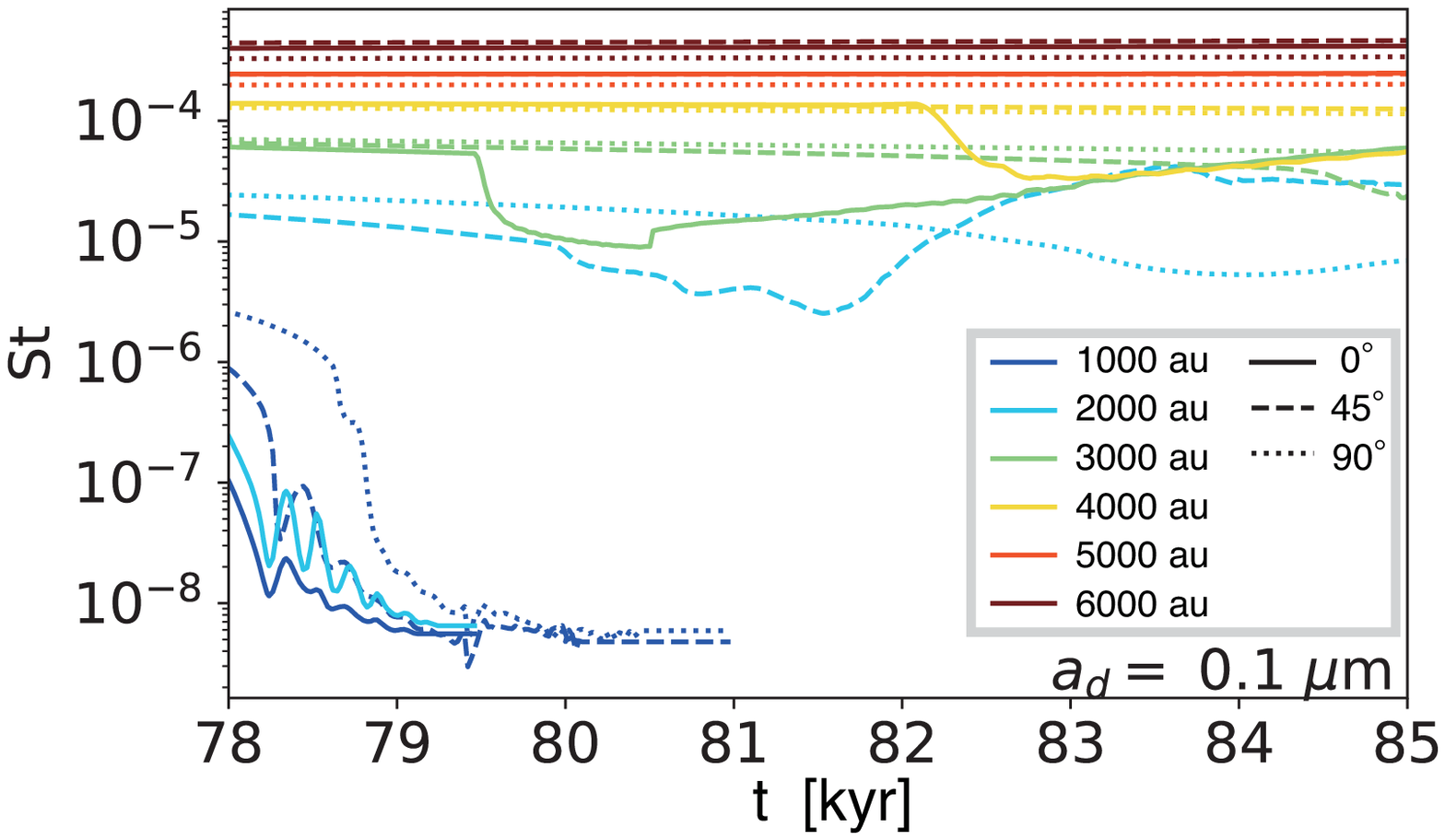}
\end{minipage}
\begin{minipage}{0.33\hsize}
\includegraphics[width=\columnwidth]{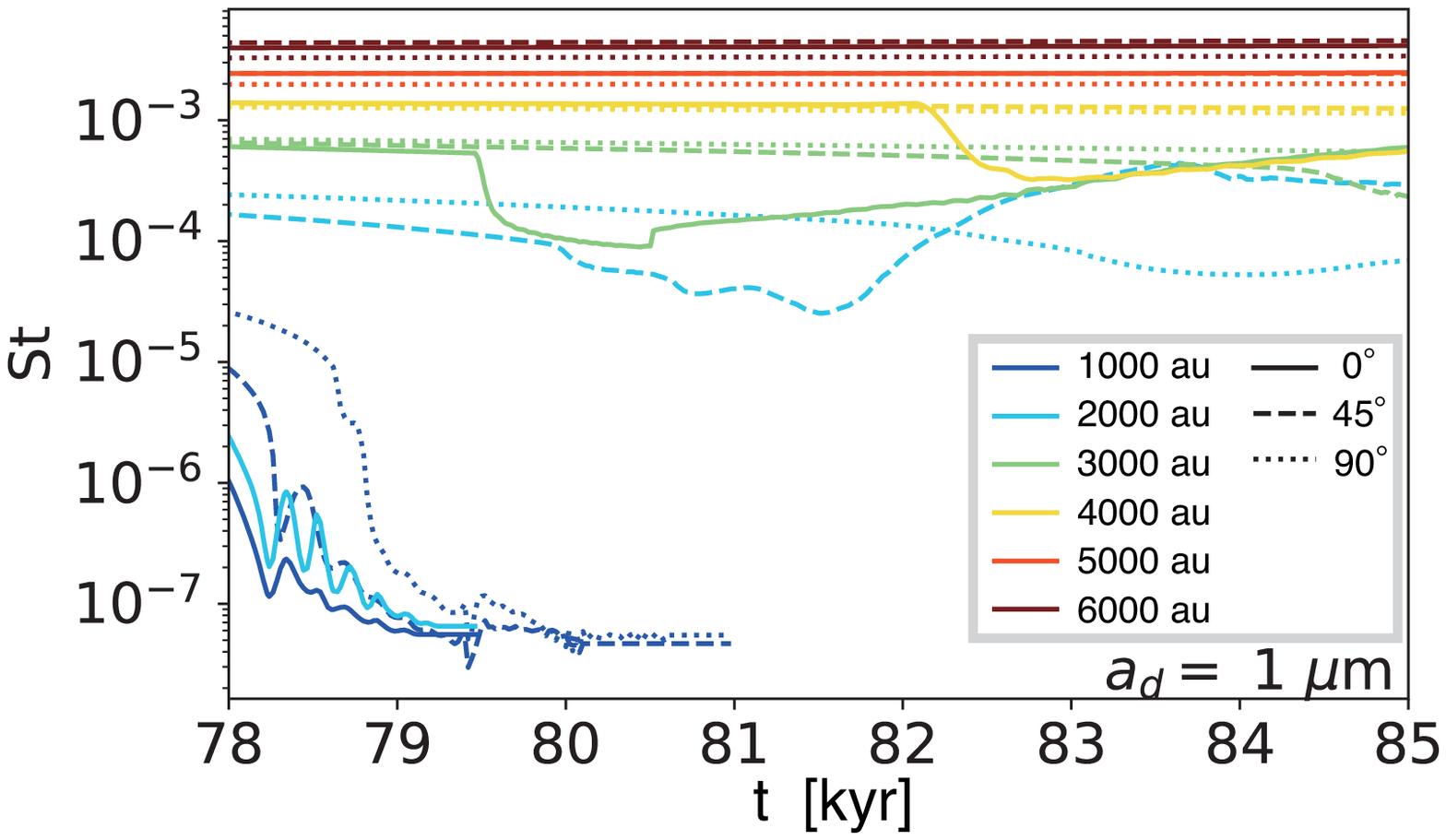}
\end{minipage}
\\
\\
\begin{minipage}{0.33\hsize}
\includegraphics[width=\columnwidth]{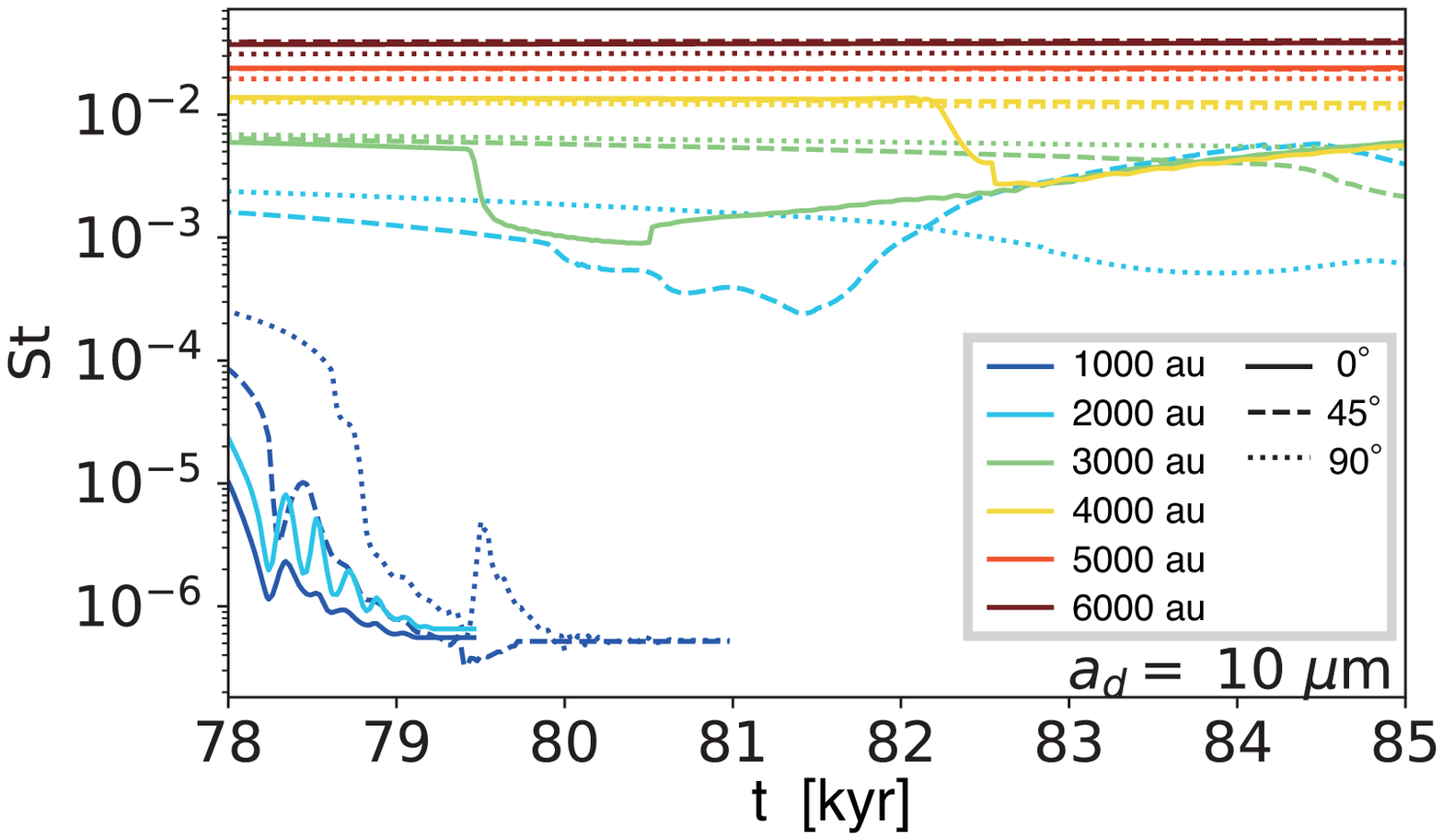}
\end{minipage}
\begin{minipage}{0.33\hsize}
\includegraphics[width=\columnwidth]{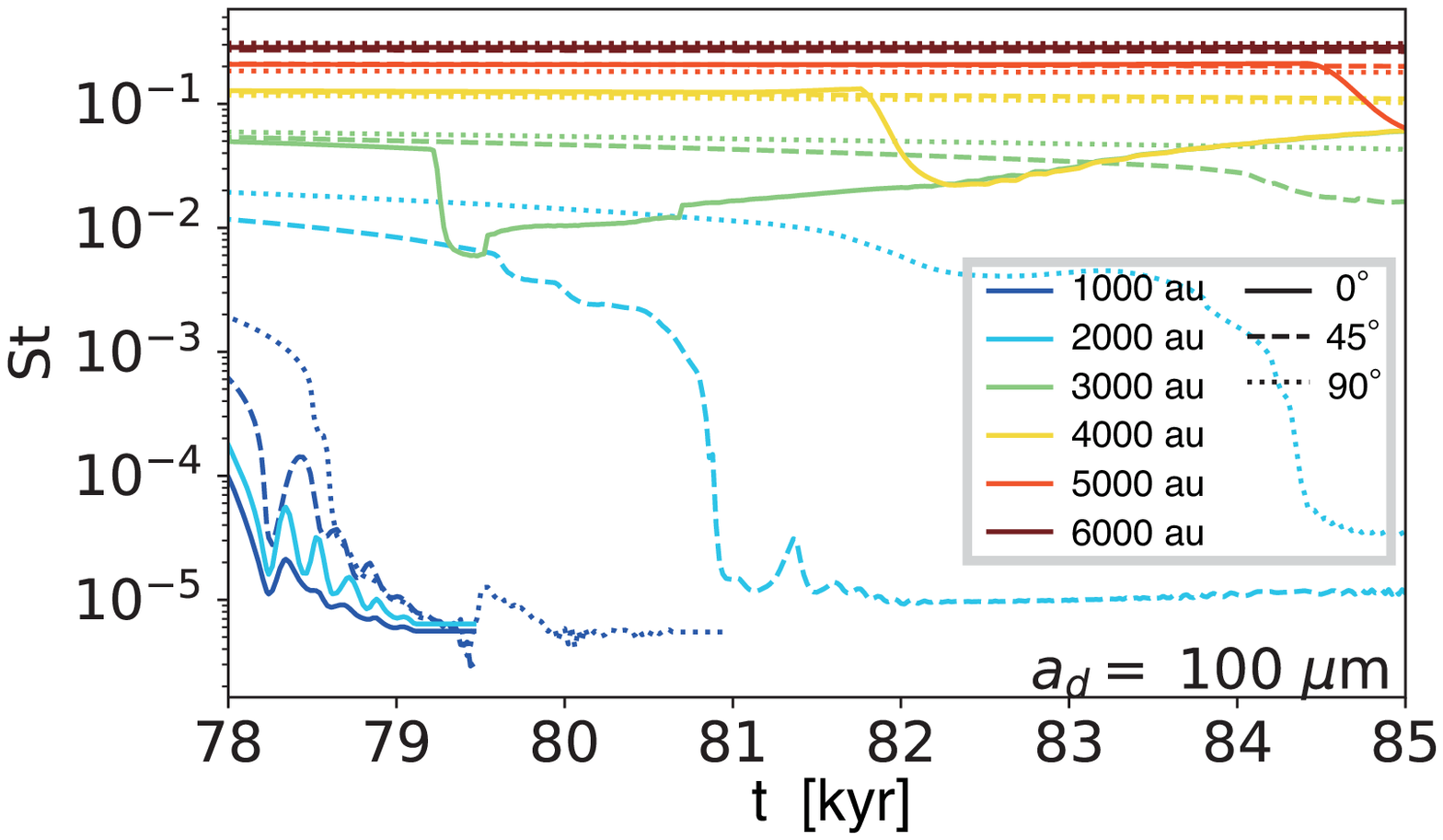}
\end{minipage}
\begin{minipage}{0.33\hsize}
\includegraphics[width=\columnwidth]{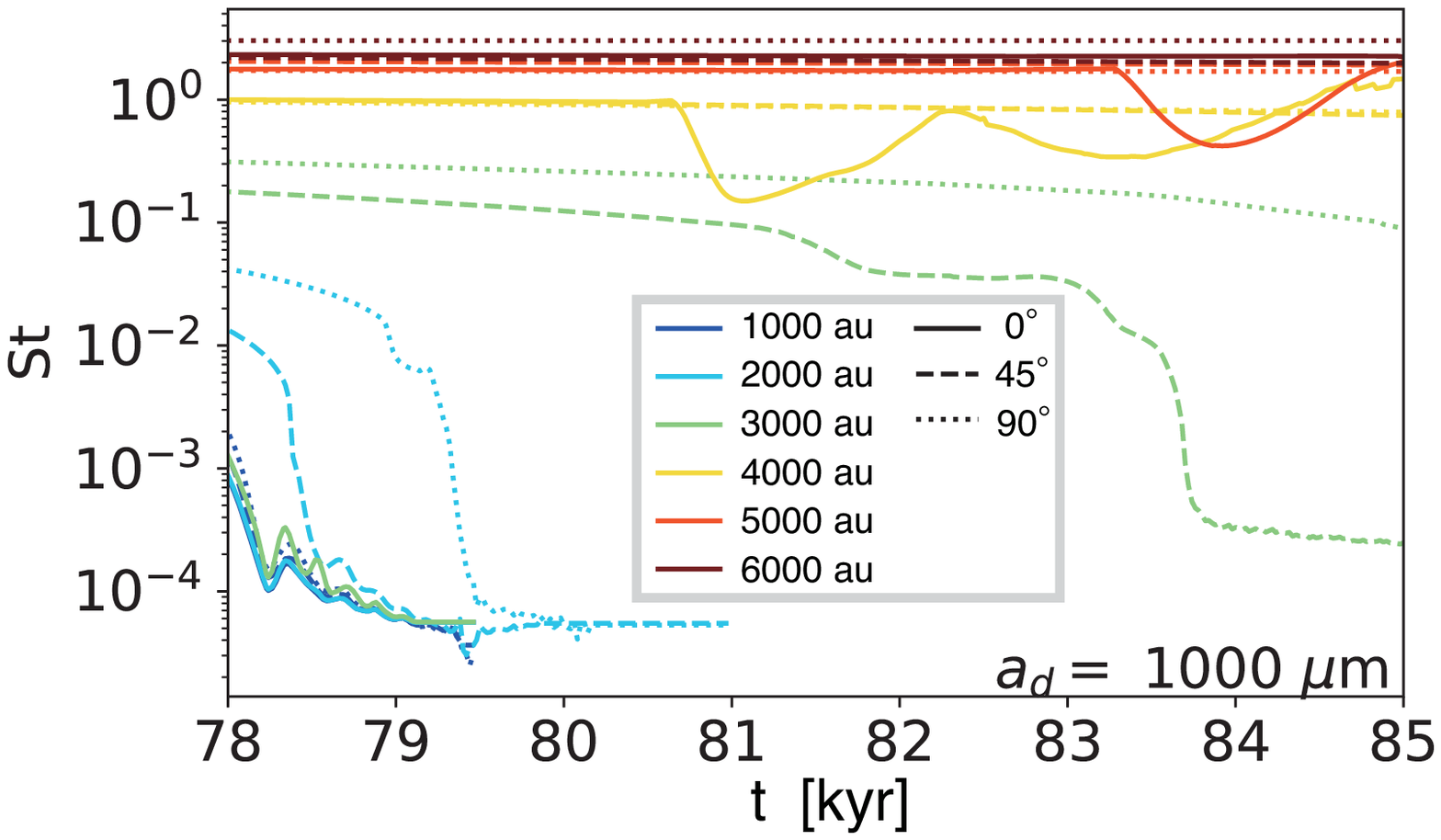}
\end{minipage}
\end{tabular}
\caption{
Stokes number St of selected particles placed at initially different locations versus elapsed time. 
In each panel, the dust size $a_{\rm d}$, shown in the bottom-right corner,  is the same. 
The initial distance from the center for each particle is represented by the color and the initial angle $\theta$ is represented by the line style.
Curves that stop before the end of the simulation are for particles that fell onto the sink.
}
\label{fig:tst}
\end{figure*}

Fig.~\ref{fig:tst} shows the time evolution of St for several selected particles placed at initially different locations.
In each panel, St decreases when the dust particle remains in a high density gas region (e.g., disk).
Dust grains with a size of $0.01 \,{\rm \mu m} \leq a_{\rm d} \leq 10 \,{\rm \mu m}$ move in the gas envelope with a relatively large St.
When the dust grains move into the disk, where the density is high, St decreases significantly, for example, at 79.5 kyr for dust particles with a size of $a_{\rm d}= 0.01\,{\rm \mu m}$ initially located at $r$ = 3000 au (green) and $\theta = 0^\circ$ (solid). 
However, as shown in each panel,  the St value of these grains never reaches unity (i.e., St $<1$).
Thus, these particles are coupled with the gas during the simulation.  

Dust grains with a size of $a_{\rm d}\ge 100\,{\rm \mu m}$ have St $\sim1$ in the envelope. 
Thus, these grains fall into the central region faster than does the gas because they are (partially) decoupled from the gas fluid. 
Therefore, the dust-to-gas mass ratio can significantly change in the envelope, as shown in Figs.~\ref{fig:fdgregion}, \ref{fig:fdgscale}, and \ref{fig:2fdgscale}. 
However, the St value for these grains significantly decreases as they approach the center or the disk. 
As shown in Fig.~\ref{fig:tst}, even for these dust grains, the Stokes number becomes St $\lesssim 10^{-3}$ within $r\lesssim 1000$--$2000$\,au, inside which the disk is embedded.
Thus, although dust grains with a size of $a_{\rm d}\ge 100\,{\rm \mu m}$ rapidly fall onto the disk, they are coupled with the gas and move together with the rotating fluid inside the disk.  
The detailed dynamics of the dust grains inside the disk will be described in the next paper.

\section{Discussion}
\label{sec:discussion}
\subsection{Comparison with previous MHD simulations}
Recently, \citet{2020A&A...641A.112L} investigated dust dynamics during the core collapse phase using three-dimensional MHD simulations that included dust as a fluid.
Here, we adopted the sink cell as a protostar; this was not done by \citet{2020A&A...641A.112L}. 
Thus, there are some differences in the treatment of dust and the protostar. 
A significant difference is the feedback from dust to gas (or the dust back reaction onto the gas). 
Though the feedback was ignored in this study, it is included in \cite{2020A&A...641A.112L}. 
Nonetheless, our results are qualitatively and quantitatively consistent with \citet{2020A&A...641A.112L}.

The features of the dust concentration presented in our study qualitatively agree with \citet{2020A&A...641A.112L}, who reported that large dust particles partially decoupled from the gas in the high-density regions near the disk and  protostar (called fragments in \citealt{2020A&A...641A.112L}), and the dust particles were depleted in the low-density regions of the outflow and envelope.
Our results are also quantitatively consistent with \citet{2020A&A...641A.112L}, in which dust particles that exceeded $10 \,{\rm \mu m}$ were decoupled from the gas in the early star and disk formation processes.

Strictly speaking, there exists a difference in the change in the dust-to-gas mass ratio ($\delta f_{\rm dg}$ in this study is described as $\epsilon$ in \citealt{2020A&A...641A.112L}) between this study and their study.
The $\delta f_{\rm dg}$ values in the present study are at most 2 or 3 times smaller. 
In \citet{2020A&A...641A.112L}, for example, dust particles with a size of $100\,{\rm \mu m}$ are enhanced to about 3 (i.e., $\delta f_{\rm dg}=3$) in the disk.
In the present study, the change in the dust-to-gas mass ratio was $\delta f_{\rm dg}=1.2$ at most.
Thus, the $\delta f_{\rm dg}$ value in our study is about 2.5 times smaller  than that in \citet{2020A&A...641A.112L}.

One possible reason for this discrepancy is the difference in the initial gas distribution or gas mass density.
The gas density in the initial gas sphere here is about 1--2 orders of magnitude lower than that in \citet{2020A&A...641A.112L}. 
When the gas density is low, dust particles tend to decouple from the gas. 
It is difficult to precisely explain why our results show less dust concentration compared with that in  \citet{2020A&A...641A.112L}. 
Other possible reasons for the discrepancy are the magnetic field strength (or mass-to-flux ratio) and the angle between the rotational angular velocity and the magnetic field. 
The inclusion of the feedback from dust may also affect the results. 

Many factors determine the dust-to-gas mass ratio, as described above.
However, the rough agreement between the two studies indicates that the decoupling of dust becomes significant when dust grains have a size of $a_{\rm d}\gtrsim 10$--$100$\,$\mu$m.

\subsection{Effects of dust properties on non-ideal MHD resistivities}
\label{subsec:eta}
Some previous researches have studied the effects of dust properties on the star formation process in terms of non-ideal MHD effects \citep{2016A&A...592A..18M,2016MNRAS.460.2050Z,2018MNRAS.478.2723Z,2021MNRAS.505.5142Z,2017A&A...603A.105D,2019MNRAS.484.2119K,2020ApJ...896..158T}.
These studies considered chemical networks that included gas molecules and dust grains and calculated the resistivities of the non-ideal MHD effects. 
In addition, although they discussed the influences of the dust properties on the star formation process using numerical simulations and analytical calculations, they did not consider the relative velocity between the gas and dust.
The present study showed that dust grains with a size of $a_{\rm d} \le 1 \,{\rm \mu m}$, which cover the MRN size distribution, are coupled with gas during star formation.
Thus, the relative velocity would not significantly change the resistivities as long as the dust size is   $\lesssim 1 \,{\rm \mu m}$.

Using one-dimensional gas evolution calculations that included ambipolar diffusion and turbulence, \citet{2020A&A...643A..17G} recently showed that dust grain size can reach $10\,{\rm \mu m}$ even when the MRN size distribution  is given as the initial distribution of the dust grains.
In particular, one of the most advanced aspects of \citet{2020A&A...643A..17G} is the inclusion of charged grains (also see \S~\ref{subsec:dustgrowth}).
The dust growth can occur efficiently in particularly low density gas region when considering the existence of charged dust grains and ambipolar diffusion (for details, see the Fig.8 of \citealt{2020A&A...643A..17G}). 
They pointed out that the resistivities change significantly. 
Our study indicates that the maximum dust grain size coupled with the gas is $10\,{\rm \mu m}$.
Thus, dust clustering in a high density gas region could change the resistivities.
However, \citet{2020A&A...643A..17G} did not include the porosity and fragmentation processes of dust, which affect the relation between dust and the magnetic field (also see the Sec. 5.1 in \citealt{2020A&A...643A..17G}). 
Further refitment treatment of dust grains is necessary to correctly understand the magnetic resistivities.

\subsection{Charged dust dynamics}
\label{subsec:chargedust}
\citet{2021ApJ...913..148T} investigated the behavior and treatment of charged dust in  star formation simulations.
Here, we assumed that dust is electrically neutral.
However, in the gas density and temperature ranges during the star formation process, dust grains of sub-${\rm \mu\,m}$ size are mostly negatively charged (for details, see \citealt{1987ApJ...320..803D}).
Therefore, to treat the dust dynamics more realistically, dust charge should  be considered. 

\citet{2021ApJ...913..148T} discussed the difficulty of simulating the evolution of charged dust, especially in numerical simulations with a two-fluid approximation of dust and a gas fluid.
Although we did not use a two-fluid approximation, simulations were still difficult.
Fig.~11 of \citet{2021ApJ...913..148T} compares the magnitude of electric current generated by charged dust motion with that generated by gas particles (e.g., electrons).
The figure shows that the current generated by charged dust dominates that generated by the gas (or electrons) when the charged dust grain size is in the range of  $\leq 1 {\rm \mu\,m}$ and the dust is embedded in a region with low density and a strong magnetic field (e.g., outflow region).

In this calculation, the current generated by charged dust $\bm{J}_{\rm d}$ cannot be simply defined using $\rho_{\rm d}$ because the dust grains  are treated as discrete particles.
However,  $\bm{J}_{\rm d}$ may be dominant, especially in the outflow region. 
Thus,  the assumption that the total current $\bm{J} = \bm{J}_{\rm d} + \bm{J}_{\rm g} \approx \bm{J}_{\rm g}$ (where $\bm{J}_{\rm g}$ is the current generated by the charged gas) does not hold. 
This means that the spatial and time evolution of the magnetic field cannot be precisely calculated with the induction equation of the gas fluid (equation~(\ref{eq:inductioneq})).
We avoid these problems in this study because dust is assumed to be electrically neutral.  
However, we need to carefully consider the treatment of charged dust in future studies.

\subsection{Dust growth in star formation process}
\label{subsec:dustgrowth}
As mentioned in \S\ref{sec:intro}, dust growth via collisions should be considered for dust dynamics. 
Currently, dust growth in the star formation process is mainly calculated using one-dimensional gas evolution calculations \citep[][]{2020A&A...643A..17G,2020A&A...641A..39S}.

\citet{2021ApJ...920L..35T} included the dust growth process in three-dimensional MHD simulations. They calculated the dust size evolution with a single-size approximation and thus the dust size distribution was ignored. 
Although the evolution of grain size used in \citet{2021ApJ...920L..35T} may be valid, the evolution of the size distribution is required for calculating chemical reactions and porosity, which significantly affect the resistivities of non-ideal MHD effects and the dust growth related to planet formation. In addition, the dust size distribution and porosity can affect dust opacity (for details, see \citealt[][]{2009A&A...502..845O,2011A&A...532A..43O}).

Very recently, for investigating dust growth, \citet{2021A&A...649A..50M} proposed a method for calculating the dust size distribution using the Lagrangian history (or trajectory) of individual dust particle.
However, their method considers only the collision rate for the gas turbulence proposed by \citet[][]{2007A&A...466..413O}. 
In their method, the coupling of dust particles with gas eddies is determined based on dust grain size and the relative velocity is obtained from collisions of grains with different sizes.
In future work, we plan to expand the method of \citet{2021A&A...649A..50M} so that we can treat other relative motions of dust produced by differences in dust grain size and various  processes, such as fragmentation due  to high-velocity collisions.

\section{Summary}
\label{sec:conclusion}
In this study, we proposed a method for calculating the trajectories of dust particles and implemented it in previously developed nested grid code in which the local gas physical quantities of the gas fluid are used to calculate dust dynamics and dust grains are treated as  Lagrangian particles. 
We performed a three-dimensional MHD simulation that included the trajectory calculation of dust particles, and investigated dust dynamics in a collapsing cloud with different-sized dust grains. 
We confirmed that our results are qualitatively and quantitatively consistent with previous studies that adopted one- or two-fluid approximation with the Eulerian approach.
 
We found that dust grains that satisfy $a_{\rm d} \leq 10 \,{\rm \mu m}$ are coupled with the gas during the gravitational collapse at least until the protostellar mass  reaches about 8 per cent of the initial cloud core mass.
This coupling condition is consistent with previous studies.
We showed that the trajectory calculation adopted in this study is appropriate for tracing dust dynamics in the star formation process.

Some dust grains are swept up by the gas outflow.
The dust grains initially located in the range of $0^\circ \leq \theta \leq 45^\circ$ are preferentially ejected by the outflow.
Those initially located in the range of $60^\circ \leq \theta \leq 90^\circ$ fall onto the disk and move within it; they can grow into planetesimals via collisions.
Unlike small grains ($a_{\rm d} \le 100\,{\rm \mu m}$ ), dust particles with a size of $a_{\rm d} \geq 100\,{\rm \mu m}$ are decoupled from the gas fluid.
Dust grains with a size of $1000 \,{\rm \mu m}$ are hardly rolled up by the outflow and are likely to fall onto the disk. 
This phenomenon was confirmed with the time evolution of $\delta f_{\rm dg}$, which is the enhancement factor of the dust-to-gas ratio normalized by the initial value $f_{\rm dg}$.
The $\delta f_{\rm dg}$ value for dust with a size of $a_{\rm d} = 1000 \,{\rm \mu m}$ significantly increases with time in a high-density-gas region, indicating a significant increase of the dust to gas mass ratio.

Dust grains with a size of $a_{\rm d} \geq 100\,{\rm \mu m}$ are decoupled from the  gas when moving in low density gas regions (i.e., envelope and outflow).
In the rotationally supported disk, even dust grains with a size of $a_{\rm d}=1000\,{\rm \mu m}$, the largest dust grain size adopted in this study, are coupled with the gas because of the high gas density. 

This study mainly focused on the implementation of the calculation method of Lagrangian dust particles. Our results were compared with previous studies to validate our method. 
We will show  the dust trajectories within rotationally supported disks obtained using three-dimensional MHD simulations in the next paper.

\section*{Acknowledgements} 
We thank the referee for very useful comments and suggestions on this paper. 
This work was supported by the Japan Society for the Promotion of Science KAKENHI (grant numbers JP20J12062: SK,  JP17H06360, JP17K05387, JP17KK0096, JP21H00046, JP21K03617: MNM).
 This research used the computational resources of the High-Performance Computing Infrastructure (HPCI) system provided by the CyberScience Center at Tohoku University, the Cybermedia Center at Osaka University, and the Earth Simulator at JAMSTEC through the HPCI System Research Project (project IDs hp190035, hp200004, hp210004). 
The simulations reported in this paper were also performed by 2020 and 2021 Koubo Kadai on the Earth Simulator (NEC SX-ACE and NEC SX-Aurora TSUBASA) at JAMSTEC.

\section*{Data Availability}
The data underlying this article are available in the article and in its online supplementary material.




\bibliographystyle{mnras}
\bibliography{koga} 

\appendix
\section{Derivation of equation of motion for dust}
\label{sec:appendix}
The equations of motion for gas fluid and dust particle in both the Lagrangian and Eulerian formulations are described as
\begin{eqnarray}
	\rho_{\rm g} \frac{\mathrm{d}_{\rm g} \boldsymbol{v}_{\rm g}}{\mathrm{\rm d} t}
	&=& \rho_{\rm g} \left( \frac{\partial \boldsymbol{v}_{\rm g}}{\partial t} + \boldsymbol{v}_{\rm g}\cdot\nabla \boldsymbol{v}_{\rm g}\right)
	= \rho_{\rm d} \frac{\boldsymbol{v}_{\rm d} - \boldsymbol{v}_{\rm g} }{t_{\rm s}} + \rho_{\rm g} \left( \boldsymbol{f}_{\rm g, p} + \boldsymbol{f}_{\rm g, em} + \boldsymbol{f}_{\rm grav} \right),
\label{eq:eom-gas} \\
	\rho_{\rm d} \frac{\mathrm{d}_{\rm d} \boldsymbol{v}_{\rm d}}{\mathrm{d} t}
	&=& \rho_{\rm d} \left( \frac{\partial \boldsymbol{v}_{\rm d}}{\partial t} + \boldsymbol{v}_{\rm d}\cdot\nabla \boldsymbol{v}_{\rm d}\right)
	= -\rho_{\rm d}\frac{\boldsymbol{v}_{\rm d} - \boldsymbol{v}_{\rm g} }{t_{\rm s}} + \rho_{\rm d} \boldsymbol{f}_{\rm grav},
\label{eq:eom-dust}
\end{eqnarray}
where $\rho_{\rm g}$ and $\rho_{\rm d}$ are the gas and dust density,  $v_{\rm g}$ and $v_{\rm d}$ are the gas and dust velocity, $t_{\rm s}$ is the stopping time,  $\boldsymbol{f}_{\rm g, p}$ is the gas pressure gradient force, $\boldsymbol{f}_{\rm g, em}$  and $\boldsymbol{f}_{\rm grav}$ are the Lorentz and gravitational force. 
The time derivative of $d_{\rm g}/dt$  and $d_{\rm d}/dt$ means the Lagrangian derivative of gas and dust, while $\partial/\partial t$ represents the Eulerian derivative. 
Since we only consider electrically neutral dust grains in this study, the Lorentz force term is ignored in equation~(\ref{eq:eom-dust}).
Then, we define the barycentric velocity $\boldsymbol{v}$ as
\begin{equation}
	\boldsymbol{v} = \frac{\rho_{\rm g} \boldsymbol{v}_{\rm g} + \rho_{\rm d} \boldsymbol{v}_{\rm d}}{\rho_{\rm g} + \rho_{\rm d}}
	= (1 - \epsilon) \boldsymbol{v}_{\rm g} + \epsilon \boldsymbol{v}_{\rm d},
\label{eq:barycentric}
\end{equation}
where $\epsilon = \rho_{\rm d}/(\rho_{\rm g} + \rho_{\rm d})$. 
We also define the relative velocity $\Delta \boldsymbol{v}$ as 
\begin{equation}
	\Delta \boldsymbol{v} = \boldsymbol{v}_{\rm d} - \boldsymbol{v}_{\rm g}.
\label{eq:dv}
\end{equation}

Using the barycentric velocity $\boldsymbol{v}$ and the relative velocity $\Delta \boldsymbol{v}$, the equations of motion (eqs.~(\ref{eq:eom-gas}) and (\ref{eq:eom-dust}))  can be transformed into \citep{2005ApJ...620..459Y, 2014MNRAS.440.2147L}
\begin{equation}
	\frac{\mathrm{d} \Delta \boldsymbol{v}}{\mathrm{d} t}
	= \frac{\partial \Delta\boldsymbol{v}}{\partial t} + \boldsymbol{v}\cdot\nabla \Delta\boldsymbol{v}
	= - \frac{\Delta\boldsymbol{v}}{t_{\rm s}} - \boldsymbol{f}_{\rm g,p} - \boldsymbol{f}_{\rm g, em}
	- \Delta\boldsymbol{v}\cdot\nabla \boldsymbol{v} + \mathcal{O}\left( \left(\Delta\boldsymbol{v}\right)^{2} \right).
\label{eq:eom-dv}
\end{equation}
In equation~(\ref{eq:eom-dv}), the Lagrangian derivative ($\mathrm{d}/\mathrm{d} t$) is differentiated along the barycentric velocity.
Considering a small dust-to-gas mass ratio (i.e. $\epsilon \ll 1$),  we can approximate the barycentric velocity as $\boldsymbol{v} \simeq \boldsymbol{v}_{\rm g}$ (see, eq.~(\ref{eq:barycentric})). 
Therefore, the barycentric frame can be approximated by the gas frame, and equation~(\ref{eq:eom-dv})  can be written as  
\begin{equation}
	\frac{\mathrm{d}_{\rm g} \Delta \boldsymbol{v}}{\mathrm{d} t}
	= \frac{\partial \Delta\boldsymbol{v}}{\partial t} + \boldsymbol{v}_{\rm g} \cdot\nabla \Delta\boldsymbol{v}
	= - \frac{\Delta\boldsymbol{v}}{t_{\rm s}} - \boldsymbol{f}_{\rm g,p} - \boldsymbol{f}_{\rm g, em}
	- \Delta\boldsymbol{v}\cdot\nabla \boldsymbol{v}_{\rm g} + \mathcal{O}\left( \left(\Delta\boldsymbol{v}\right)^{2} \right).
\label{eq:eom-dvg}
\end{equation}
Note that, in this study, since the dust-to-gas mass ratio was $\sim 0.03$ at the maximum during the calculation (see, \S\ref{sec:results}), and thus the condition $\epsilon \ll 1$ is fulfilled.

To further simplify equation~(\ref{eq:eom-dvg}), we use the Stoke number that is defined as ${\rm St} = t_{\rm s}/ t_{\rm dyn}$, where $t_{\rm dyn}$ is the dynamical timescale. 
We can ignore the terms of ($\Delta\boldsymbol{v}\cdot\nabla\boldsymbol{v}_{\rm g}$) and $\mathcal{O}\left((\Delta\boldsymbol{v})^{2}\right)$ in equation~(\ref{eq:eom-dvg}) with  ${\rm St} \ll 1$ because of $|(\Delta\boldsymbol{v}\cdot\nabla\boldsymbol{v})/(\Delta\boldsymbol{v}/t_{s})| = \mathcal{O}({\rm St})$ and $|\Delta\boldsymbol{v}^{2}/\boldsymbol{v}^{2}| = \mathcal{O}({\rm St}^2)$
\citep{2014MNRAS.440.2147L, 2020A&A...641A.112L}. 

\begin{figure}
  \includegraphics[width=\linewidth]{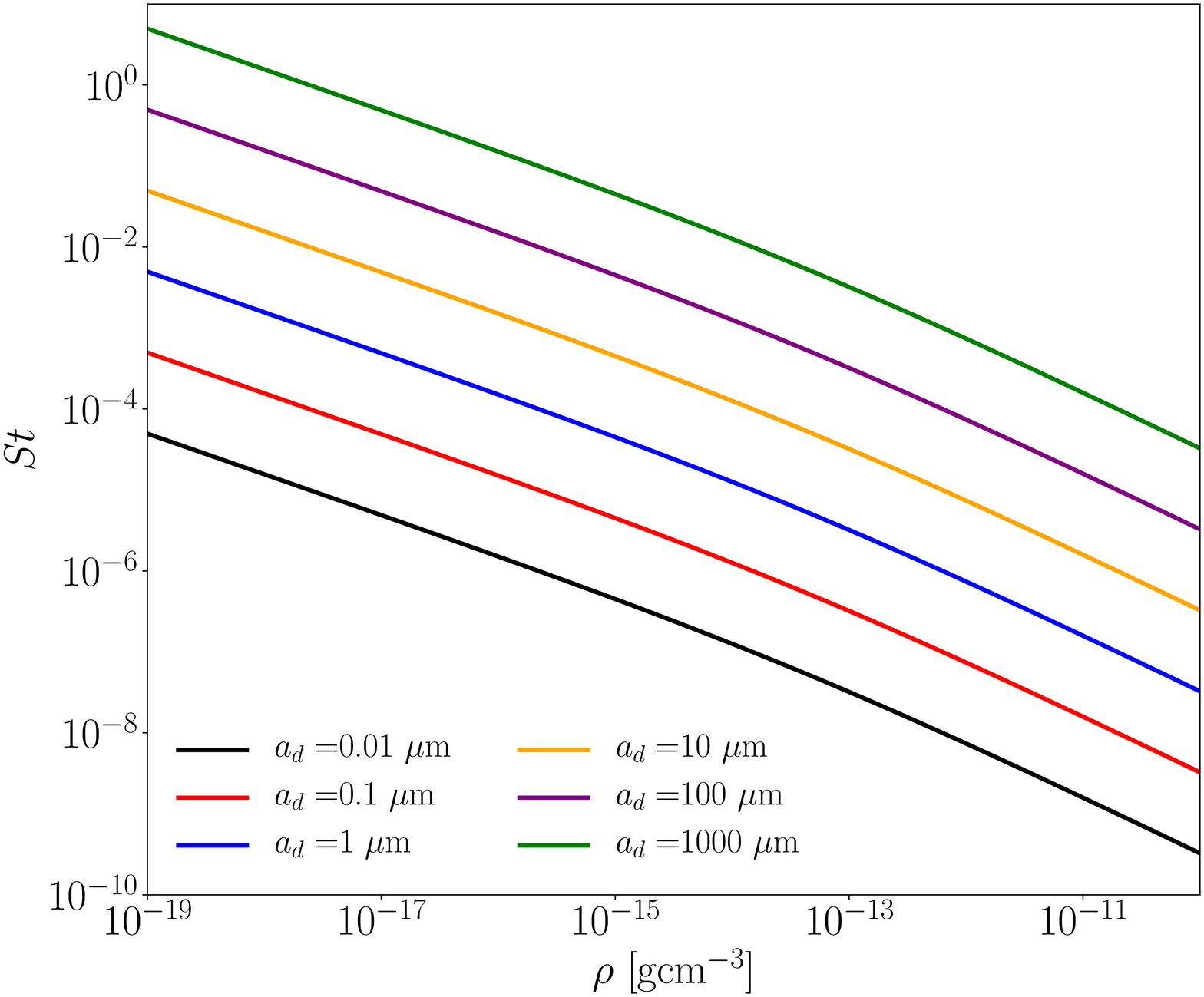}
  \caption{Stokes number against gas density with different dust sizes $a_{\rm d}$}.
  \label{fig:stokos_rho}
\end{figure}

Figure~~\ref{fig:stokos_rho} plots the Stokes number for dust grains with different sizes ($a_{\rm d}=0.01-1000$\,$\mu$m) against the gas density, in which the freefall timescale $t_{\rm ff}$ ($=(3\pi/32G\rho)^{1/2}$) is adopted as the dynamical timescale (i.e., $t_{\rm dyn}= t_{\rm ff}$). 
The figure shows that the Stokes number is less than unity (${\rm St}<1$) in all density range as long as the dust size is smaller than $a_{\rm d}\le 100$\,$\mu$m. 
When the dust size is $a_{\rm d}=1000$\,$\mu$m, the Stokes number exceeds unity only in the range of $\rho \lesssim 10^{-18}$\,g\,cm$^{-3}$.   
However, ${\rm St}<1$ is realized in the range of $\rho \gtrsim 10^{-18}$\,g\,cm$^{-3}$ even with $a_{\rm d}=1000$\,$\mu$m.  
Although we require attention for the dust with $a_{\rm d}=1000$\,$\mu$m especially in the low density region of $\rho \lesssim 10^{-18}$\,g\,cm$^{-3}$, we can ignore the two terms of ($\Delta\boldsymbol{v}\cdot\nabla\boldsymbol{v}_{\rm g}$) and $\mathcal{O}\left((\Delta\boldsymbol{v})^{2}\right)$ in equation~(\ref{eq:eom-dvg}) in the high density region. 
Therefore, the equation of the relative velocity can be approximated as  
\begin{equation}
\frac{\mathrm{d} \Delta \boldsymbol{v}}{\mathrm{d} t}
	= - \frac{\Delta\boldsymbol{v}}{t_{\rm s}} - \boldsymbol{f}_{\rm g,p} - \boldsymbol{f}_{\rm g, em}. 
\label{eq:eom-dv2}
\end{equation}

Next, we consider the equation of motion for dust. 
Using equations~(\ref{eq:eom-dust})--(\ref{eq:dv}), the Lagrangian derivative of dust can be transformed into  
\begin{equation}
	\frac{\mathrm{d}_{\rm d} \boldsymbol{v}_{\rm d}}{\mathrm{d} t} 
	= \frac{\mathrm{d} \boldsymbol{v}_{\rm d}}{\mathrm{d} t} + \left( 1- \epsilon \right) \Delta\boldsymbol{v}\cdot \Delta \boldsymbol{v}_{\rm d}
	\simeq  \frac{\mathrm{d} \boldsymbol{v}_{\rm d}}{\mathrm{d} t} + \Delta\boldsymbol{v}\cdot\nabla \boldsymbol{v}_{\rm d}. 
\label{eq:lag-dust}
\end{equation}
In the rightmost side of equation~(\ref{eq:lag-dust}), the term of ($\epsilon \Delta\boldsymbol{v}\cdot \Delta$) is neglected with $\epsilon \ll 1$.
Thus, using the ralative velocity $ \Delta \boldsymbol{v} $, the equation of motion for dust particles in the Lagrangian formulation can be described as 
\begin{equation}
	\frac{\mathrm{d} \boldsymbol{v}_{\rm d}}{\mathrm{d} t}
	= -\frac{\Delta\boldsymbol{v}}{t_{\rm s}}  - \Delta\boldsymbol{v}\cdot\nabla \boldsymbol{v}_{\rm d} + \boldsymbol{f}_{\rm grav}. 
\label{eq:eom-dust1}
\end{equation}
In the right hand side of equation~(\ref{eq:eom-dust1}), we compare the second term ($\Delta\boldsymbol{v}\cdot\nabla \boldsymbol{v}_{d} $) with the first term ($\Delta\boldsymbol{v}/t_{\rm s}$) as 
\begin{equation}
	\mathcal{O}\left(\frac{ \Delta\boldsymbol{v}\cdot\nabla \boldsymbol{v}_{\rm d} }{\Delta\boldsymbol{v}/t_{\rm s}} \right)
	= \mathcal{O}\left( \frac{|\Delta\boldsymbol{v}|\boldsymbol{v}_{\rm d}/L}{|\Delta\boldsymbol{v}|/t_{\rm s}} \right)
	= \mathcal{O}\left( t_{\rm s}\frac{|\boldsymbol{v}_{\rm d}|}{L} \right)
	= \mathcal{O}\left( \frac{t_{\rm s}}{t_{\rm dyn}}\right)
	= \mathcal{O}\left({\rm St}\right).
\end{equation}
Thus, we can ignore the term ($\Delta\boldsymbol{v}\cdot\nabla \boldsymbol{v}_{\rm d}$) in equation~(\ref{eq:eom-dust1}) when ${\rm St} \ll 1$ is hold. 
As a result, the equation of motion for dust particle is described as
\begin{equation}
 \frac{\mathrm{d} \boldsymbol{v}_{\rm d}}{\mathrm{d} t} = -\frac{\Delta\boldsymbol{v}}{t_{\rm s}} + \boldsymbol{f}_{\rm grav}.
\label{eq:eom-dust2} 
\end{equation}

\bsp	
\label{lastpage}
\end{document}